\pdfoutput=1
\documentclass[12pt,a4paper]{article}
\usepackage{ifthen} % for conditional statements
\newboolean{pdflatex}
\setboolean{pdflatex}{true} % False for eps figures 

\newboolean{articletitles}
\setboolean{articletitles}{true} % False removes titles in references

\newboolean{uprightparticles}
\setboolean{uprightparticles}{false} %True for upright particle symbols

\def\paperauthors{LHCb collaboration} % Leave as is for PAPER, CONF and FIGURE
\def\paperasciititle{} % Set ASCII title here !! MAKE sure it's only ASCII characters !! 
\def\papertitle{Study of charmonium decays to $K^0_{\rm S} K \pi$ in the $B \to (K^0_{\rm S} K \pi) K$ channels} % Latex formatted title
\def\paperkeywords{{High Energy Physics}, {LHCb}} % Comma separated list
\def\papercopyright{\the\year\ CERN for the benefit of the LHCb collaboration} % new since 9/Apr/2018
\def\paperlicence{CC BY 4.0 licence}
\def\paperlicenceurl{https://creativecommons.org/licenses/by/4.0/}

\usepackage[top=1in, bottom=1.25in, left=1in, right=1in]{geometry}

\usepackage{microtype}
\usepackage{lineno}  % for line numbering during review
\usepackage{xspace} % To avoid problems with missing or double spaces after
\usepackage{caption} %these three command get the figure and table captions automatically small

\usepackage{graphicx}  % to include figures (can also use other packages)
\usepackage{color}
\usepackage{colortbl}
\graphicspath{{./figs/}} % Make Latex search fig subdir for figures
\usepackage{amsmath} % Adds a large collection of math symbols
\usepackage{amssymb}
\usepackage{amsfonts}
\usepackage{upgreek} % Adds in support for greek letters in roman typeset

\newcommand*\patchAmsMathEnvironmentForLineno[1]{%
\expandafter\let\csname old#1\expandafter\endcsname\csname #1\endcsname
\expandafter\let\csname oldend#1\expandafter\endcsname\csname
end#1\endcsname
 \renewenvironment{#1}%
   {\linenomath\csname old#1\endcsname}%
   {\csname oldend#1\endcsname\endlinenomath}%
}
\newcommand*\patchBothAmsMathEnvironmentsForLineno[1]{%
  \patchAmsMathEnvironmentForLineno{#1}%
  \patchAmsMathEnvironmentForLineno{#1*}%
}
\AtBeginDocument{%
\patchBothAmsMathEnvironmentsForLineno{equation}%
\patchBothAmsMathEnvironmentsForLineno{align}%
\patchBothAmsMathEnvironmentsForLineno{flalign}%
\patchBothAmsMathEnvironmentsForLineno{alignat}%
\patchBothAmsMathEnvironmentsForLineno{gather}%
\patchBothAmsMathEnvironmentsForLineno{multline}%
\patchBothAmsMathEnvironmentsForLineno{eqnarray}%
}

\usepackage{hyperxmp}

\usepackage[pdftex,
            pdfauthor={\paperauthors},
            pdftitle={\paperasciititle},
            pdfkeywords={\paperkeywords},
            pdfcopyright={Copyright (C) \papercopyright},
            pdflicenseurl={\paperlicenceurl}]{hyperref}
\usepackage[colorinlistoftodos,textsize=scriptsize]{todonotes}

\usepackage[bottom,flushmargin,hang,multiple]{footmisc}

\usepackage[all]{hypcap} % Internal hyperlinks to floats.

\usepackage{xspace} 
\usepackage{upgreek}

\def\lhcb   {\mbox{LHCb}\xspace}

\def\MagUp {\mbox{\em Mag\kern -0.05em Up}\xspace}

\ifthenelse{\boolean{uprightparticles}}%
{

 \def\Peta        {\ensuremath{\upeta}\xspace}

 \def\Ppi         {\ensuremath{\uppi}\xspace}

 \def\Pchi        {\ensuremath{\upchi}\xspace}                 
 \def\Ppsi        {\ensuremath{\uppsi}\xspace}

 \def\PDelta      {\ensuremath{\Delta}\xspace}                 
 \def\PXi         {\ensuremath{\Xi}\xspace}                 
 \def\PLambda     {\ensuremath{\Lambda}\xspace}                 
 \def\PSigma      {\ensuremath{\Sigma}\xspace}                 
 \def\POmega      {\ensuremath{\Omega}\xspace}                 
 \def\PUpsilon    {\ensuremath{\Upsilon}\xspace}
 \let\oldPi\Pi
 \def\PPi         {\ensuremath{\oldPi}\xspace}

 \def\PB      {\ensuremath{\mathrm{B}}\xspace}                 
                  
 \def\PD      {\ensuremath{\mathrm{D}}\xspace}

 \def\PJ      {\ensuremath{\mathrm{J}}\xspace}                 
 \def\PK      {\ensuremath{\mathrm{K}}\xspace}

 \def\Pb      {\ensuremath{\mathrm{b}}\xspace}                 
 \def\Pc      {\ensuremath{\mathrm{c}}\xspace}

 \def\Pi      {\ensuremath{\mathrm{i}}\xspace}

 \def\Ps      {\ensuremath{\mathrm{s}}\xspace}

 \def\thebaroffset{0.0em}
}
{

 \def\Peta        {\ensuremath{\eta}\xspace}

 \def\Ppi         {\ensuremath{\pi}\xspace}

 \def\Pchi        {\ensuremath{\chi}\xspace}                 
 \def\Ppsi        {\ensuremath{\psi}\xspace}                 
                  
 \mathchardef\PDelta="7101
 \mathchardef\PXi="7104
 \mathchardef\PLambda="7103
 \mathchardef\PSigma="7106
 \mathchardef\POmega="710A
 \mathchardef\PUpsilon="7107
 \mathchardef\PPi="7105
                  
 \def\PB      {\ensuremath{B}\xspace}                 
                  
 \def\PD      {\ensuremath{D}\xspace}

 \def\PJ      {\ensuremath{J}\xspace}                 
 \def\PK      {\ensuremath{K}\xspace}

 \def\Pb      {\ensuremath{b}\xspace}                 
 \def\Pc      {\ensuremath{c}\xspace}

 \def\Pi      {\ensuremath{i}\xspace}

 \def\Ps      {\ensuremath{s}\xspace}

 \def\thebaroffset{0.18em}
}
\newcommand{\offsetoverline}[2][\thebaroffset]{\kern #1\overline{\kern -#1 #2}}%

\makeatletter
\ifcase \@ptsize \relax% 10pt
  \newcommand{\miniscule}{\@setfontsize\miniscule{4}{5}}% \tiny: 5/6
\or% 11pt
  \newcommand{\miniscule}{\@setfontsize\miniscule{5}{6}}% \tiny: 6/7
\or% 12pt
  \newcommand{\miniscule}{\@setfontsize\miniscule{5}{6}}% \tiny: 6/7
\fi
\makeatother

\DeclareRobustCommand{\optbar}[1]{\shortstack{{\miniscule (\rule[.5ex]{1.25em}{.18mm})}
  \\ [-.7ex] $#1$}}

\def\squark    {{\ensuremath{\Ps}}\xspace}

\def\cquark    {{\ensuremath{\Pc}}\xspace}

\def\bquark    {{\ensuremath{\Pb}}\xspace}

\def\pion   {{\ensuremath{\Ppi}}\xspace}

\def\pip    {{\ensuremath{\pion^+}}\xspace}
\def\pim    {{\ensuremath{\pion^-}}\xspace}

\def\kaon    {{\ensuremath{\PK}}\xspace}
\def\Kbar    {{\ensuremath{\offsetoverline{\PK}}}\xspace}
\def\Kb      {{\ensuremath{\Kbar}}\xspace}
\def\KorKbar {\kern \thebaroffset\optbar{\kern -\thebaroffset \PK}{}\xspace}
\def\Kz      {{\ensuremath{\kaon^0}}\xspace}

\def\Kp      {{\ensuremath{\kaon^+}}\xspace}
\def\Km      {{\ensuremath{\kaon^-}}\xspace}

\def\KS      {{\ensuremath{\kaon^0_{\mathrm{S}}}}\xspace}

\newcommand{\etapr}{\ensuremath{\Peta^{\prime}}\xspace}

\def\Dbar    {{\ensuremath{\offsetoverline{\PD}}}\xspace}
\def\D       {{\ensuremath{\PD}}\xspace}

\def\DorDbar {\kern \thebaroffset\optbar{\kern -\thebaroffset \PD}\xspace}
\def\Dz      {{\ensuremath{\D^0}}\xspace}
\def\Dzb     {{\ensuremath{\Dbar{}^0}}\xspace}
\def\Dp      {{\ensuremath{\D^+}}\xspace}
\def\Dm      {{\ensuremath{\D^-}}\xspace}

\def\DpDm    {\ensuremath{\Dp {\kern -0.16em \Dm}}\xspace}

\def\Ds      {{\ensuremath{\D^+_\squark}}\xspace}
\def\Dsp     {{\ensuremath{\D^+_\squark}}\xspace}
\def\Dsm     {{\ensuremath{\D^-_\squark}}\xspace}

\def\B       {{\ensuremath{\PB}}\xspace}

\def\BorBbar {\kern \thebaroffset\optbar{\kern -\thebaroffset \PB}\xspace}

\def\Bd      {{\ensuremath{\B^0}}\xspace}

\def\BdorBdbar {\kern \thebaroffset\optbar{\kern -\thebaroffset \Bd}\xspace}
\def\Bu      {{\ensuremath{\B^+}}\xspace}

\def\Bp      {{\ensuremath{\Bu}}\xspace}

\def\Bs      {{\ensuremath{\B^0_\squark}}\xspace}

\def\BsorBsbar {\kern \thebaroffset\optbar{\kern -\thebaroffset \Bs}\xspace}

\def\jpsi     {{\ensuremath{{\PJ\mskip -3mu/\mskip -2mu\Ppsi}}}\xspace}

\def\etac     {{\ensuremath{\Peta_\cquark}}\xspace}

\def\chiczero {{\ensuremath{\Pchi_{\cquark 0}}}\xspace}
\def\chicone  {{\ensuremath{\Pchi_{\cquark 1}}}\xspace}
\def\chictwo  {{\ensuremath{\Pchi_{\cquark 2}}}\xspace}

\def\Y#1S{\ensuremath{\PUpsilon{(#1S)}}\xspace}

\def\LorLbar     {\kern \thebaroffset\optbar{\kern -\thebaroffset \PLambda}\xspace}

\newcommand{\decay}[2]{\ensuremath{#1\!\to #2}\xspace} 

\def\to                 {\ensuremath{\rightarrow}\xspace}

\def\eps   {{\ensuremath{\varepsilon}}\xspace}

\def\AT#1     {\ensuremath{A_{\mathrm{T}}^{#1}}\xspace}           % 2

\def\C#1      {\ensuremath{\mathcal{C}_{#1}}\xspace}                       % 9
\def\Cp#1     {\ensuremath{\mathcal{C}_{#1}^{'}}\xspace}                    % 7
\def\Ceff#1   {\ensuremath{\mathcal{C}_{#1}^{\mathrm{(eff)}}}\xspace}        % 9  
\def\Cpeff#1  {\ensuremath{\mathcal{C}_{#1}^{'\mathrm{(eff)}}}\xspace}       % 7
\def\Ope#1    {\ensuremath{\mathcal{O}_{#1}}\xspace}                       % 2
\def\Opep#1   {\ensuremath{\mathcal{O}_{#1}^{'}}\xspace}                    % 7

\newcommand{\aunit}[1]{\ensuremath{\text{\,#1}}}       
\newcommand{\tev}{\aunit{Te\kern -0.1em V}\xspace}
\newcommand{\gev}{\aunit{Ge\kern -0.1em V}\xspace}
\newcommand{\mev}{\aunit{Me\kern -0.1em V}\xspace}
\newcommand{\kev}{\aunit{ke\kern -0.1em V}\xspace}
\newcommand{\ev}{\aunit{e\kern -0.1em V}\xspace}
 
\newcommand{\mevc}{\ensuremath{\aunit{Me\kern -0.1em V\!/}c}\xspace}
\newcommand{\gevc}{\ensuremath{\aunit{Ge\kern -0.1em V\!/}c}\xspace}
\newcommand{\mevcc}{\ensuremath{\aunit{Me\kern -0.1em V\!/}c^2}\xspace}
\newcommand{\gevcc}{\ensuremath{\aunit{Ge\kern -0.1em V\!/}c^2}\xspace}
\newcommand{\gevgevcccc}{\ensuremath{\gev^2\!/c^4}\xspace} % for q^2

\def\fb   {\ensuremath{\aunit{fb}}\xspace}
\def\invfb   {\ensuremath{\fb^{-1}}\xspace}

\newcommand{\chisqndf}{\ensuremath{\chi^2/\mathrm{ndf}}\xspace}

\def\gsim{{~\raise.15em\hbox{$>$}\kern-.85em
          \lower.35em\hbox{$\sim$}~}\xspace}
\def\lsim{{~\raise.15em\hbox{$<$}\kern-.85em
          \lower.35em\hbox{$\sim$}~}\xspace}

\def\pt         {\ensuremath{p_{\mathrm{T}}}\xspace}

\def\evtgen     {\mbox{\textsc{EvtGen}}\xspace}

\def\geant      {\mbox{\textsc{Geant4}}\xspace}

\def\photos     {\mbox{\textsc{Photos}}\xspace}

\def\pythia     {\mbox{\textsc{Pythia}}\xspace}

\def\tell1  {TELL1\xspace}
\def\ukl1   {UKL1\xspace}

\newcommand{\ie}{\mbox{\itshape i.e.}\xspace}

\newcommand{\lhcborcid}[1]{\href{https://orcid.org/#1}{\hspace*{0.1em}\raisebox{-0.45ex}{\includegraphics[width=1em]{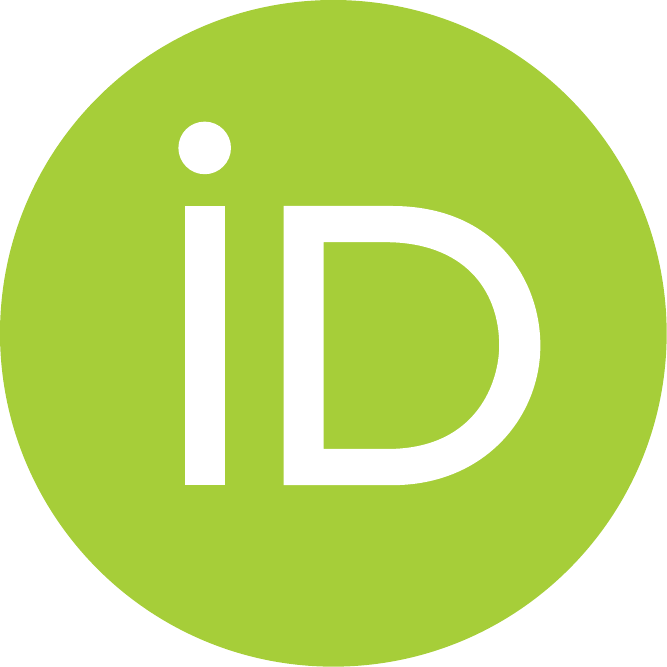}}}}

\usepackage{cite} % Allows for ranges in citations
\usepackage{mciteplus}
\usepackage{lscape}
\usepackage{longtable} % only for template; not usually to be used in PAPERs
\def\calL         {{\ensuremath{\cal L}\xspace}}
\def\KSLL      {{\ensuremath{\kaon^0_{\mathrm{SLL}}}}\xspace}
\def\KSDD      {{\ensuremath{\kaon^0_{\mathrm{SDD}}}}\xspace}
\def\calB         {{\ensuremath{\cal B}\xspace}}
\def\calR         {{\ensuremath{\cal R}\xspace}}
\def\kskkpi {{\ensuremath{\KS\Kp\Km\pi}}\xspace}
\def\kskkpip {{\ensuremath{\KS\Kp\Km\pip}}\xspace}
\def\kskkpim {{\ensuremath{\KS\Kp\Kp\pim}}\xspace}
\def\bkskkpip {{\ensuremath{B^+\to\KS\Kp\Km\pip}}\xspace}
\def\bkskkpim {{\ensuremath{B^+\to\KS\Kp\Kp\pim}}\xspace}

\def\bkzkkpip {{\ensuremath{B^+\to\Kz\Kp\Km\pip}}\xspace}
\def\bkzkkpim {{\ensuremath{B^+\to\Kz\Kp\Kp\pim}}\xspace}
\def\bkskkpi {{\ensuremath{B\to\KS K K\pi}}\xspace}
\def\kskpi {{\ensuremath{\KS K\pi}}\xspace}

\def\etactwo {{\ensuremath{\eta_c(2S)}}\xspace}
\def\Kpisk {{\ensuremath{(K\pi)_S K}}\xspace}
\newcommand{\alp}{\ensuremath{\kern 1.0em }}
\newcommand{\al}{\ensuremath{\kern 0.5em }}
\newcommand{\all}{\ensuremath{\kern 0.25em }}
\newcommand{\allm}{\ensuremath{\kern 0.15em }}
\newcommand{\almm}{\ensuremath{\kern -1.00em }}
\newcommand{\aln}{\ensuremath{\kern -0.25em }}
\newcommand{\alm}{\ensuremath{\kern -0.50em }}
\mathchardef\myhyphen="2D
\begin{document}

%%%%%%%%%%%%%%%%%%%%%%%%%
%%%%% Title     %%%%%%%%%
%%%%%%%%%%%%%%%%%%%%%%%%%
\renewcommand{\thefootnote}{\fnsymbol{footnote}}
\setcounter{footnote}{1}

% %%%%%%% CHOOSE TITLE PAGE--------
%\onecolumn
%\input{title-LHCb-INT}
%\input{title-LHCb-ANA}
%\input{title-LHCb-CONF}
%\input{title-LHCb-FIGURE}
% ===============================================================================
% Purpose: LHCb-PAPER journal paper title page template
% Author: 
% Created on: 2010-09-25
% ===============================================================================

%%%%%%%%%%%%%%%%%%%%%%%%%
%%%%%  TITLE PAGE  %%%%%%
%%%%%%%%%%%%%%%%%%%%%%%%%
\begin{titlepage}
\pagenumbering{roman}

% Header ---------------------------------------------------
\vspace*{-1.5cm}
\centerline{\large EUROPEAN ORGANIZATION FOR NUCLEAR RESEARCH (CERN)}
\vspace*{1.5cm}
\noindent
\begin{tabular*}{\linewidth}{lc@{\extracolsep{\fill}}r@{\extracolsep{0pt}}}
\ifthenelse{\boolean{pdflatex}}% Logo format choice
{\vspace*{-1.5cm}\mbox{\!\!\!\includegraphics[width=.14\textwidth]{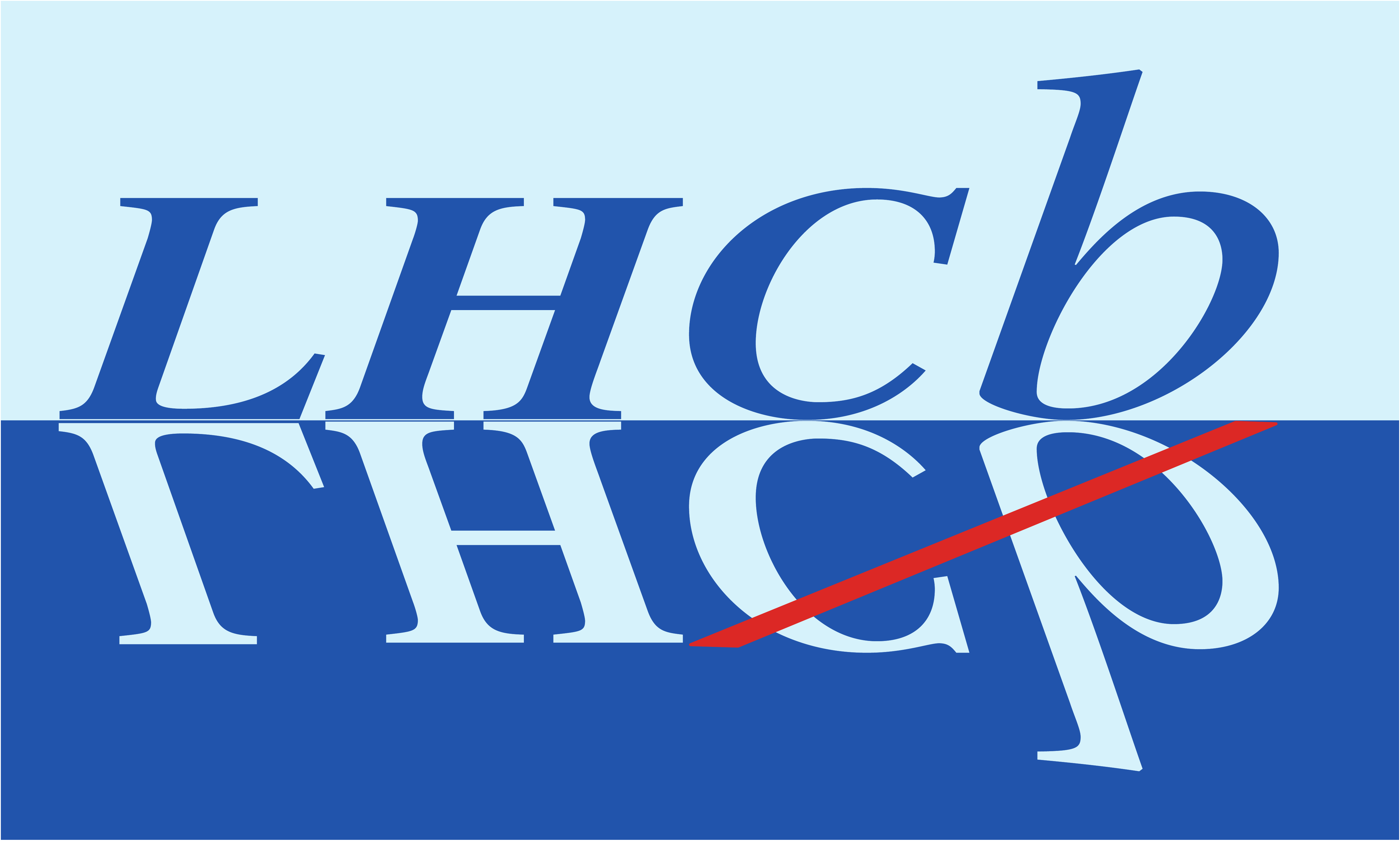}} & &}%
{\vspace*{-1.2cm}\mbox{\!\!\!\includegraphics[width=.12\textwidth]{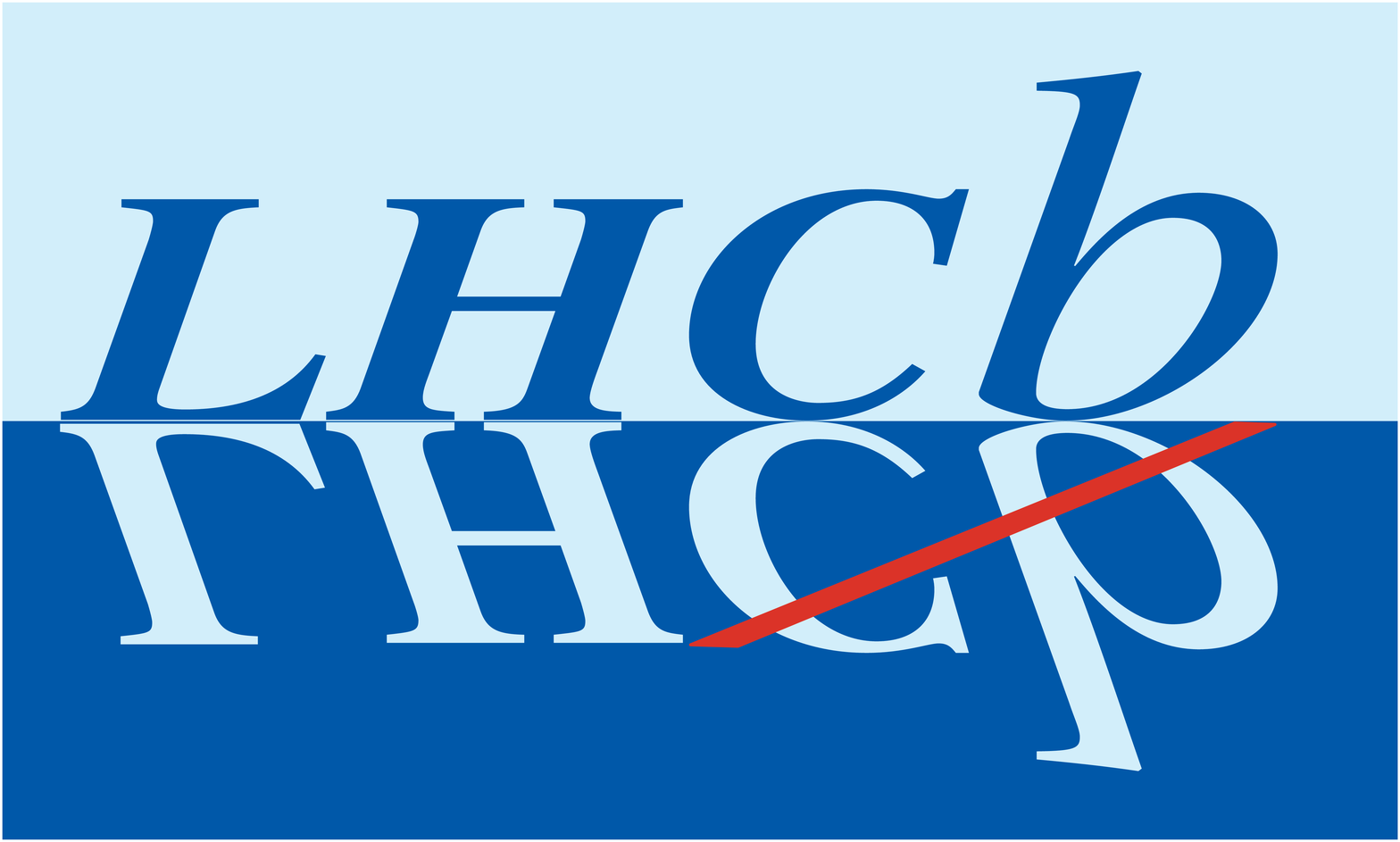}} & &}%
\\
 & & CERN-EP-2023-060 \\  % ID 
& & LHCb-PAPER-2022-051\\
% & & DRAFT-V1\\  % ID 
 & & 17 August 2023 \\ % Date - Can also hardwire e.g.: 23 March 2010
 & & \\
% not in paper \hline
\end{tabular*}

\vspace*{4.0cm}

% Title --------------------------------------------------
{\normalfont\bfseries\boldmath\huge
\begin{center}
% DO NOT EDIT HERE. Instead edit macro in main.tex to keep metadata correct
  \papertitle 
\end{center}
}

\vspace*{2.0cm}

% Authors -------------------------------------------------
\begin{center}
%In the footnote, replace 'paper' by 'Letter' in case of submission to PRL or PLB 
% Edit macro in main.tex to keep metadata correct
\paperauthors\footnote{Authors are listed at the end of this paper.}
\end{center}

\vspace{\fill}

% Abstract -----------------------------------------------
\begin{abstract}
  \noindent

A study of the \bkskkpip and \bkskkpim decays is performed using proton-proton collisions at center-of-mass energies of 7, 8 and 13\tev at the \lhcb experiment. The \kskpi invariant mass spectra from both decay modes reveal a rich content of charmonium resonances. New precise measurements of the \etac and \etactwo resonance parameters are performed and 
branching fraction measurements are obtained for \Bp decays to \etac, \jpsi, \etactwo and \chicone resonances. In particular, the first observation and branching fraction measurement of $\Bp \to \chiczero K^0 \pip$ is reported as well as first measurements of the \bkzkkpip and \bkzkkpim  branching fractions. Dalitz plot analyses of \decay{\etac}{\kskpi} and \decay{\etactwo}{\kskpi} decays are performed. A new measurement of the amplitude and phase of the $K \pi$ {\it S}-wave as functions of the $K \pi$ mass is performed, together with measurements of the $K^*_0(1430)$, $K^*_0(1950)$ and $a_0(1700)$ parameters.
Finally, the branching fractions of \chicone decays to $K^*$ resonances are also measured.
  
\end{abstract}

%\vspace*{2.0cm}

\begin{center}
  Published in Phys.Rev.D108 (2023) 032010  
\end{center}

\vspace{\fill}

{\footnotesize 
% Edit macro in main.tex to keep metadata correct
\centerline{\copyright~\papercopyright. \href{\paperlicenceurl}{\paperlicence}.}}
\vspace*{2mm}

\end{titlepage}

%%%%%%%%%%%%%%%%%%%%%%%%%%%%%%%%
%%%%%  EOD OF TITLE PAGE  %%%%%%
%%%%%%%%%%%%%%%%%%%%%%%%%%%%%%%%

%  empty page follows the title page ----
\newpage
\setcounter{page}{2}
\mbox{~}
%\newpage
%
%% Author List ----------------------------
%%  You need to get a new author list!
%\input{LHCb_authorlist.tex}
%
%The author list for journal publications is provided by the Membership Committee shortly after 'approval to go to paper' has been given.
%%It will be made available on the page
%%\verb!http://www.physik.uzh.ch/~strauman/forMemCo/LHCb-PAPER-XXXX-XXX/! .
%It will be sent to you by email shortly after a paper number has beens assigned.
%The author list should be included already at first circulation, 
%to allow new members of the collaboration to verify whether they have been included correctly.
%Occasionally a misspelled name is corrected or associated institutions become full members.
%In that case, a new author list will be sent to you.
%In case line numbering doesn't work well after including the authorlist, try moving the \verb!\bigskip! after the last author to a separate line.
%
%
%The authorship for Conference Reports should be ``The LHCb
%  collaboration'', with a footnote giving the name(s) of the contact
%  author(s), but without the full list of collaboration names.

%\twocolumn
% %%%%%%%%%%%%% ---------

\renewcommand{\thefootnote}{\arabic{footnote}}
\setcounter{footnote}{0}

%%%%%%%%%%%%%%%%%%%%%%%%%%%%%%%%
%%%%%  Table of Content   %%%%%%
%%%%%%%%%%%%%%%%%%%%%%%%%%%%%%%%
%%%% Uncomment if desired
%\tableofcontents
\cleardoublepage

%%%%%%%%%%%%%%%%%%%%%%%%%
%%%%% Main text %%%%%%%%%
%%%%%%%%%%%%%%%%%%%%%%%%%

\pagestyle{plain} % restore page numbers for the main text
\setcounter{page}{1}
\pagenumbering{arabic}

%% Uncomment during review phase. 
%% Comment before a final submission.
%\linenumbers

%% This is the main body
%% It is useful to have a single file so comemnts are not missed in overleaf.
\section{Introduction}
\label{sec:Introduction}

Understanding strong interaction effects in exclusive weak decays of heavy hadrons is of importance to gain information on 
fundamental aspects of the phenomenology of strong and weak interactions.  In two-body nonleptonic decays such as
$B \to (\bar c c) K$, a
simple factorization method has been adopted to compute nonleptonic decay amplitudes~\cite{Neubert:1997uc}.
The method involves expressing the  hadronic matrix elements of four-quark  operators in the effective Hamiltonian inducing the decay as the product
of two matrix elements of quark currents. It has been successful, such as in the description of the $ B \to \eta_c K$ and $J/\psi K$ decays~\cite{Suzuki:2002sq}. However, it clearly misses
important effects, since it fails to describe the $ B \to \chi_{c0} K$ mode.
Indeed, in this  mode the factorized amplitude involves the matrix element of the $(\bar c c)_{V,A}$ vector (V) and axial (A) currents
between the vacuum and  $\chi_{c0}$ resonance, which vanishes due to  charmed vector current conservation and parity conservation.
In contrast, the measured $ B \to \chi_{c0} K$
branching fraction is sizable~\cite{Workman:2022ynf}, clearly indicating  that the nonfactorizable part of the amplitude plays an important role~\cite{Colangelo:2002mj}.
Additional observations of new $B \to \chiczero X$ decay modes are therefore of interest.

The only established strange scalar meson is the  $K^*_0(1430)$ resonance,
whose parameters are yet to be precisely measured~\cite{Workman:2022ynf}. Scalar resonances decaying to $K \pi$ are particularly interesting, since many amplitude analyses of heavy-flavor decays involve a $K \pi$ system~\cite{LHCb:2019maw}, whose theoretical description is a source of large systematic uncertainty.
A widely used method of modelling the $K \pi$ $S$-wave relies on the results from Ref.~\cite{Aston:1987ir}, which consists of a large threshold enhancement described by a scattering length term and a relativistic Breit--Wigner (BW) function describing the $K^*_0(1430)$ resonance. A similar behaviour is observed in the 
$K \pi$ $S$-wave measured in \Dp decays~\cite{E791:2005gev,CLEO:2008jus,FOCUS:2007mcb}.
Still unresolved is the possible existence of a broad scalar resonance, $\kappa/K^*_0(700)$, claimed by several experiments~\cite{Workman:2022ynf}; its existence would suggest the possible presence of tetraquark states in the light meson system~\cite{tHooft:2008rus}.

Further information has been obtained from the Dalitz plot analysis of \etac decays to $K \Kb \pi$ with the \etac meson produced in two-photon interactions, and in an extended $K \pi$ mass region~\cite{BaBar:2015kii}.
Due to its large width, the description of the lineshape of the $K^*_0(1430)$ resonance could be complicated by the effects of the opening of the $K \eta$ and $K \eta'$ thresholds.
The decay of the $K^*_0(1430)$ resonance to $K\eta$ has been observed in a Dalitz-plot analysis of $\etac \to \Kp \Km \eta$~\cite{BaBar:2014asx}, and its branching fraction has been found to be small. Its decay to $K \eta'$ has been observed in Refs.~\cite{BESIII:2014dlb,BaBar:2021fkz}.
Another resonance, $K^*_0(1950)$, seen in the $K \pi$ decay mode~\cite{Aston:1987ir}, is still to be confirmed.
Evidence for the $K^*_0(1950) \to K \etapr$ decay mode has been found in a Dalitz-plot analysis of \decay{\etac}{\Kp \Km \etapr} decays~\cite{BaBar:2021fkz}.

In the Dalitz-plot analysis of the \decay{\etac}{\eta \pip \pim} decay, a new $a_0(1700)$ resonance has been observed in the $\eta \pi$ mass spectrum~\cite{BaBar:2021fkz} and recently confirmed in the $a_0(1700) \to \KS K$ decay mode~\cite{BESIII:2022npc}.
The $a_0(1700)$ decay to $\KS K$ is therefore expected to contribute to $\etac/\etactwo \to \kskpi$ decays.
To date, no Dalitz-plot analysis of \etactwo has been performed.

The $\chicone \to\KS K \pi$ decay has been studied in Ref.~\cite{Ablikim:2006vm} in $\psi(2S) \to \gamma \KS K \pi$ decays with $220 \pm 16$ events and a low background. By fitting to the $K \pi$ mass projections, partial branching fractions to
$K^*$ resonances have been measured. Given the small dataset, only upper limits have been obtained for some $\chicone \to K^*X$ decay modes, and therefore further measurements of these branching fractions are useful.

The \bkskkpip and \bkskkpim decays have been previously studied in Refs.~\cite{BaBar:2008fle,Vinokurova:2011dy}, but their branching fractions are yet to be measured. New large datasets may therefore help in clarifying several of the above-listed issues related to light-meson spectroscopy and $B$ to charmonium decays. 

This paper is organized as follows: Sec.~\ref{sec:lhcb} describes the LHCb detector; Sec.~\ref{sec:evsel}, the signal candidate selection procedure; Sec.~\ref{sec:mass}, the study of various mass spectra; Sec.~\ref{sec:charmrespar}, the measurement of the charmonium-resonance parameters; Sec.~\ref{sec:effy}, the efficiency evaluation; Secs.~\ref{sec:etac}~and~\ref{sec:etactwo}, the Dalitz plot analysis of the \etac and \etactwo mesons, respectively; Sec.~\ref{sec:chicone}, the study of \chicone decays; and Sec.~\ref{sec:br}, the measurements of various
branching fractions. Finally, Sec.~\ref{sec:summary} summarizes the results.

\section{Detector, simulation and analysis}
\label{sec:lhcb}
The LHCb 
detector~\cite{LHCb-DP-2008-001,LHCb-DP-2014-002} is a
single-arm forward spectrometer covering the pseudorapidity range $2 < \eta < 5$, designed for
the study of particles containing \bquark\ or \cquark\ quarks. The detector elements that are particularly
relevant to this analysis are: a silicon-strip vertex detector (VELO)~\cite{Aaij:2014zzy} surrounding the $pp$ interaction
region that allows \cquark\ and \bquark\ hadrons to be identified from their characteristically long
flight distance; a tracking system that provides a measurement of the momentum, $p$, of charged
particles; and two ring-imaging Cherenkov detectors that are able to discriminate between
different species of charged hadrons. 
Muons are
identified by a system composed of alternating layers of iron and multiwire proportional
chambers.
The entire dataset collected with the LHCb experiment during Runs 1 and 2 is used,
corresponding to center-of-mass energies $\sqrt s$=7, 8 and 13 \tev and comprising an integrated luminosity of 9 \invfb.
The online event selection starts with a trigger~\cite{Aaij:2012me}, which consists of a hardware
stage, based on information from the calorimeter and muon systems, followed by a software
stage, which applies a full event reconstruction. During offline selection, trigger signatures
are associated with reconstructed particles. Since the trigger system uses the transverse momentum of the
charged particles with respect to the beam axis, \pt, the phase-space and time acceptance is different for events where signal
tracks were involved in the trigger decision (called trigger-on-signal or TOS throughout)
and those where the trigger decision was made using information from the rest of the
event only (noTOS).
Data from both trigger conditions are used and studied separately for consistency tests and the evaluation of systematic uncertainties.

Simulation is required to model the effects of the detector acceptance and the
  imposed selection requirements.
  In the simulation, $pp$ collisions are generated using
  \pythia~\cite{Sjostrand:2007gs,*Sjostrand:2006za} 
  with a specific \lhcb configuration~\cite{LHCb-PROC-2010-056}.
  Decays of unstable particles
  are described by \evtgen~\cite{Lange:2001uf}, in which final-state
  radiation is generated using \photos~\cite{davidson2015photos}.
  The interaction of the generated particles with the detector, and its response,
  are implemented using the \geant
  toolkit~\cite{Allison:2006ve, *Agostinelli:2002hh} as described in
  Ref.~\cite{LHCb-PROC-2011-006}.
  
  Two types of simulations are performed: (a) where the \Bp is decayed according to a phase-space model and (b) where the \Bp decays as $\Bp \to (c \bar c) K$, where $(c \bar c)$ indicates a charmonium resonance decaying to \kskpi by phase space.

\section{Event selection}
\label{sec:evsel}

The present work reports a study of the two \Bp decays \footnote{The inclusion of charge-conjugate processes is implied throughout the paper.}
\begin{equation}
  \Bp \to \KS \Km \Kp \pip
  \label{eq:reac1}
\end{equation}
and 
\begin{equation}
  \Bp \to \KS \Kp \Kp \pim,
  \label{eq:reac2}
\end{equation}
with $\KS \to \pip \pim$.
Decays of the \KS  are reconstructed in two categories: the first involving \KS
mesons that decay early enough for the pions to be reconstructed inside the VELO;
and the second containing \KS mesons that decay later such that track segments from
the pions are outside the VELO. These categories are referred to as long \KS (indicated in the following with \KSLL) and
downstream \KS (indicated in the following with \KSDD), respectively. While the \KSLL category has better mass, momentum and vertex
resolution, there are approximately twice as many \KSDD candidates.
Candidate \Bp particles are formed by combining the \KS candidate with three other charged tracks, having a total charge of one, performing a global vertex fit to the decay tree
and requiring the \Bp candidate to originate from one of the  primary $pp$ collision vertices in the event. Selection criteria and efficiency measurements are performed separately for each \KS category.
To suppress backgrounds, in particular combinatorial background formed from random
combinations of unrelated tracks, the events satisfying the trigger requirements are filtered
by a loose preselection, followed by a multivariate selection optimized separately for each
data sample. Selection requirements are tuned to minimize correlation of the signal
efficiency with kinematic variables, resulting in better control of the corresponding
systematic uncertainties. Consequently, the selection relies minimally on the kinematics of
the final-state particles and instead exploits the topological features that arise
from the detached vertex of the \Bp candidate. These include: the impact parameters of
the \Bp candidate and its decay products, the quality of the decay vertices of the \Bp and
\KS candidates and the separation of these vertices from each other and from the
primary vertex.

The preselection of \KS and \Bp candidates requires, for each track, the presence of appropriate particle identification (PID) information and imposes 
invariant mass selections around the known \KS and \Bp particle masses.
The separation of signal from combinatorial background is achieved by means of a 
boosted decision tree~(BDT) classifier~\cite{Breiman,AdaBoost}, implemented using the TMVA
toolkit~\cite{Hocker:2007ht,*TMVA4}.
The multivariate classifier chosen for this analysis is a BDT with a gradient boosting algorithm~\cite{Roe:2004na}, separated for \KSLL and \KSDD data.
The classifiers are trained using simulated signal decays, composed of samples (a) and (b) described in Sec.~\ref{sec:lhcb} with proportions corresponding to the resonance composition observed in the data (see Sec.~\ref{sec:mass}). 
The simulation also matches the relative proportions of the dataset at the various center-of-mass energies. It is assumed that the efficiencies for the reconstruction of the decays displayed in Eq.~(\ref{eq:reac1}) and Eq.~(\ref{eq:reac2})
are the same. For the background input to the training, data in the lower and upper sidebands of the \Bp signal region are used. The composition of the background sample reflects the data-taking conditions and the decays in Eq.~(\ref{eq:reac1}) and Eq.~(\ref{eq:reac2}) are used in equal proportions. 

\begin{figure}[tb]
\centering
\small
\includegraphics[width=0.45\textwidth]{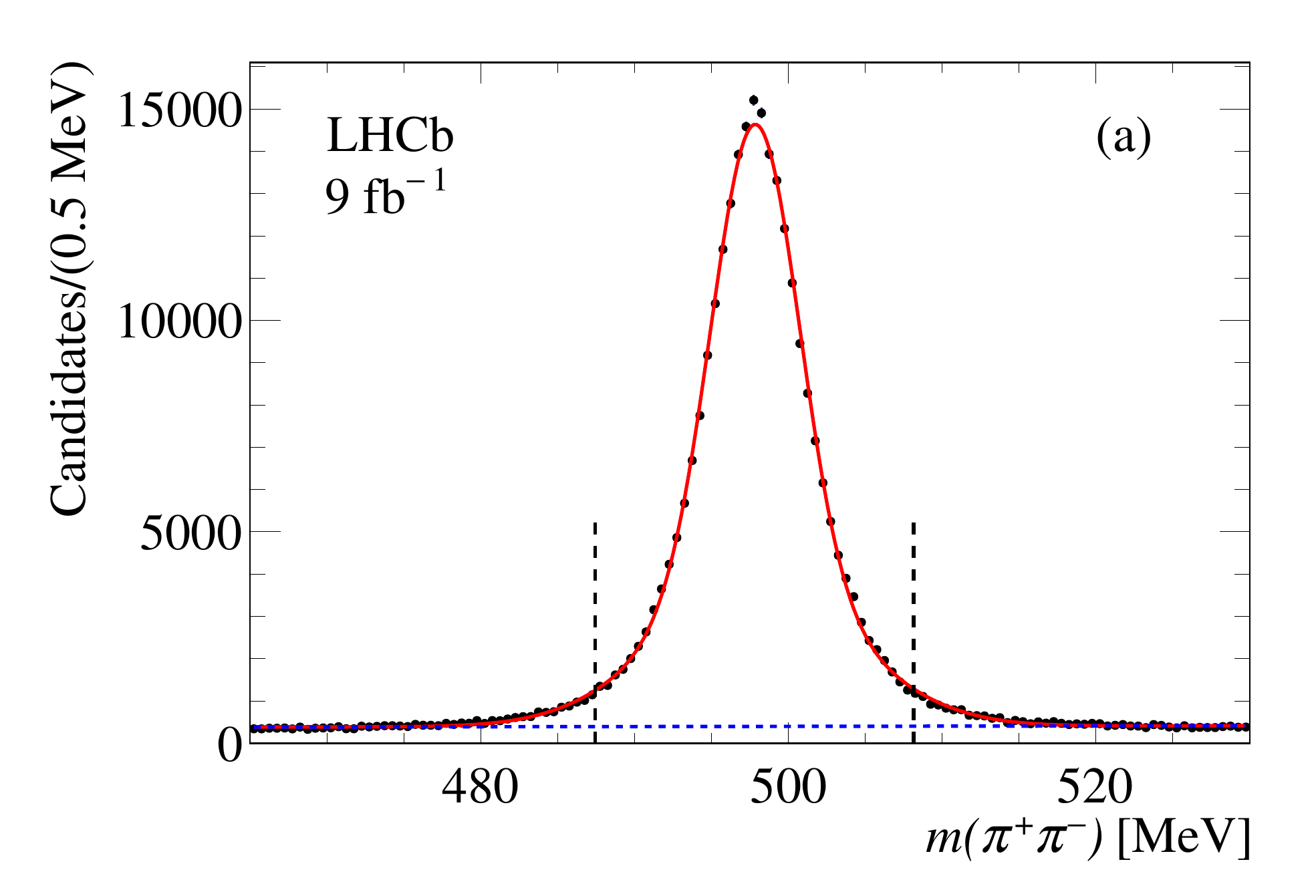}
\includegraphics[width=0.45\textwidth]{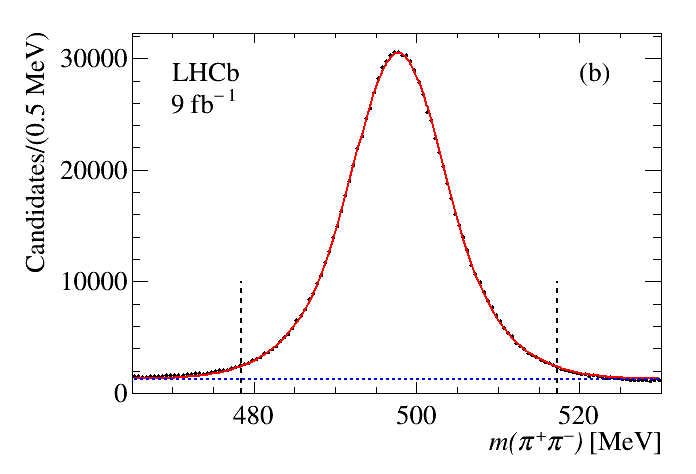}
\caption{\small\label{fig:fig1} Invariant $\pip \pim$ mass distribution for (a) \KSLL and (b) \KSDD candidates. The full (red) line indicates the fit results and the dashed (blue) line the background contribution. The vertical dashed lines indicate the region used to select the \KS signal.}
\end{figure}

The optimization of the BDT classifier working point is performed by scanning the figure of merit
\begin{equation}
  S=\frac{N_{\rm sig}}{\sqrt{N_{\rm sig} + N_{\rm bkg}}},
 \label{eq:sign} 
  \end{equation}
where $N_{\rm sig}$ ($N_{\rm bkg}$) represents the signal (combinatorial background) yield in the signal region.
The yields are evaluated in a $\pm 2.5\sigma$ window around the \Bp mass, where $\sigma$ is the mass resolution, through fits to the \kskkpi mass spectra using two Gaussian functions sharing the same mean for the signal and a linear function for the background.
Figure~\ref{fig:fig1} shows the $\pip \pim$ invariant-mass distributions at the \KS candidate vertex, separated for \KSLL and \KSDD, after selecting candidates using the optimized figure of merit $S$ defined in Eq.~(\ref{eq:sign}). 

The two $\pip \pim$ invariant mass distributions are fitted using the sum of two Gaussian functions sharing the same mean, with $\sigma_1$ and $\sigma_2$ resolutions, and a linear function for the background.
An effective resolution is computed as $\sigma=f\sigma_1+(1-f)\sigma_2$ where $f$ is the fraction of the first Gaussian contribution.
The resulting effective resolutions for the LL and DD categories are $\sigma_{\rm LL} = 2.53\mev$ and $\sigma_{\rm DD} =6.46\mev$.
The \KS signals are selected within $3.0\sigma$ of the fitted \KS mass of $497.8\mev$.\footnote{Natural units with $\hbar = c = 1$ are used throughout this paper.}

In order to facilitate the extraction of the \Bp and \KS signal and combinatorial background components from these invariant-mass spectra, no kinematic mass constraint is applied to the \KS and \Bp signals. To improve the mass resolution (see Sec.~\ref{sec:reso}), the 
energy of the selected candidate \KS is computed as
\begin{equation}
E_{\KS}=\sqrt{p^2_{\KS}+m^2_{\KS}},
\end{equation}
where $p_{\KS}$ is the reconstructed \KS momentum and $m_{\KS}$ the known \KS mass.
Compared with the resolution obtained from the use of the \KS mass constraint, this method gives the same resolution for the $\KS K K \pi$ invariant mass and a slightly worse resolution, by $\approx 6$\%, for the $\KS K \pi$ invariant mass.

Particle identification of the three charged hadrons is performed using the output of a probabilistic neural network (NN) trained on the output of all the subdetectors. The figures of merit are expressed as \mbox{$P_{K} = \mathit{NN}_K(1 - \mathit{NN}_{\pi})$} for kaon identification and
\mbox{$P_{\pi} =\mathit{NN}_{\pi}(1 - \mathit{NN}_K)$} for pion identification, where $\mathit{NN}_{\pi}$ and $\mathit{NN}_K$ are the NN probabilities for pion and kaon identification, respectively. Thresholds are applied to these quantities to maximize the significance of the \Bp-candidate invariant-mass peak as a function of $P_K$ or $P_{\pi}$.

Open charm production in the \Bp decays is significant, with the presence of several signals of \Dp, \Dz and \Dsp in two-body and three-body mass combinations. The largest contributions are due to $\Dz \to \KS \Kp \Km$ decays in \bkskkpip decays \mbox{($17.4 \pm 0.2$)\% } and $\Dzb \to \Kp \pim$ decays ($6.4 \pm 0.1$)\% in \bkskkpim decays. As both \Dz signals have large signal to background
ratios, they are removed from the \Bp-candidate samples. The $\Dz \to \KS \Kp \Km$ contribution is removed by requiring $|m - m_0|>3.5\sigma$, where $m$ indicates the $\KS \Kp \Km$ invariant mass, with $m_0=1864.6\mev$ and $\sigma=4.5\mev$ for \KSLL data and $\sigma=6.3\mev$ for \KSDD data.
The background from $\Dzb \to \Kp \pim$ decays is removed by requiring $|m - m_0|>3.5\sigma$, where $m$ indicates the $\Kp \pim$ mass, with $m_0=1864.5\mev$ and $\sigma=8.4\mev$.
In both cases the \Dz parameters are extracted from a fit to the data, using a Gaussian function and a linear polynomial for signal and background, respectively.

\begin{figure}[tb]
\centering
\small
\includegraphics[width=0.45\textwidth]{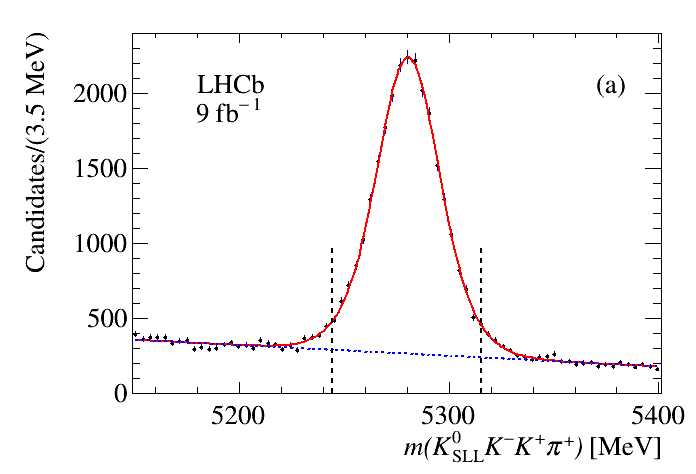}
\includegraphics[width=0.45\textwidth]{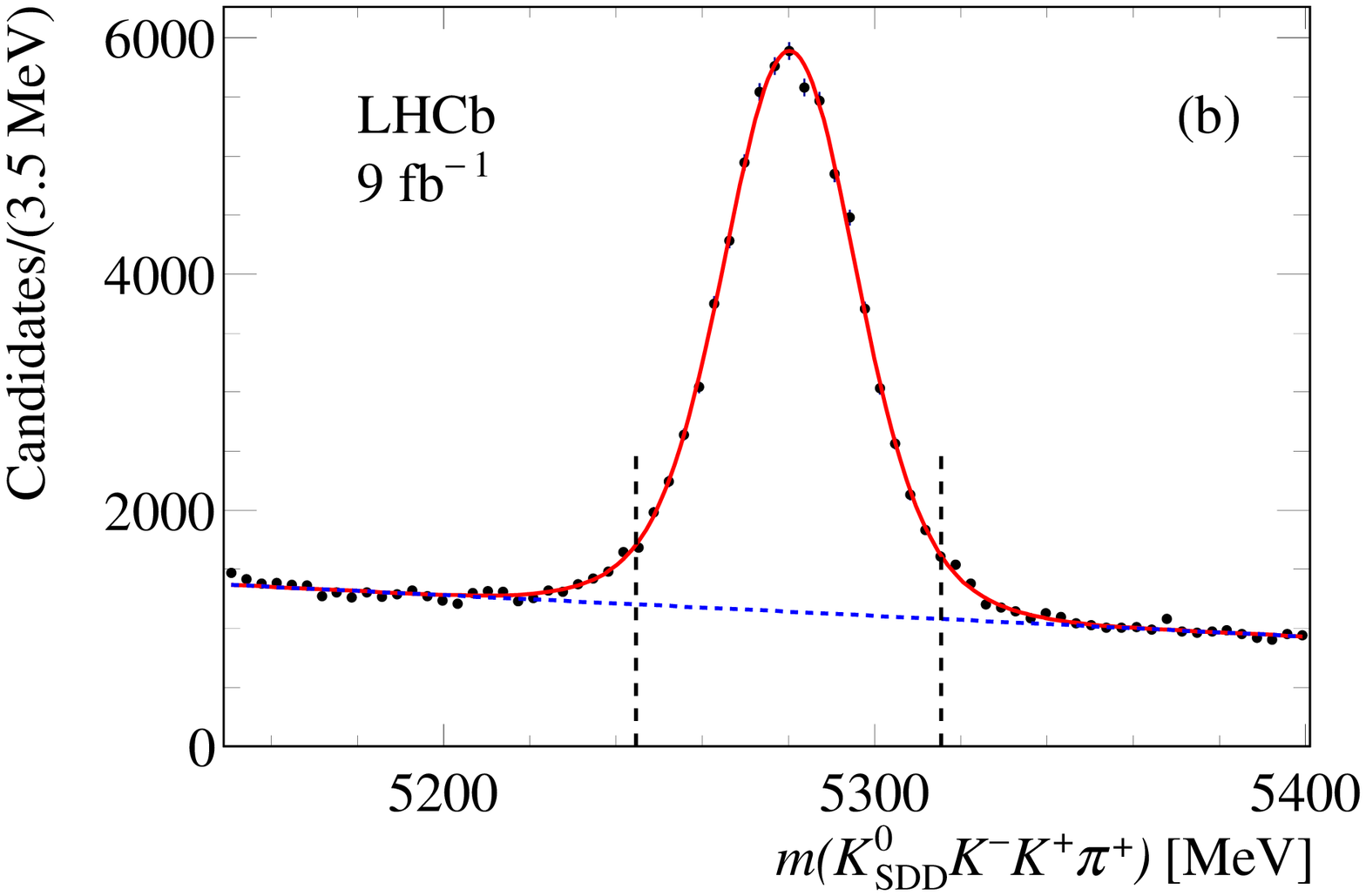}
\includegraphics[width=0.45\textwidth]{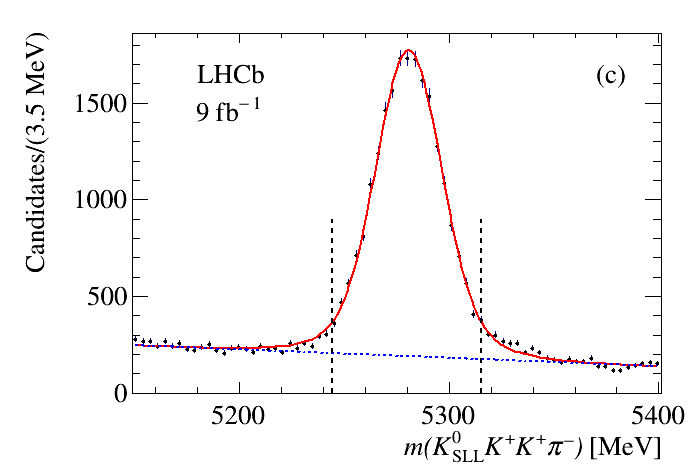}
\includegraphics[width=0.45\textwidth]{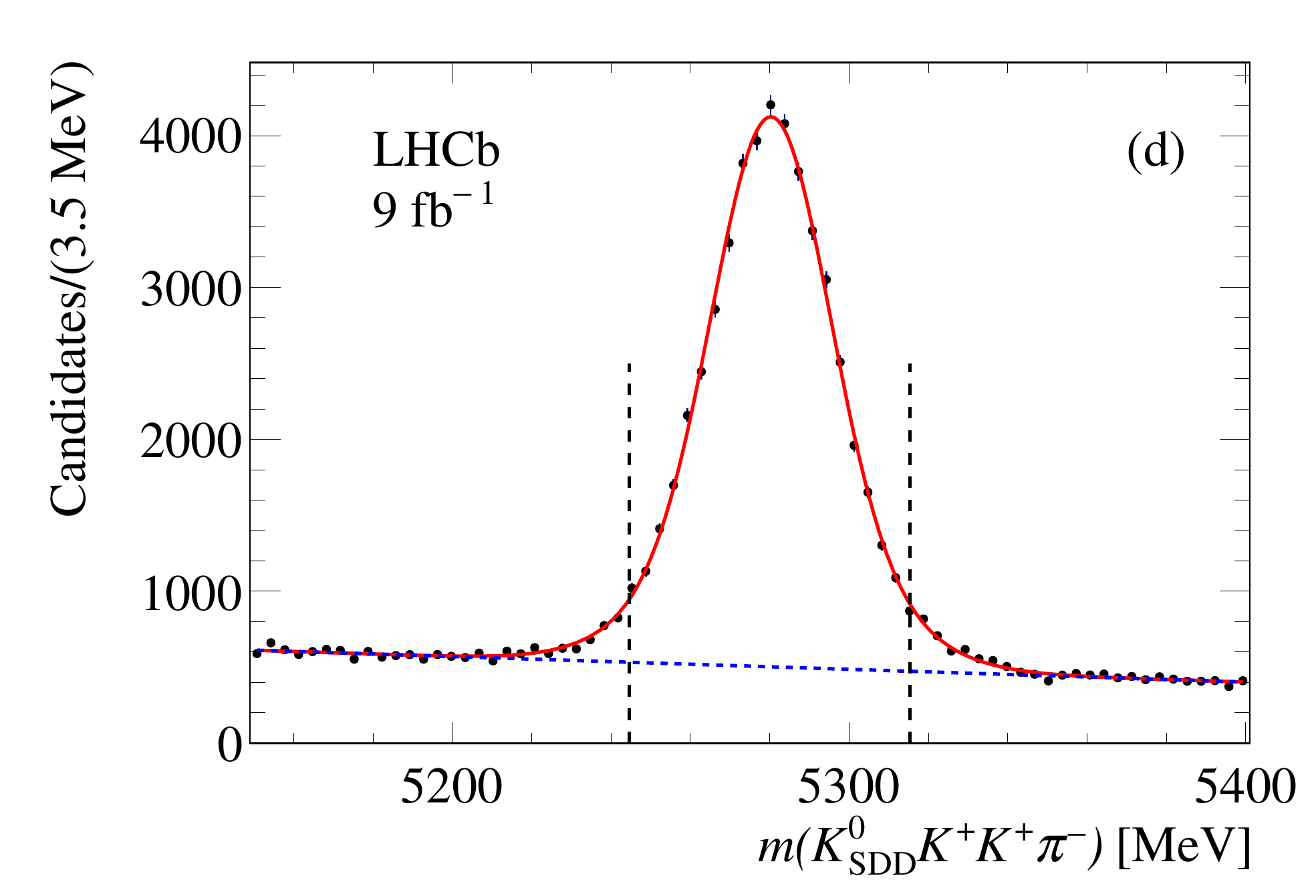}
\caption{\small\label{fig:fig2} 
Distributions of \kskkpip invariant mass for (a) \KSLL and (b) \KSDD candidates and distributions of \kskkpim invariant mass for (c) \KSLL and (d) \KSDD candidates.
  The full (red) lines indicate the signal component and the dashed (blue) lines the background. The vertical dashed lines indicate the regions used to select the \Bp signals.}
\end{figure}

Figure~\ref{fig:fig2} shows the \kskkpip and \kskkpim invariant mass spectra, after the optimized offline selection requirements, separated by \KS category. The distributions are fitted to a sum of two Gaussian functions sharing the same mean values and a linear background function. 
The fits give an average \Bp-mass value of 5280\mev and an average width of \mbox{$\sigma= 17.7$\mev}.
Signal candidates are selected in a window of $\pm 2\sigma$ of the fitted \Bp mass, common to the four datasets.
Table~\ref{tab:tab1} lists the fitted yields and purities ($P$)
in the \Bp signal region for the different datasets, where the purity is defined as $P=N_{\rm sig}/(N_{\rm sig}+N_{\rm bkg})$.
\begin{table} [tb]
  \centering
  \caption{\small\label{tab:tab1} Fitted \Bp signal yield and purity for \bkskkpip and \bkskkpim final states separated by \KS type.}
\begin{tabular}{lcc}
\hline
Final state & \Bp signal yield & \Bp purity  [\%]\cr
\hline\\ [-2.3ex]
\kskkpip &                &       \cr
\KSLL & $21460\pm220$ & $80.0 \pm 0.2$ \cr
\KSDD &  $52690\pm420$ & $69.5 \pm 0.2$ \cr
\hline\\ [-2.3ex]
\kskkpim &                &       \cr
\KSLL & $17730\pm220$ & $82.0\pm0.3$\cr
\KSDD & $40730\pm320$ & $79.9\pm0.2$\cr
\hline
\end{tabular}
\end{table}
The dependence of $P$ on the collision energy and data-taking conditions is approximately uniform for all the four datasets, simplifying the Dalitz-plot analyses reported in the following. 
Approximately 0.02\% of events contain multiple \Bp decay candidates,  as selected with the above procedure, and therefore their effect is considered negligible. 
An inspection of the $\pip \pim$ mass spectra for \KSLL and \KSDD candidates after all selections, and in the \Bp signal region, shows \KS\ signals with negligible background.

\section{Mass spectra}
\label{sec:mass}

\begin{figure}[tb]
\centering
\small
\includegraphics[width=0.80\textwidth]{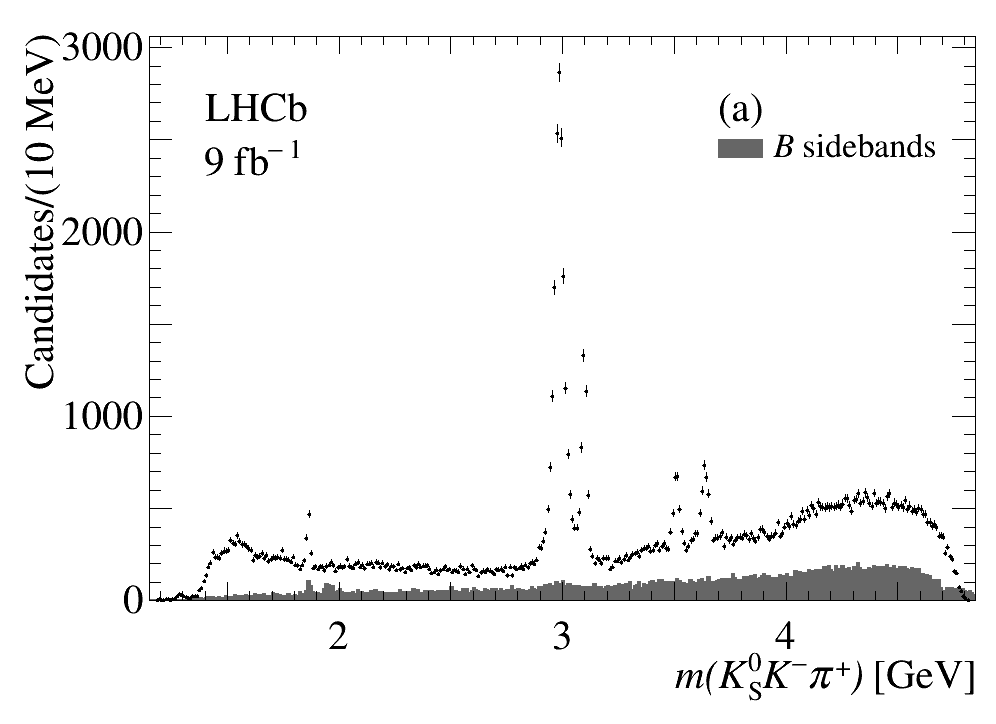}
\includegraphics[width=0.80\textwidth]{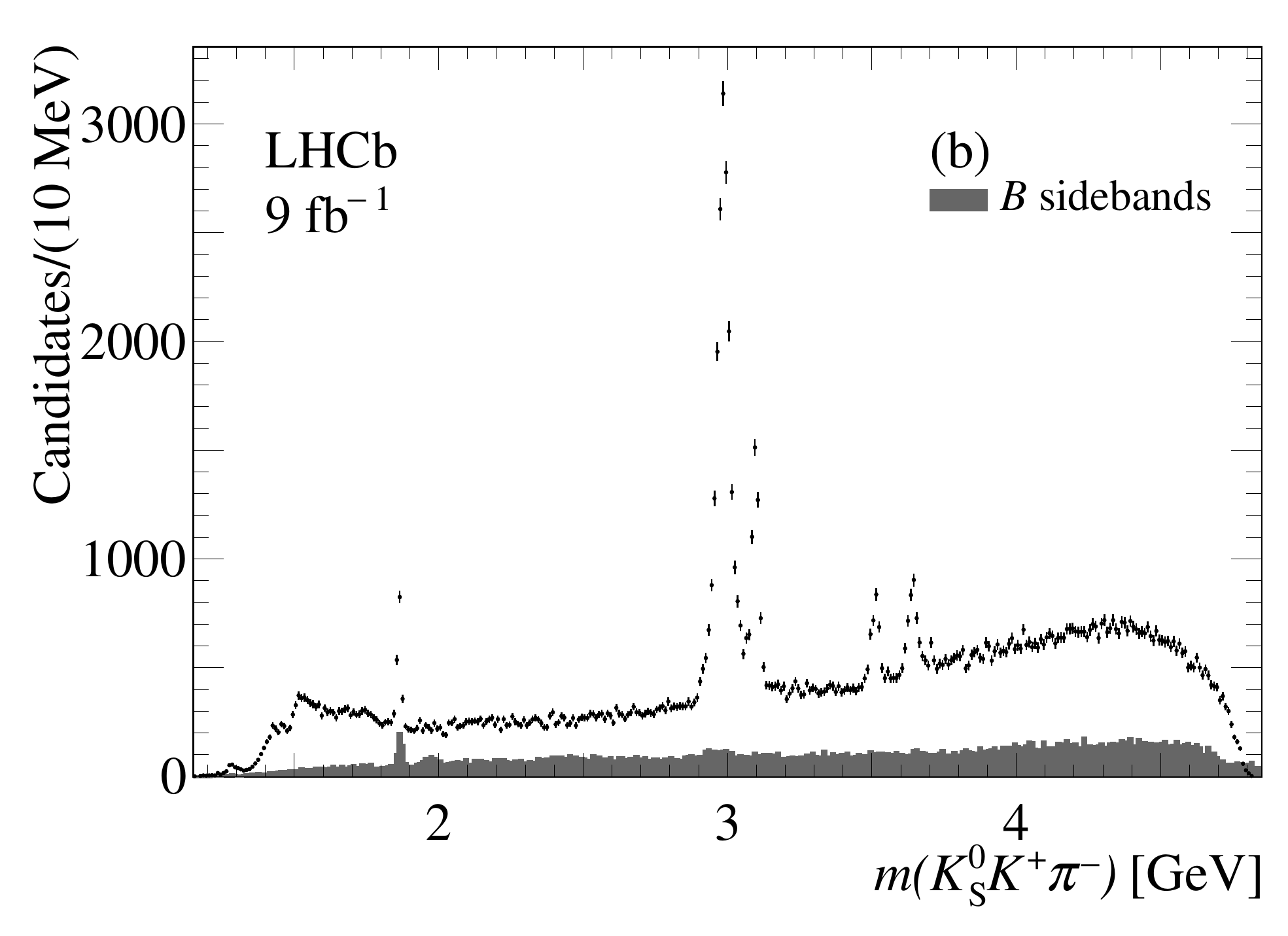}
\caption{\small\label{fig:fig3} Invariant \kskpi mass distributions for (a) \bkskkpip and (b) \mbox{\bkskkpim} candidates (two entries per event).
}
\end{figure}

The \kskpi invariant-mass spectra for events in the \Bp signal region, summed over the \KSLL and \KSDD datasets, are shown in Fig.~\ref{fig:fig3} for \bkskkpip and \mbox{\bkskkpim} final states.
The lower and upper mass sidebands around the \Bp signal peak are defined in the ranges [$-6\sigma,-4\sigma$] and [$4\sigma,6\sigma$], respectively, with $\sigma=17.7\mev$. The corresponding \kskpi invariant mass spectra, representative of background candidates, are superimposed onto the
\kskpi invariant mass spectrum from the \Bp signal region. The \bkskkpim final state has two kaons with the same charge and therefore both combinations are included in the \kskpi invariant-mass spectrum.

There are 119,198 entries and 159,694 combinations in the \kskpi invariant mass spectra from the two \Bp decay modes, respectively.
The \kskpi invariant-mass distributions show a signal at threshold in the position of the $f_1(1285)$ resonance and a broad complex structure in the $1.5 \gev$ mass region. A $\Dz \to \kskpi$ signal can be
observed, due to the open-charm final state $\Bp \to \Dzb \Kp$. Prominent signals of \etac, \jpsi, \chicone and \etactwo can be observed in both invariant-mass spectra.
A broad enhancement in the \etac mass region is present in the \kskpi mass spectrum from \Bp sidebands. 
The effect can be understood as due to the presence of prompt $\etac \to \kskpi$ decays~\cite{LHCb-PAPER-2019-024} reconstructed using an incorrect decay chain.  
The structure above $1.9\gev$ that appears  mostly in the sidebands is due to the reflection from $\Dz \to \KS \pip \pim$ decays, where one pion is misidentified as a kaon. The strong \etac signal present in the data allows for a test of the agreement between data and simulation for the PID.
Particle identification is removed for each final-state kaon and pion in turn on both data and simulation, and the \etac event losses due to the kaon (3\%) and the pion (0.4\%) identification are compared. 
The results for data and simulation agree within $2 \sigma$ and this effect is therefore ignored.

\subsection{Mass resolution}
\label{sec:reso}

The mass resolution is obtained from simulated data.
In this analysis it is used to perform fits to the invariant-mass spectra and obtain resonance parameters in which the effects of the mass resolution are expected to be significant, in particular in fitting narrow charmonium states.
As the resolution functions are mass dependent, they are evaluated in specific mass intervals, namely the \etac--\jpsi mass region, defined in the
$[2.90\text{--}3.15]\gev$ region, and the \chicone--\etactwo mass region, defined in the $[3.46\text{--}3.70]\gev$ region. Due to the presence of two types of reconstructed
\KS with different resolutions, the mass resolution is computed separately for \KSLL and \KSDD data.
The resulting mass-difference distributions are well described by the sum of a Gaussian and a Crystal Ball function. The width ($\sigma$) of the dominant
Gaussian contribution is $8.8\mev$ ($9.9\mev$) in the \etac--\jpsi mass region and $10.3\mev$ ($12.0\mev$) in 
the \chicone--\etactwo mass region for the \KSLL (\KSDD) data.

\section{Measurement of charmonium-resonance parameters}
\label{sec:charmrespar}

\begin{figure}[tb]
\centering
\small
\includegraphics[width=0.75\textwidth]{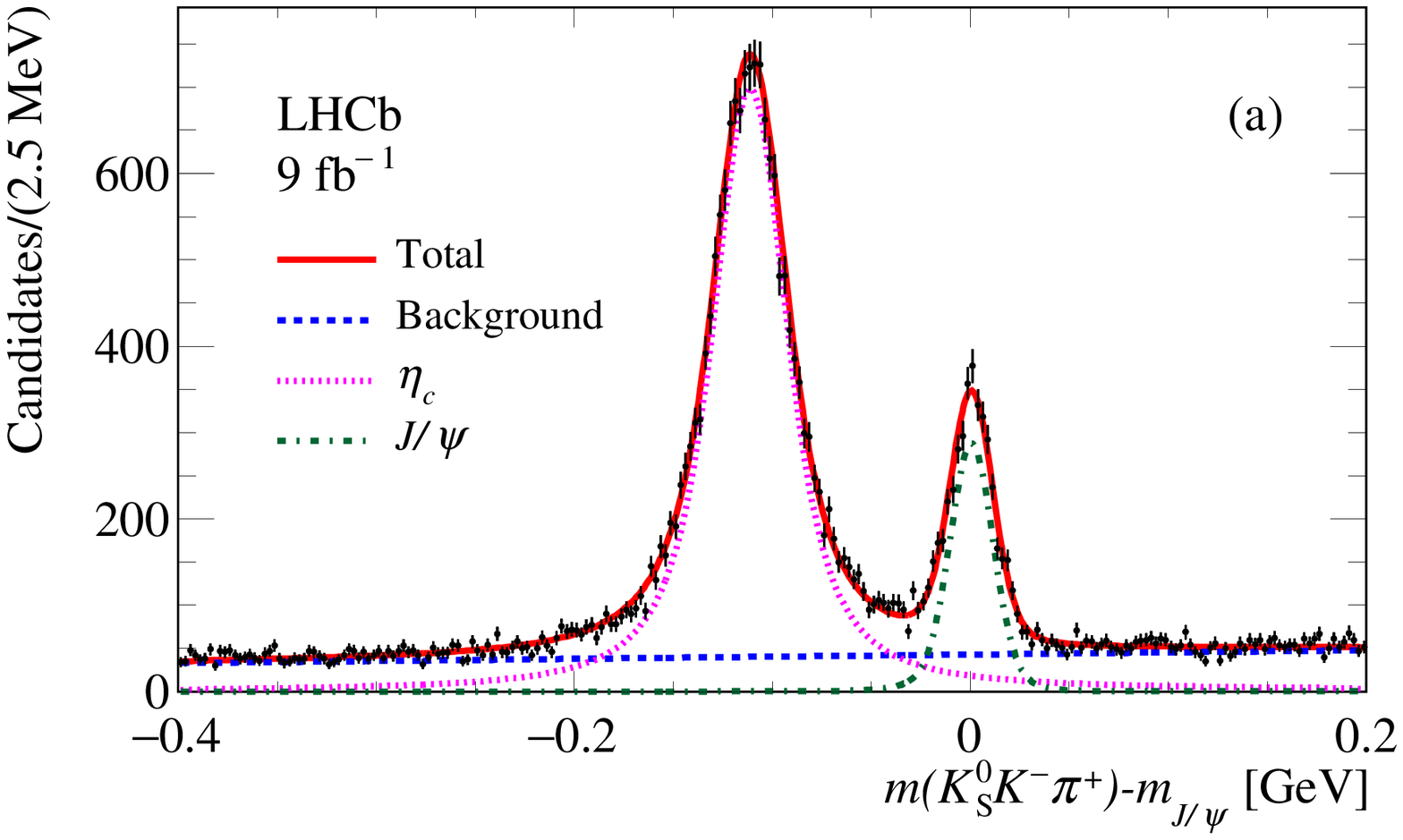}
\includegraphics[width=0.75\textwidth]{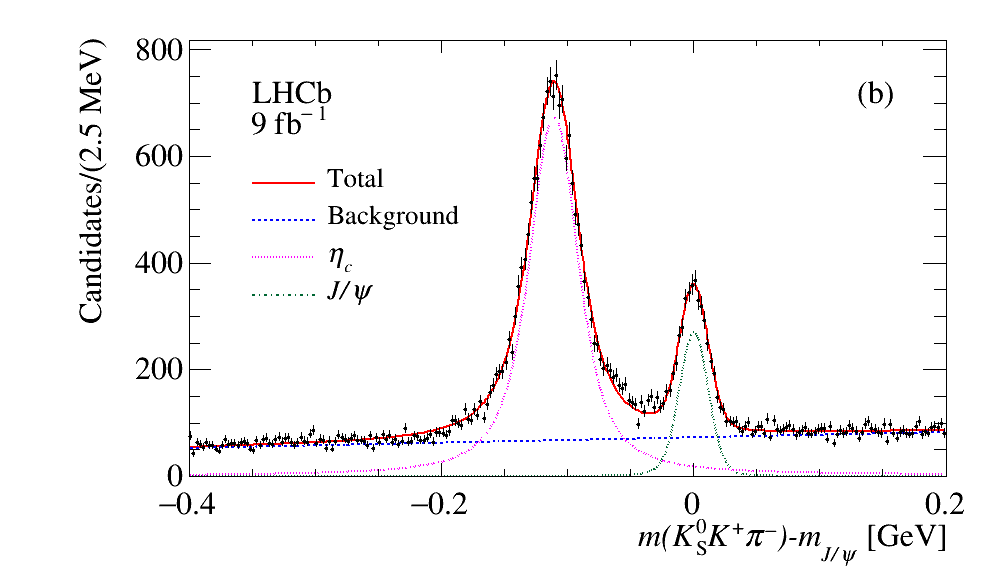}
\caption{\small\label{fig:fig4} Invariant \kskpi invariant mass distributions, with the known \jpsi mass subtracted, in the \etac--\jpsi mass region for (a) \bkskkpip and (b) \bkskkpim candidates. The results of the fits are overlayed.}
\end{figure}

\begin{figure}[tb]
\centering
\small
\includegraphics[width=0.75\textwidth]{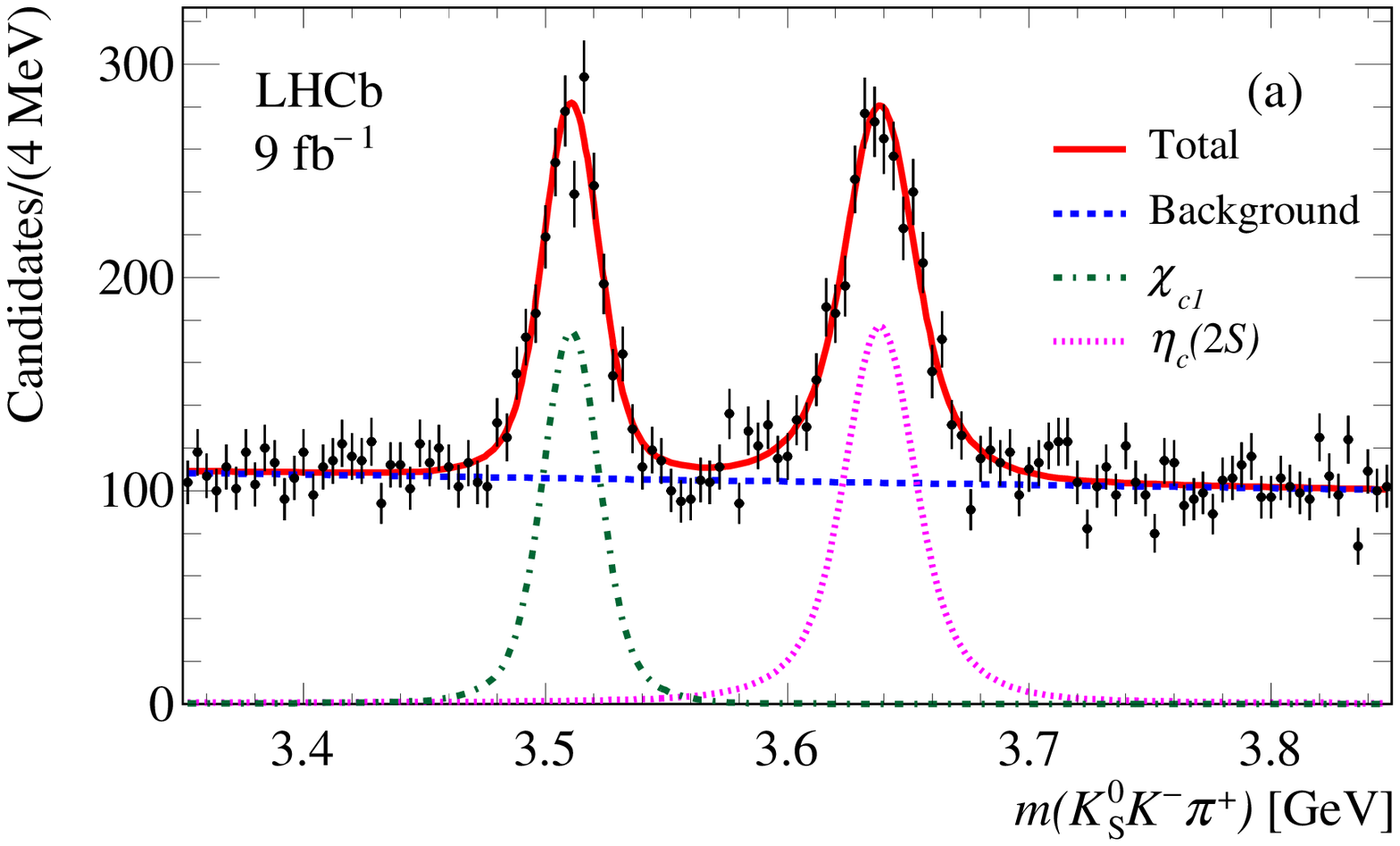}
\includegraphics[width=0.75\textwidth]{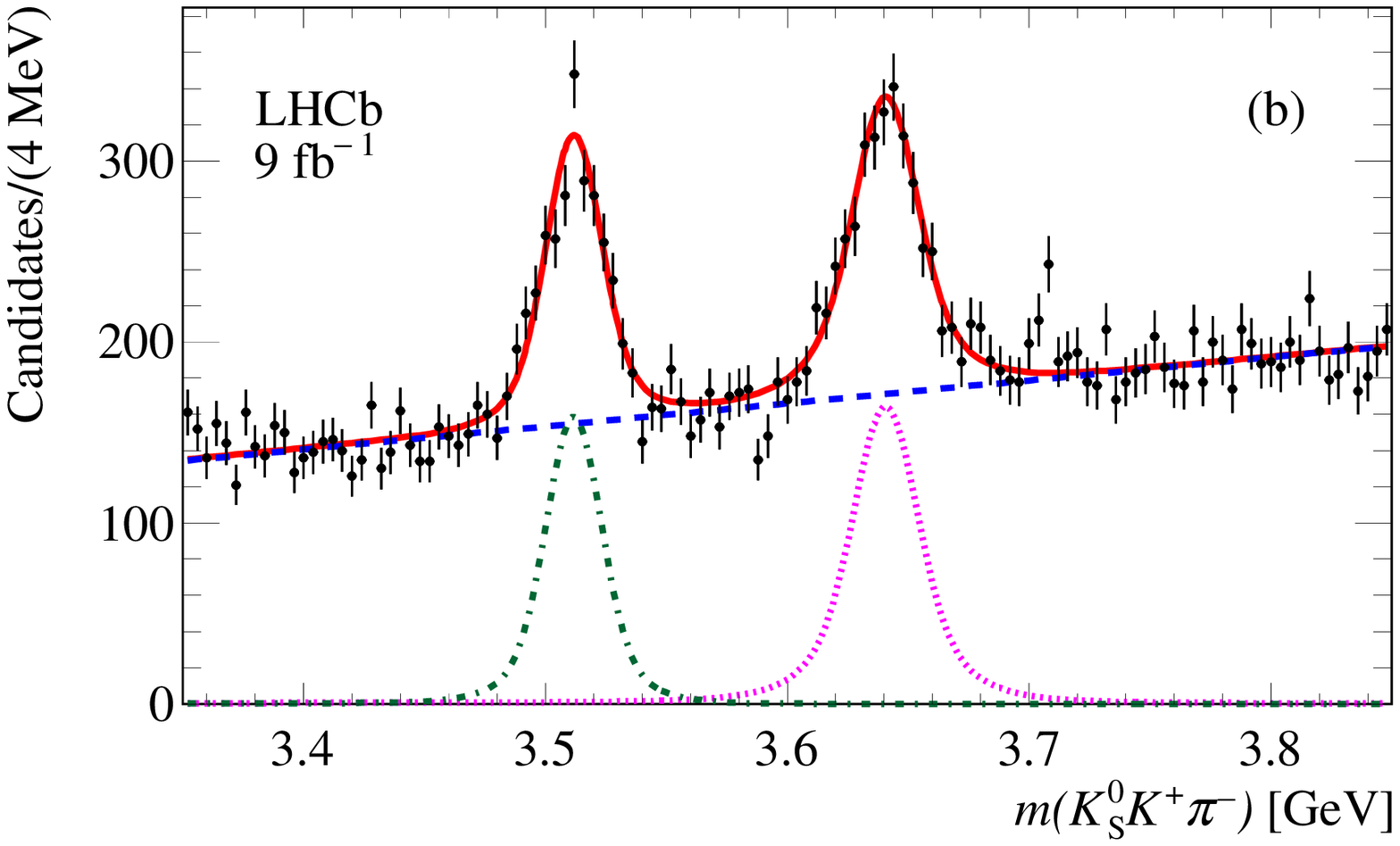}
\caption{\small\label{fig:fig5} Invariant \kskpi mass distributions in the \chicone--\etactwo mass region for (a) \mbox{\bkskkpip} and (b) \bkskkpim candidates.
The results of the fits are overlayed.}
\end{figure}

The measurements of charmonium-resonance parameters are performed with binned fits to the \kskpi invariant-mass spectra separately in the \etac--\jpsi and the \chicone--\etactwo mass regions in the ranges shown in Figs.~\ref{fig:fig4} and~\ref{fig:fig5}, respectively.
In both invariant-mass regions, the \bkskkpip and \bkskkpim data are fitted separately. Since the experimental resolutions for \KSLL and \KSDD data are different, a simultaneous fit to the two invariant-mass spectra is performed, sharing only the resonance parameters.
In the fit to the \etac--\jpsi mass region, the \jpsi width is fixed to the known value~\cite{Workman:2022ynf} and the \kskpi mass value is shifted by the \jpsi mass: $m' = m - m_{\jpsi}$, where $m_{\jpsi}$ is fixed to the known value~\cite{Workman:2022ynf}.
Here, all resonances are described by simple non-relativistic BW functions
\begin{equation}
 BW(m) = \frac{1}{(m_0 - m) - i \Gamma/2},  
 \label{eq:bw}
\end{equation}
convolved with the appropriate mass resolution functions.
The backgrounds, which are due to a combination of \bkskkpi decays and incoherent \kskpi
production, are represented by first-order polynomials.
Table~\ref{tab:tab2} lists the fitted resonance parameters and yields, together with fit p-values. A comparison with PDG~\cite{Workman:2022ynf} averages shows improvements both in the mass and width of the \etac resonance, with the \jpsi mass in good agreement with the known value. A small negative shift, consistent with zero, can be seen on the fitted \jpsi mass. The good description of the \jpsi lineshape, whose observed width is dominated by the mass resolution, demonstrates the good agreement 
of the mass resolution in data and simulation.

\begin{table} [tb]
  \centering
  \caption{\small\label{tab:tab2} Fitted \etac, \jpsi, \etactwo and \chicone parameters. 
  For the \jpsi the $m'=m-m_{\jpsi}$ value is reported. The first uncertainty is statistical, the second systematic.}
  {\small
\begin{tabular}{lclccr}
\hline\\ [-2.3ex]
Final state & p-val. [\%]& Res. & Mass [$\mev$] & Width [$\mev$]& Yield \all\cr
\hline\\ [-2.3ex]
\kskkpip & 16.3 & \etac & $2984.84 \pm 0.23 \pm 1.01$ & $30.0 \pm 0.7 \pm 0.2$ & $17700 \pm 190$ \cr
          &     & \jpsi & $\al\all -0.27 \pm 0.11 \pm 0.61$ & 0.0929 (fixed) & $ 3386  \pm  \al 70$ \cr
\kskkpim & 1.5 & \etac & $2985.19 \pm 0.24 \pm 1.88$ & $29.4 \pm 0.8 \pm 0.8$ & $17210 \pm 210$ \cr
          &     & \jpsi & $\al\all -0.81 \pm 0.11 \pm 0.67$ &  & $3310 \pm  \al 80$ \cr
\hline\\ [-2.3ex]
Average   &     & \etac & $2985.01 \pm 0.17 \pm 0.89$ & $29.7 \pm 0.5 \pm 0.2$ & \cr
&     & \jpsi & $\al\all -0.54 \pm 0.08 \pm 0.45$  &                        & \cr
\hline\\ [-2.3ex]
\kskkpip & 46.6 & \etactwo & $3636.92 \pm 0.71 \pm 1.50$ & $11.70 \pm 2.04 \pm 1.39$ & $1960 \pm \al 80$ \cr
&     & \chicone & $3509.32 \pm 0.70 \pm 0.84$ & 0.88 (fixed)  & $1300 \pm  \al 50$ \cr
\kskkpim & 5.3 & \etactwo & $3639.28 \pm 0.84 \pm 3.83$ & $9.18 \pm 2.67 \pm 1.70$ & $1720 \pm 100$ \cr
&     & \chicone & $3510.35 \pm 0.69 \pm 1.00$ &        & $1460 \pm \al 70$ \cr
\hline\\ [-2.3ex]
Average &   & \etactwo &  $3637.90 \pm 0.54 \pm 1.40$ & $10.77 \pm 1.62 \pm 1.08$ &  \cr
&     & \chicone & $3509.84 \pm 0.69 \pm 0.64$ & & \cr
\hline
\end{tabular}
}
\end{table}

Systematic uncertainties include the following sources. The bin width is varied from 2.5 to $3.0\mev$ and the background shape is changed from linear to a second-order polynomial. The \etac component is allowed to interfere with the background that has a significant contribution from the \Bp decay.
The interference is parameterized as
\begin{equation}
      f(m) = |A_{\rm nres}|^2+|A_{\rm res}|^2+c\cdot 2Re(A_{\rm nres} A^*_{\rm res})
      \label{eq:int}
    \end{equation}   
where $A_{\rm nres}$ is the non-resonant amplitude and $|A_{\rm nres}|^2$ is described by a linear function. The resonant contribution is constructed as $A_{\rm res}=\alpha\cdot {\rm BW}(m) \cdot \text{exp}(i\phi)$, where $\alpha$ and $\phi$ are free parameters and 
BW$(m)$ is the Breit--Wigner function of Eq.~(\ref{eq:bw}) describing the \etac lineshape convolved with the experimental resolution. The coherence factor, $c$, is a free parameter. It is found that the interference model produces a small improvement in the description of the data with fitted phases
of $\phi=1.596\pm 0.009$ rad and $\phi=1.631 \pm 0.011$ rad, both close to $\pi/2$, for \bkskkpip and \bkskkpim data, respectively.
The deviations of the fitted resonance parameters from those obtained without interference are included as systematic uncertainties.

The systematic uncertainty associated with the background model is evaluated by varying the BDT classifier selection working point, resulting in a variation of the \Bp purity of around $\pm5\%$.
The average value of the absolute values of the two resulting variations of the resonance parameters is used to quantify the systematic uncertainty associated to the background level. Starting with the fitted functions from the reference fit, 400 pseudoexperiments are generated and fitted. The average value of the deviations 
of the fitted parameters is included as a systematic uncertainty.

The momentum-scale uncertainty is evaluated as $0.03Q$~\cite{LHCb:2013fwo}, where $Q$ is evaluated as the difference between the \etac mass and the sum of the masses of the decay particles.
The uncertainties on the resonance parameters arising from the limited sizes of the simulation samples, used to obtain the resolution functions, are obtained by fitting the \kskpi invariant-mass spectrum from 400 pseudoexperiments where, in each fit, all the parameters describing the resolution functions are varied randomly from a Gaussian distribution defined by
their statistical uncertainties. 
Table~\ref{tab:tab3} gives the resulting contributions to the systematic uncertainties which are then added in quadrature.

\begin{table} [tb]
  \centering
  \caption{\small\label{tab:tab3} 
  Summary of systematic uncertainties on \jpsi and \etac resonance parameters.}
          {\small
\begin{tabular}{lccc}
\hline
Contribution & $m(\etac)$ [\mev] & $\Gamma(\etac)$ [\mev] & $\Delta m(\jpsi)$ [\mev]\cr
\hline\\ [-2.3ex]
\bkskkpip & & & \cr
\hline\\ [-2.3ex]
Bin width & 0.03 & 0.12 & 0.08 \cr
Background    &   0.04 & 0.01 & 0.01 \cr
Interference  &  0.84 & 0.09 & 0.15 \cr
Fit bias & 0.01 & 0.02 & 0.04 \cr
Momentum scale & 0.56 & - & 0.56 \cr
BDT variation &   0.07 & 0.12 & 0.16 \cr
Resolution & 0.00 & 0.01 & 0.00 \cr
\hline
Squared sum  &  1.01 & 0.19 & 0.61 \cr
\hline\\ [-2.3ex]
\bkskkpim & & & \cr
\hline
Bin width & 0.02 & 0.01 & 0.03 \cr
Background    &   0.11 & 0.09 & 0.05 \cr
Interference  &   1.79 & 0.74 & 0.36 \cr
Fit bias & 0.00 & 0.09 & 0.07 \cr
Momentum scale & 0.56 & - & 0.56 \cr
BDT variation &   0.10 & 0.06 & 0.05 \cr
Resolution & 0.00 & 0.01 & 0.00 \cr
\hline\\ [-2.3ex]
Squared sum     &   1.88 & 0.75 & 0.67 \cr
\hline
\end{tabular}
}
\end{table}

\begin{table} [tb]
  \centering
  \caption{\small\label{tab:tab4} 
  List of the systematic uncertainties on the \chicone and \etactwo parameters.
  }
  {\small
\begin{tabular}{lcccc}
\hline\\ [-2.3ex]
Final state & $m(\etactwo)$ [$\mev$]& $\Gamma(\etactwo)$ [$\mev$]& $m(\chicone)$ [$\mev$]\cr
\hline
\bkskkpip & & & \cr
\hline\\ [-2.3ex]
Bin width & 0.13 & 0.17 & 0.09 \cr
Background    &   0.01 & 0.99 & 0.05 \cr
Interference     &   1.30 & 0.09 & 0.29 \cr
Fit bias & 0.02 & 0.22 & 0.00 \cr
Momentum scale & 0.75 & - & 0.75 \cr
BDT variation &   0.09 & 0.93 & 0.25 \cr
\hline\\ [-2.3ex]
Squared sum  &  1.50 & 1.39 & 0.84 \cr
\hline\\ [-2.3ex]
\bkskkpim & & & \cr
\hline\\ [-2.3ex]
Bin width & 0.22 & 0.15 & 0.10 \cr
Background    &   0.14 & 1.15 & 0.09 \cr
Interference  &   3.75 & 0.70 & 0.64 \cr
Fit bias & 0.05 & 0.55 & 0.04 \cr
Momentum scale & 0.75 & - & 0.75 \cr
BDT variation &   0.10 & 0.87 & 0.11 \cr
\hline\\ [-2.3ex]
Squared sum &   3.83 & 1.70 & 1.00 \cr
\hline
\end{tabular}
}
\end{table}

A similar model is used to fit the \chicone--\etactwo mass region, shown in Fig.~\ref{fig:fig5}, with the fit results summarized in Table~\ref{tab:tab2} together with the inverse-variance-weighted  averages of the fitted parameters from the two \Bp decay modes. In this case, the \chicone width is fixed to the known value~\cite{Workman:2022ynf}. The inverse-variance method is also used, here and in the following, to evaluate average systematic uncertainties.
A comparison of the results listed in Table~\ref{tab:tab2} with PDG~\cite{Workman:2022ynf} measurements shows improvements both in the mass and width of the \etactwo resonance, with the \chicone mass in good agreement with the known value. 
Systematic uncertainties on the fit parameters are evaluated in a similar way as for the \etac--\jpsi fit except now the \etactwo is allowed to interfere with the background. The fit with interference returns relative phases of $\phi=1.52 \pm 0.05$ rad and $\phi=1.82 \pm 0.09$ rad for the \bkskkpip and \bkskkpim final states, respectively (similarly close to $\pi/2$). The presence of signals corresponding to the $h_c(1P)$ and $\chi_{c2}(1P)$ resonances is explored by adding additional components to the fit function and resonance parameters fixed to the known values~\cite{Workman:2022ynf};
their yields are found to be consistent with zero.
A summary of the systematic uncertainties, together with their quadratic sum,
is given in Table~\ref{tab:tab4}.

\subsection{First observation of {\boldmath$\Bp \to \chiczero \KS \pip$}}

\begin{figure}[tb]
\centering
\small
\includegraphics[width=0.75\textwidth]{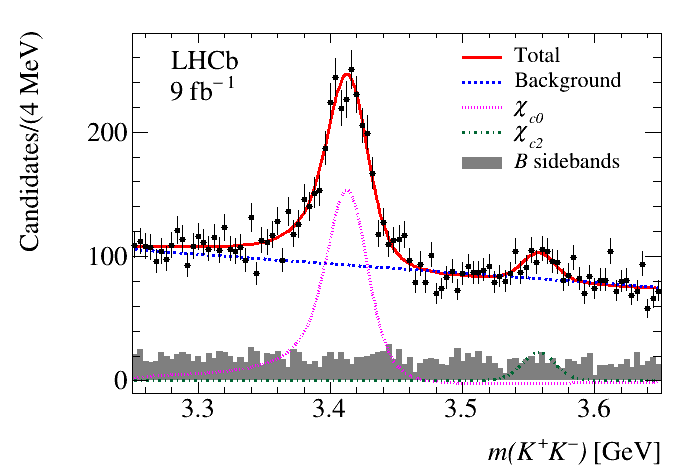}
\caption{\small\label{fig:fig6} Invariant $\Kp \Km$ mass distribution in the \chiczero-\chictwo mass region for \bkskkpip decays. The two \KS datasets are combined. The results of the fit are overlayed. The curves also include interference terms and therefore the \chiczero lineshape takes slightly negative values.}
\end{figure}

Figure~\ref{fig:fig6} shows the $\Kp \Km$ invariant-mass spectrum for \bkskkpip candidates in the \chiczero--\chictwo mass region, where a prominent \chiczero signal can be seen, together with a weaker signal at the \chictwo mass. Also superimposed is the $\Kp \Km$ invariant-mass spectrum from the \Bp sidebands, where no \chiczero signal can be seen. For this \Bp decay mode, the PDG~\cite{Workman:2022ynf} reports a branching-fraction upper limit of $\calB(\Bp \to \chi_{c0}K^*)<2.1 \times 10^{-4}$~\cite{BaBar:2008ccs}.

Using a similar method as for the fit to the other charmonium resonances, a simultaneous binned fit of the $\Kp \Km$ mass spectrum  for the \KSLL and \KSDD data is performed. The background is parameterized by a second-order polynomial; the resonance parameters for the $\chi_{c0}$ state are unconstrained, while those for the $\chi_{c2}$ component are fixed to their known values. The $\chi_{c0}$ signal is described by the relativistic spin-0 Breit--Wigner function

\begin{equation}
    BW(m) = \frac{1}{m_0^2 - m^2 - im_0\Gamma},
    \end{equation}
where $m_0$ is the resonance mass. The mass dependent width $\Gamma$ is written as
\begin{equation}
    \Gamma = \Gamma_0\frac{q}{q_0}\frac{m_0}{m},
\end{equation}
where $\Gamma_0$ is the resonance width and $q$ ($q_0$) is the momentum of either decay particle in the two-body (resonance) rest frame.
The  $\chi_{c2}$ signal, due to its narrow width, is described by a simple Breit--Wigner function (Eq.~\ref{eq:bw}); both distributions are convolved with the experimental resolution function modelled, as described in Sec.~\ref{sec:reso}, by the sum of a Crystal Ball function and a Gaussian function having $\sigma=11.9\mev$.
In this fit, the measured $\chi_{c0}$ mass and width are both shifted by approximately $5\mev$ with respect to their known values~\cite{Workman:2022ynf}.

Allowing for interference of the $\chi_{c0}$ resonance with a non-resonant $\Kp \Km$ component, as described in Eq.~\ref{eq:int}, results in the data description shown in Fig.~\ref{fig:fig6}, with a p-value=27.9\% and \chiczero parameters
\begin{equation}
  m(\chi_{c0}) = 3413.6 \pm 1.3 \mev, \ \Gamma(\chi_{c0}) = 12.8 \pm 2.8  \mev,
\end{equation}
consistent with PDG averages~\cite{Workman:2022ynf}. The \chiczero yield is $N_{\chi_{c0}}=1920\pm90$ and the relative phase $\phi=-1.290 \pm 0.073$ rad.
The fit also returns a \chictwo yield of $N_{\chi_{c2}}=190\pm30$. To evaluate the significance of the \chictwo signal, a fit without its contribution is performed. This results in a $\chi^2$ variation of $\Delta\chi^2=22.5$ for the difference
of one parameter, which gives a statistical significance of $4.6\sigma$.
  
\section{Efficiency}
\label{sec:effy}

Two types of efficiencies are evaluated, total and local.
The total efficiency describes the effects of the reconstruction on the \Bp decay to the 4-body final state, needed to evaluate the relative charmonium branching fractions.
The local efficiencies are evaluated in specific \kskpi mass regions, \ie the \etac--\jpsi and \chicone--\etactwo mass regions, where Dalitz-plot analyses are performed and descriptions of the \kskpi detection efficiency are needed.
The efficiencies are evaluated by generating simulation samples that undergo the same reconstruction and analysis selections as the data. The efficiency is evaluated as the ratio of selected over generated
distributions projected over the relevant kinematic variables.

The choice of the phase-space variables which describe the efficiency is somewhat arbitrary, and a mixture of two-body and three-body invariant-mass projections is used, together with variables related to the angular distributions. A comparison between the \pt distributions of the  \Bp candidates in simulated samples and in data shows a small disagreement, which is corrected by weighting the former to match the latter.

\subsection{Total efficiency}

To study efficiencies, simulated events are generated according to a four-body phase-space model.
Since the physics of the \kskpi system is of interest, the charged kaon participating in the decay under study is labeled 
$K_1$, while the second charged kaon behaves as a spectator and is labeled $K_2$. Therefore, the reaction can be written
\begin{equation}
  B \to (\KS K_1 \pi) K_2.
  \label{eq:k1k2}
\end{equation}
Only \bkskkpip decays are generated because the efficiency for \mbox{\bkskkpim} is expected to be the same. The generated angular distributions are all uniform; therefore, any observed variation is due to inefficiency. The efficiencies are evaluated separately for the \KSLL and \KSDD samples. In both cases the resulting distributions have only a mild dependence on the \kskpi invariant mass.

\begin{figure} [tb]
\centering
\includegraphics[width=5.0cm]{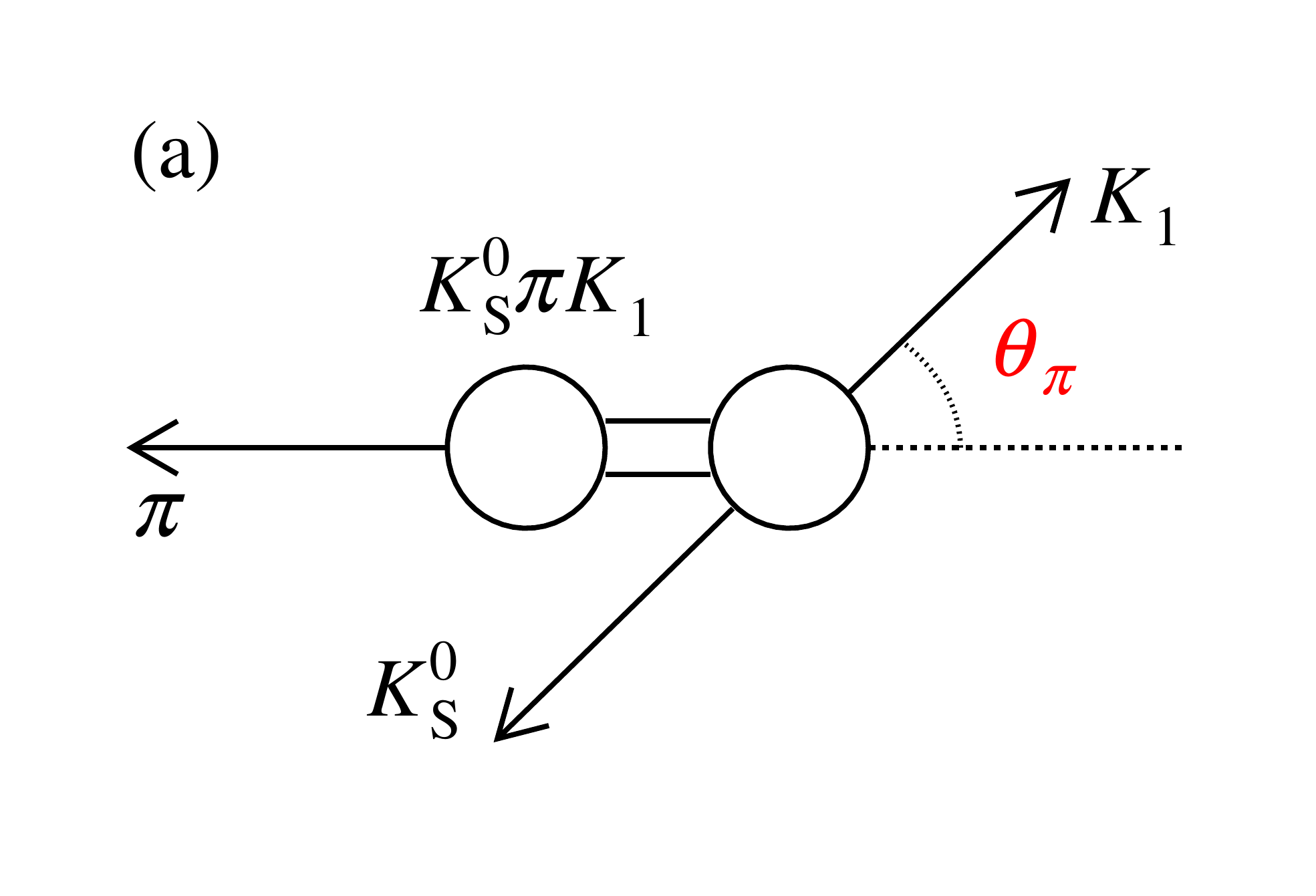}
\includegraphics[width=5.0cm]{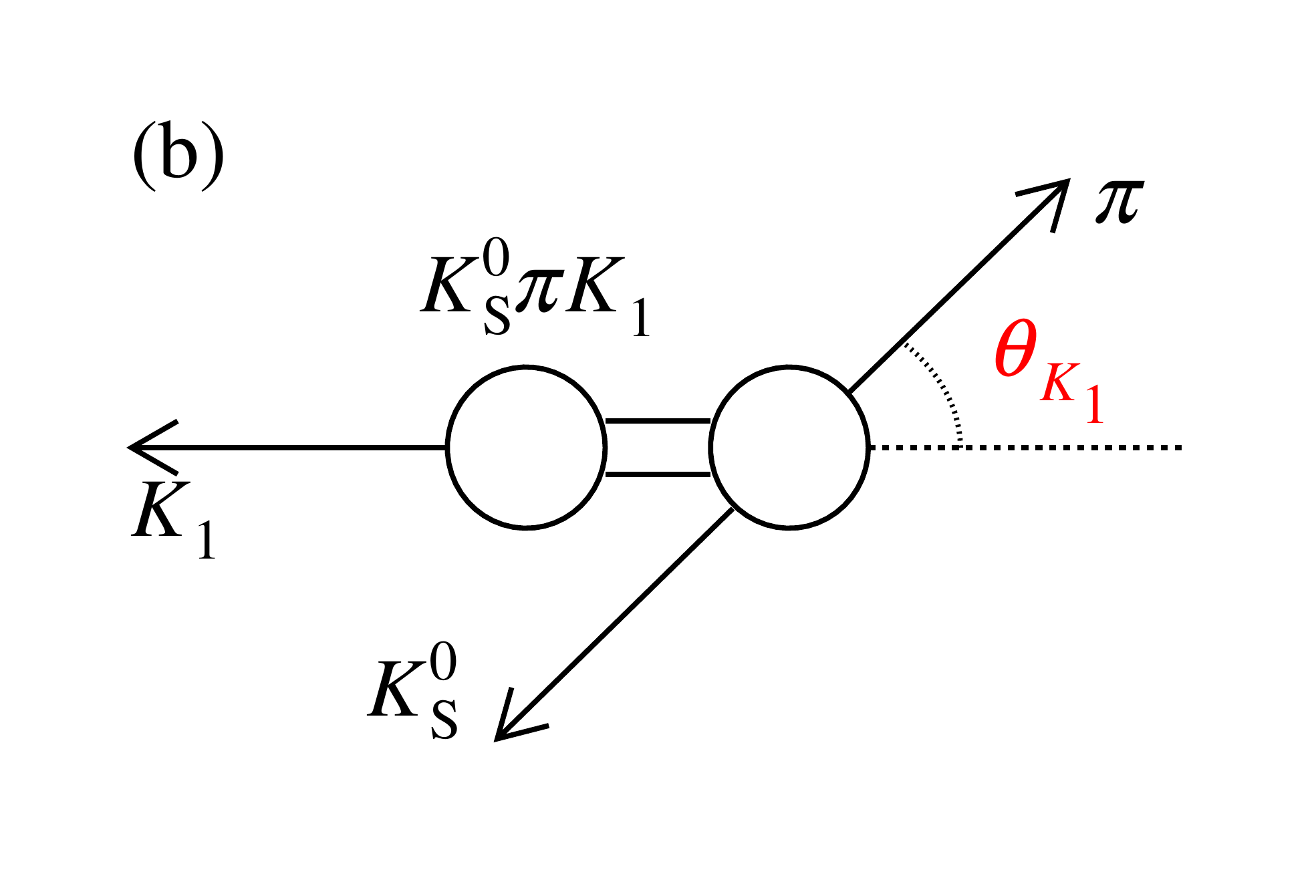}
\includegraphics[width=5.0cm]{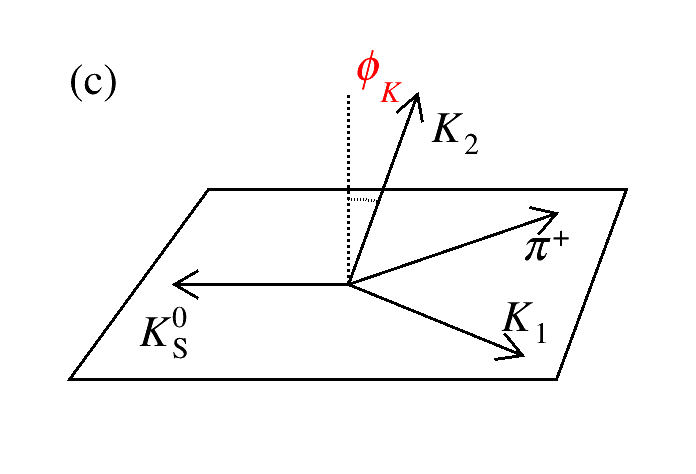}
\caption{\small\label{fig:fig7} Definition of the angles (a) $\theta_{\pi}$, (b) $\theta_{K_1}$ and (c) $\phi_K$.}
       \end{figure}

The kinematics of a four-body decay are fully described by five independent variables. 
Different invariant-mass combinations have different kinematic bounds, so mass-reduced variables are used instead because they always range between 0 and 1. They are defined as~\cite{LHCb:2019vww}

\begin{equation}
   m_x = \frac{1}{\pi} {\text{arcos}}(2\frac{m - m_{\rm min}}{m_{\rm max}-m_{\rm min}}-1),
\end{equation}
where $m$, $m_{\rm min}$ and $m_{\rm max}$ indicate the invariant mass and its minimum and maximum kinematically allowed values, respectively. 
As an example in the above equation, $m_x(\KS K)$ for the three-body \kskpi system is computed as
 
 \begin{equation}
   m_x(\KS K) = \frac{1}{\pi} {\text{arcos}}(2\frac{m(\KS K) - m_{\rm min}}{m_{\rm max}-m_{\rm min}}-1),
 \end{equation}
and $m_{\rm min}=m_{\KS}+m_{K}$ and $m_{\rm max}=m(\kskpi)-m_{\pi}$,
similarly for other two-body mass combinations. 
The angular distributions make use of helicity angles defined as follows.
In the \kskpi rest frame, shown in Fig.~\ref{fig:fig7}(a) and (b), $\theta_{\pi}$ ($\theta_{K_1}$) indicates the angle
formed by the $K_1$ ($\pi$) with respect to the $\KS K_1$ ($\KS \pi$) direction in the $\KS K_1$ ($\KS \pi$) rest frame.
Similarly, the angle $\theta_{K_2}$ is defined by exchanging $K_1$ with $K_2$.
Figure~\ref{fig:fig7}(c) shows the definition of $\phi_K$, the angle formed by the spectator $K_2$ with the normal to the $\KS K_1 \pi$ plane.
       
\begin{figure}[tb]
\centering
\small
\includegraphics[width=0.45\textwidth]{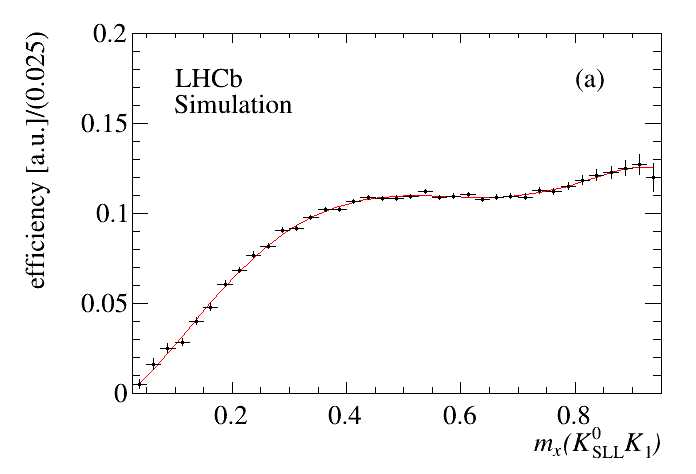}
\includegraphics[width=0.45\textwidth]{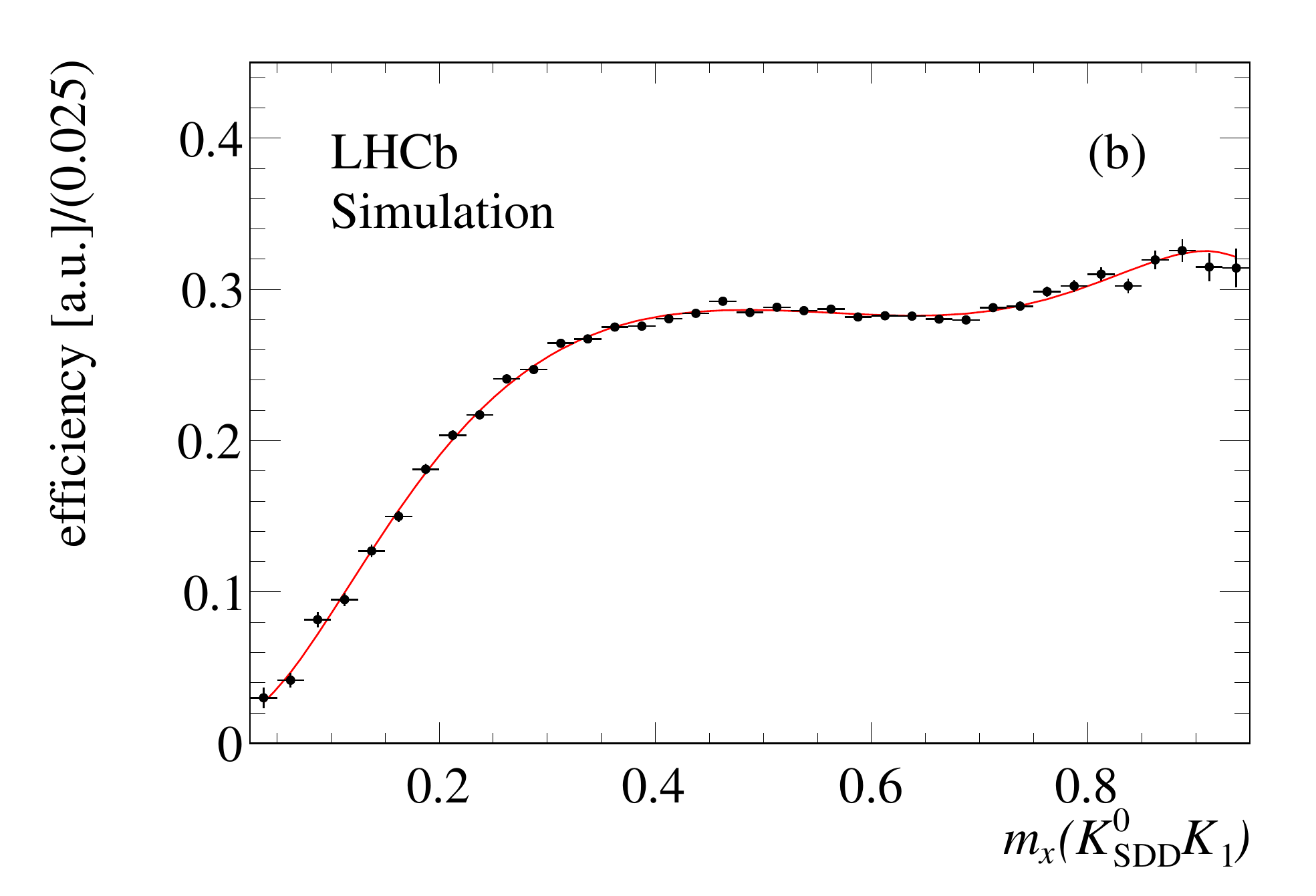}
\caption{\small\label{fig:fig8} Efficiency projections (in arbitrary units) on $m_x(\KS K_1)$ for the (a) \KSLL and (b) \KSDD samples for the total \bkskkpip dataset. The curves are the
  results of fits made using seventh-order polynomials.}
\end{figure}

The model used to describe the efficiency is obtained in an iterative manner. First, the variables showing the strongest deviation from uniformity in the simulation are chosen. The efficiency projection as a function of the first chosen variable, $m_x(\KS K_1)$, is fitted using a seventh-order polynomial labeled as $\epsilon_1(m_x(\KS K_1))$. 
Figure~\ref{fig:fig8} shows the efficiency projected on $m_x(\KS K_1)$ separately for the \KSLL and \KSDD samples.

The events are then weighted by the inverse of the efficiency $1/\epsilon_1(m_x(\KS K_1))$ and a second variable ($m_x(\KS \pi)$) is chosen which is itself fitted by a seventh-order polynomial labeled as $\epsilon_2(m_x(\KS \pi))$. The events are then weighted by $1/(\epsilon_1(m_x(\KS K_1))\cdot\epsilon_2(m_x(\KS \pi)))$. The process continues in this fashion, terminating when, after weighting, the efficiency is consistent with being uniform across all the nine considered variables, both on one- and two-dimensional projections.
The total efficiency for each \KS category, $\epsilon_{LL}$ and $\epsilon_{DD}$, is found to be well described by
\begin{equation}
  \begin{split}
    \epsilon_{LL} & = \epsilon_1(m_x(\KS K_1))\cdot \epsilon_2(m_x(\KS \pi))\cdot \epsilon_3(m(\KS K_1 \pi)) \cdot \epsilon_4(m_x(K_1 \pi)) \cdot\epsilon_5(\cos \theta_{K_2})),\\
    \epsilon_{DD} & = \epsilon_1(m_x(\KS K_1))\cdot \epsilon_2(m_x(\KS \pi))\cdot \epsilon_3(m(\KS K_1 \pi)) \cdot \epsilon_4(m_x(K_1 \pi)).
  \end{split}
\end{equation}

\subsubsection{Efficiency for different trigger conditions}

\begin{figure}[tb]
\centering
\small
\includegraphics[width=0.48\textwidth]{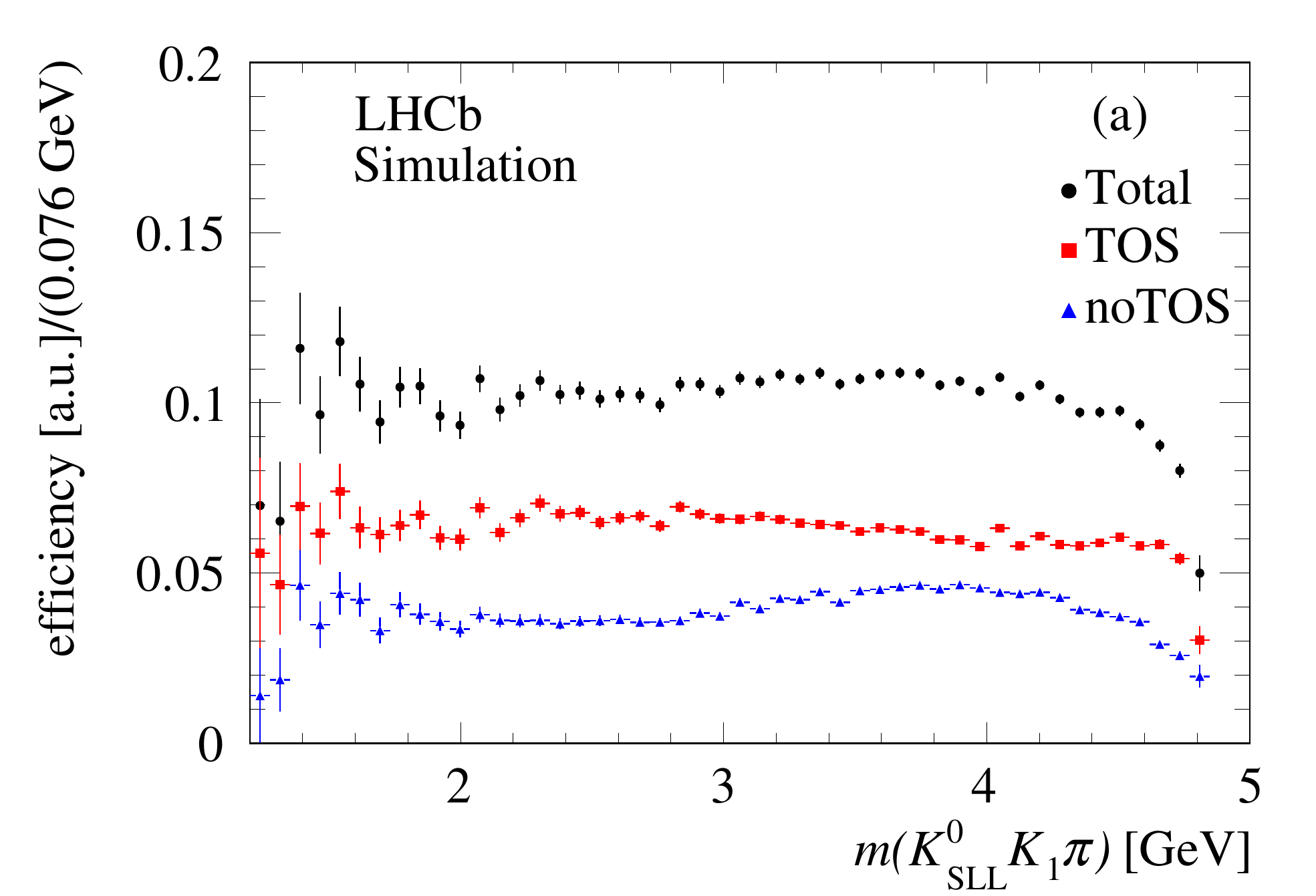}
\includegraphics[width=0.48\textwidth]{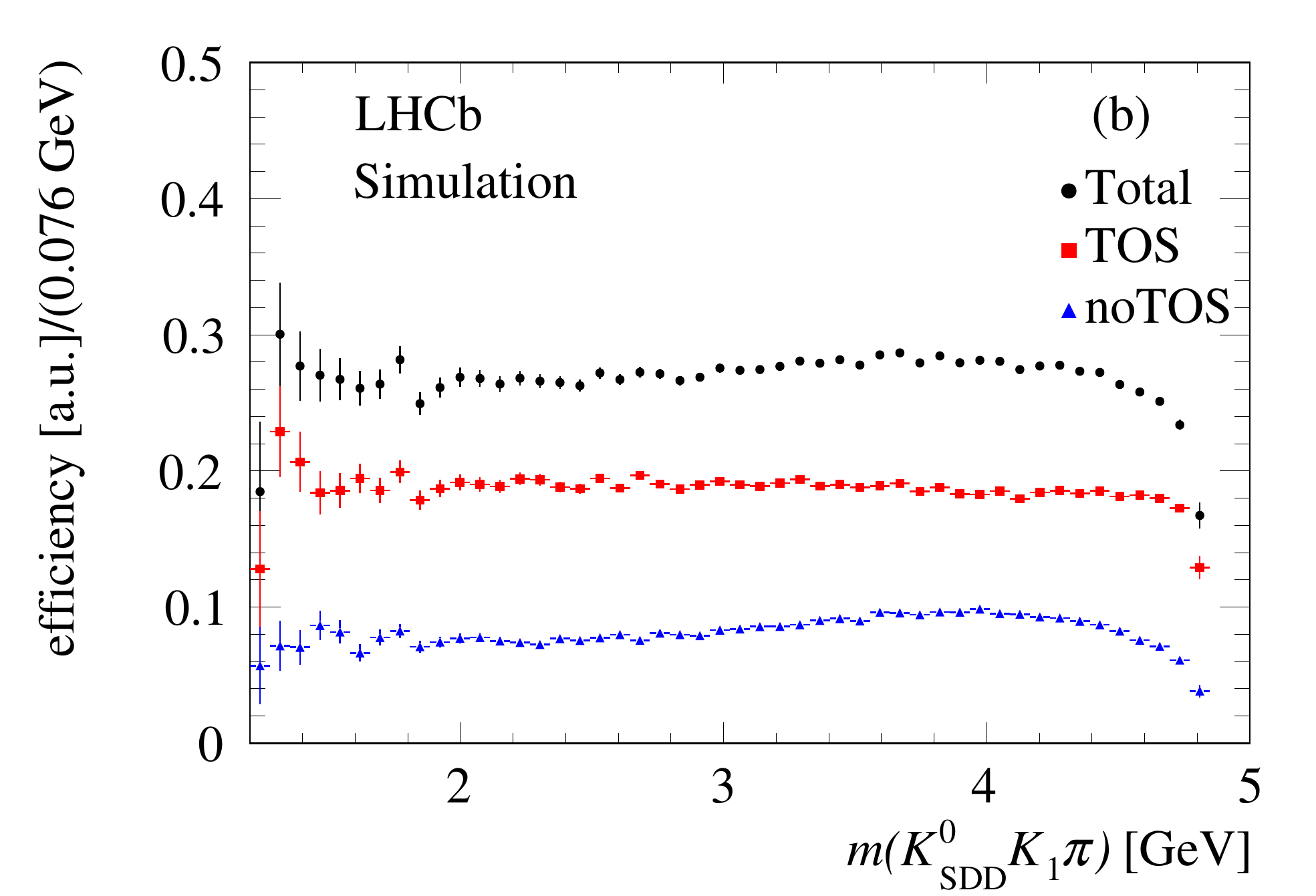}
\caption{\small\label{fig:fig9} Efficiency projections (in arbitrary units) on $m(\KS K_1 \pi)$ for the (a) \KSLL and (b) \KSDD simulation samples separated for trigger conditions.}
\end{figure}

As described in Sec.~\ref{sec:lhcb}, the reconstructed events belong to two trigger categories: TOS and noTOS.
Separate efficiencies are needed for each trigger condition, further divided into \KSLL and \KSDD categories. Figure~\ref{fig:fig9} shows the efficiency distributions as functions of the \kskpi invariant mass, separated by trigger condition and \KS type.
The efficiency evaluations are performed using the same sample of generated events and therefore the scale of the distributions also gives the fraction
of the simulation samples belonging to each category. It is found that the efficiency has a weak dependence on $m(\KS K_1 \pi)$ for all the considered samples.

\subsection{Local efficiencies}

\subsubsection{Efficiency in the \boldmath{\etac--\jpsi}  and \boldmath{\chicone--\etactwo} mass regions}
\label{sec:effloc}

\begin{figure}[tb]
\centering
\small
\includegraphics[width=0.95\textwidth]{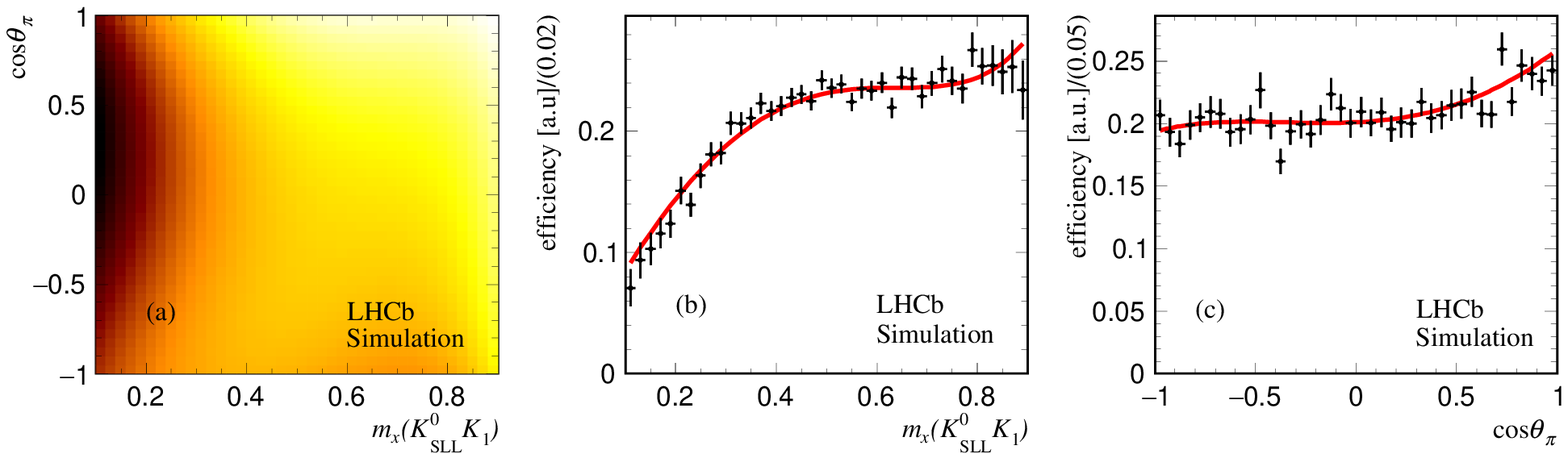}
\includegraphics[width=0.95\textwidth]{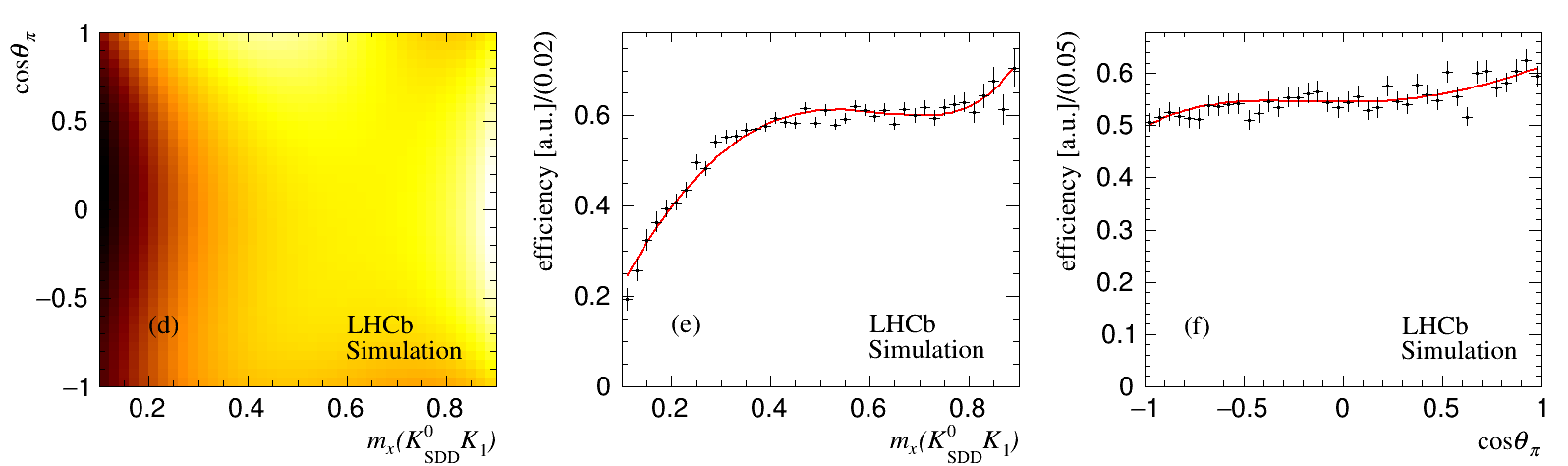}
\caption{\small\label{fig:fig10} Two-dimensional efficiency fitted distributions in the \etac mass region for the (a) \KSLL and (d) \KSDD samples. Efficiency projections (in arbitrary units) for (b)-(c) \KSLL and (e)-(f) \KSDD simulation. The curves are the results from the fits according to Eq.~\ref{eq:2dp}.}
\end{figure}

The efficiency in the \etac--\jpsi mass region is evaluated from a dedicated sample of simulated $\Bp \to \etac \Kp$ decays, where the $\etac$ decays to a $\kskpi$ state, in which the \etac is generated according to a BW function and decays uniformly in its three-body phase space. These simulations are used to evaluate the efficiency across the Dalitz plot, which can be described in terms of two independent variables, chosen to be $m_x(\KS K_1)$ and $\cos \theta_{\pi}$.
Labeling $x=m_x(\KS K_1)$ and $y=\cos \theta_{\pi}$, the efficiency map is smoothed by fitting with a two-dimensional polynomial
\begin{equation}
  \begin{split}
    \eps(x,y) = c_0(1 + c_1x + c_2y + c_3xy + c_4x^2 + c_5y^2 + c_6x^2y + c_7xy^2+ c_8x^3 + \\
    c_9y^3 + c_{10}x^4 +
    c_{11}y^4 + c_{12}x^3y + c_{13}x^2y^2 + c_{14}xy^3 + c_{15}x^5 + \\
    c_{16}x^6 + c_{17}x^6y),
    \end{split}
    \label{eq:2dp}
\end{equation}
separately for the \KSLL and \KSDD samples.
Figure~\ref{fig:fig10} shows the fitted two-dimensional efficiency distributions with the fit projections on $m_x(\KS K_1)$ and $\cos \theta_{\pi}$.

\begin{figure}[tb]
\centering
\small
\includegraphics[width=0.75\textwidth]{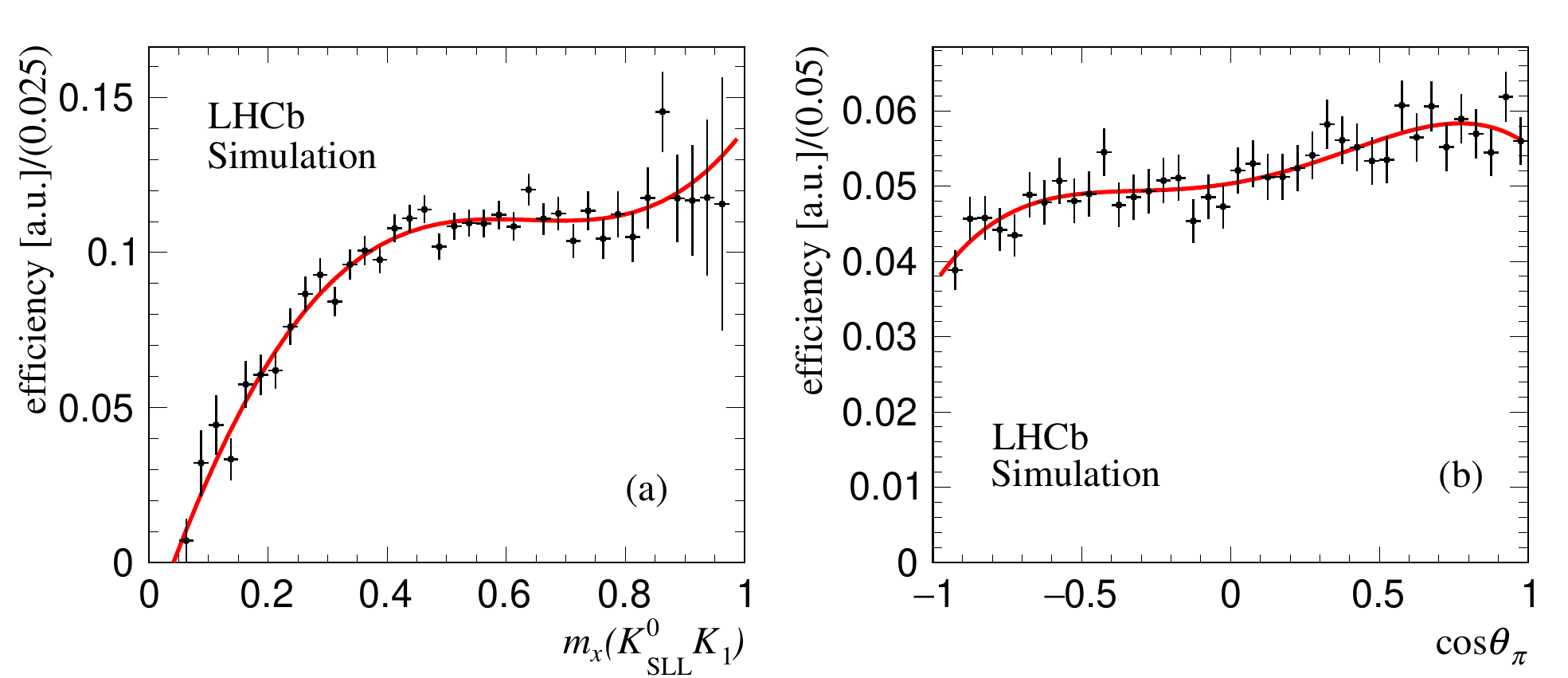}
\includegraphics[width=0.75\textwidth]{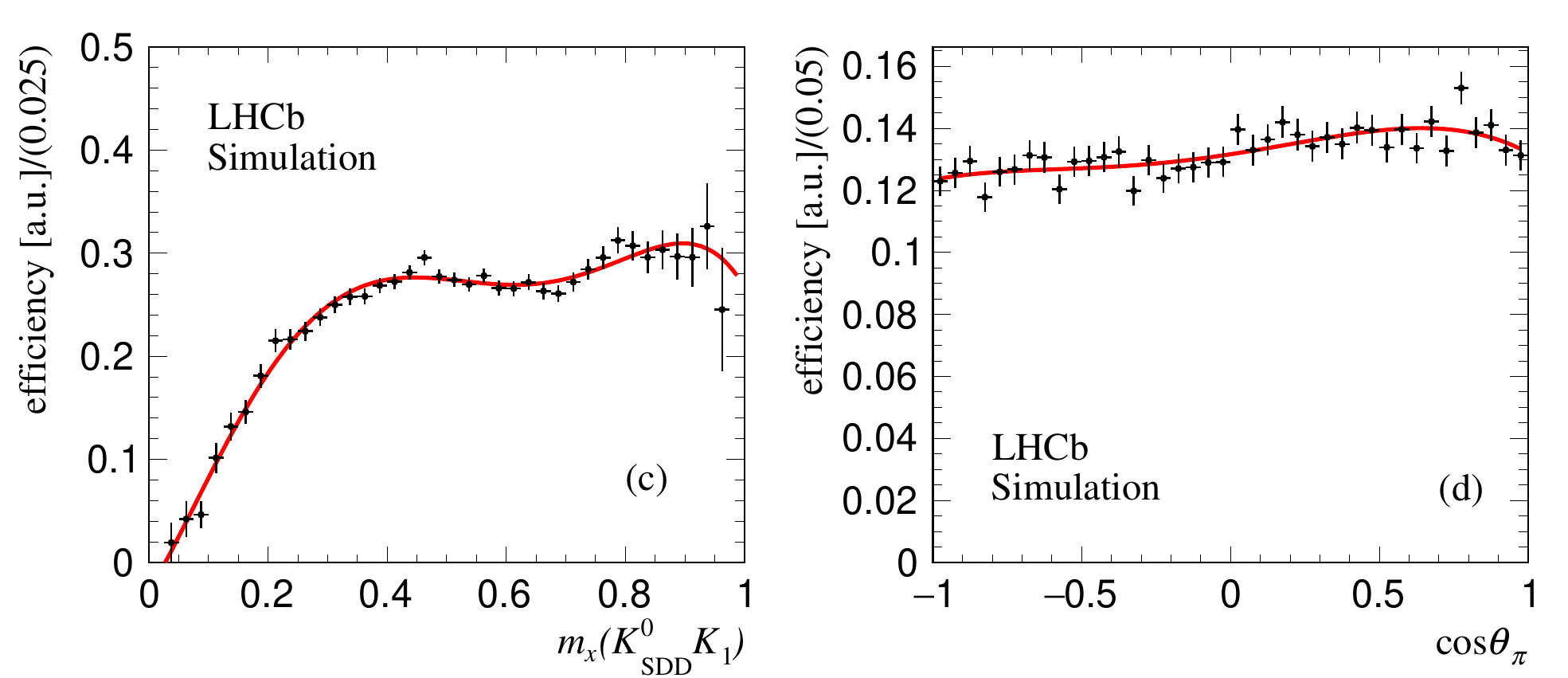}
\caption{\small\label{fig:fig11} Efficiency projections (in arbitrary units) in the \chicone--\etactwo mass region for the (a)-(b) \KSLL and (c)-(d) \KSDD simulation. The curves are the results from fits performed using seventh-order polynomials.}
\end{figure}

The efficiency in the \chicone--\etactwo mass region is computed from the sample of simulated  \bkskkpi decays by selecting events in the 
$[3.46\text{--}3.70]\gev$ mass region. Due to the size of the simulated sample and because of the weak dependence of the efficiency on $\cos \theta_{\pi}$, the efficiency model is obtained from fits with seventh-order polynomials to the efficiency projections on $m_x(\KS K_1)$ and $\cos \theta_{\pi}$, as shown in Fig.~\ref{fig:fig11}. The efficiency is then parameterized as $\epsilon= \epsilon_1(m_x(\KS K_1))\cdot\epsilon_2(\cos \theta_{\pi})$.

\section{Dalitz plot analysis of the \boldmath{\etac} decay to \boldmath{\kskpi}}
\label{sec:etac}

\subsection{Data selection for the \boldmath{\bkskkpip} final state}

This section is devoted to the study of the decay
\begin{equation}
  \begin{split}
    \Bp &\to  \etac \Kp,\\
    \etac \al &\to \KS \Km \pip,
  \end{split}
  \label{eq:etacp}
  \end{equation}
where the \Kp meson is considered a spectator. For simplicity here and in the following, kaons are labeled using their charge. 

Possible backgrounds originating from charm decays to particles amongst the \mbox{$\etac \to \KS \Km \pip$} decay products
and the spectator $\Kp$ are investigated, and no open charm-production is observed in any two-body or three-body mass combinations.
The absence of significant structures ensures that the \etac signal is decoupled from the \Kp and the
\etac analysis can be considered as a simple study of the three-body \etac decay.

The $\pim \Kp$ invariant-mass spectrum, where the \pim comes from the \KS decay, shows no evidence of a \Dz background.
Possible backgrounds from pions misidentified as kaons are considered by assigning the pion mass to each of the two kaons in the final state. Except for the small $D^0 \to \KS \pip \pim$ signal observed in the \Bp sidebands, discussed in Sec.~\ref{sec:mass}, no charm signal is observed in the resulting two-body or three-body mass combinations. Also, no evidence for the \Bp decay to the $\KS \Kp \pip \pim$ state is observed.

\begin{figure}[tb]
\centering
\small
\includegraphics[width=0.48\textwidth]{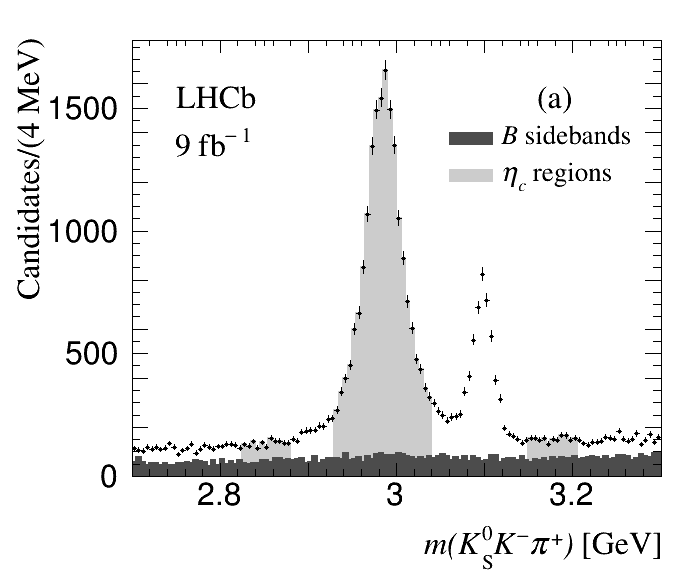}
\includegraphics[width=0.48\textwidth]{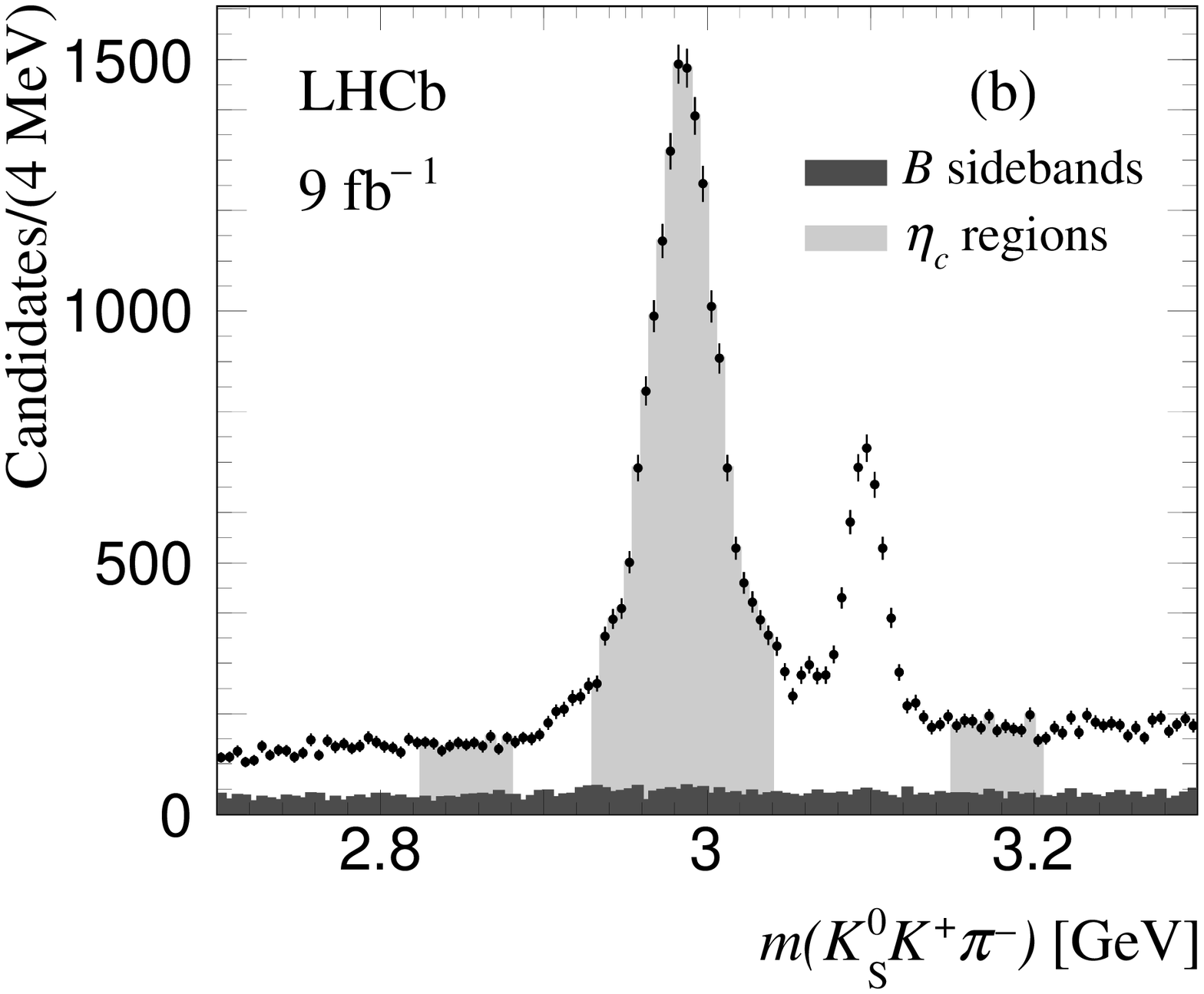}
\caption{\small\label{fig:fig12} Invariant \kskpi mass distributions in the \etac--\jpsi signal region combining the \KSLL and \KSDD data for (a) \bkskkpip and (b) \bkskkpim decays. The dark-gray area represents the \kskpi invariant-mass spectrum from the \Bp sideband; the light-gray areas indicate the \etac signal and sideband regions.}
\end{figure}

The \etac signal region is defined as $[2.935\text{--}3.035]\gev$ while the sideband regions are defined as $[2.83\text{--}2.88]\gev$ and $[3.15\text{--}3.20]\gev$ as shown in Fig.~\ref{fig:fig12}.
The background under the \etac signal peak has two components: (a) from \Bp background, estimated from \Bp sidebands, labeled in the following as incoherent and shown as dark-gray areas in Fig.~\ref{fig:fig12}; (b) from \Bp signal, labeled as coherent and indicated by light-gray shading in Fig.~\ref{fig:fig12}.
The relative fraction of these two components is estimated by 
integrating the contents of the dark- and light-gray areas in the \etac lower and higher sidebands.
The resulting incoherent backgrounds fractions, labeled as $f_B$, are listed in 
Table~\ref{tab:tab5} together with the \etac event yields in the signal region and purities, separated for the \KSLL and \KSDD data.
\begin{table} [tb]
  \centering
  \caption{\small\label{tab:tab5} Candidate events and purities in the \etac signal region and fractions of \Bp sideband contributions in the \etac sideband regions for \bkskkpip and \bkskkpim decays. 
  Low and high indicate the lower and higher sideband candidates.
  Because of the limited statistics, the values of $f_B$ are summed for the \KSLL and \KSDD data.}
  {\small
\begin{tabular}{llrccc}
\hline\\ [-2.3ex]
Final state & \KS type & Candidates & Purity [\%] & low $f_B$ [\%] & high $f_B$ [\%]\cr
\hline\\ [-2.3ex]
\bkskkpip & \KSLL & 4622 & $91.8 \pm 0.4$ &  & \cr
          & \KSDD & 11101 & $89.0 \pm 0.3$ & & \cr
          & Combined \KS & & & $29.6 \pm 2.0$ & $31.5 \pm 2.0$\cr
          \hline\\ [-2.3ex]
\bkskkpim & \KSLL & 5034 & $84.3 \pm 0.5$ & & \cr
          &\KSDD & 11439 & $83.5 \pm 0.3$ & & \cr
          & Combined \KS & & & $25.5 \pm 1.5$ & $22.0 \pm 1.2$\cr
\hline
\end{tabular}
}
\end{table}

\subsection{Data selection for the \boldmath{\bkskkpim} final state}

This section is devoted to the selection of the decay
\begin{equation}
  \begin{split}
    \Bp &\to  \etac \Kp,\\
    \etac \al &\to \KS \Kp \pim,
  \end{split}
  \label{eq:etacm}
  \end{equation}
where two identical \Kp mesons are present. Because of this, the two \Kp mesons are alternately considered as part of the \etac signal decay or as a spectator to it, and every candidate decay therefore appears twice in the sample under study. This final state is affected by a significant background from $\Dzb \to \Kp \pim$ decays,
which is removed as discussed in Sec.~\ref{sec:evsel}.
Labeling the spectator \Kp as $K^+_2$, the $\KS K^+_2\pim$ invariant-mass spectrum shows a small \Dz signal, and events are removed if they lie within $\pm 26\mev$ of the \Dz mass. No other open-charm signal is observed in other two-body or three-body mass combinations. The \kskpi mass spectrum is shown in Fig.~\ref{fig:fig12}(b) with indications of background, signal and sideband regions. Table~\ref{tab:tab5} gives information about the event yields, purities and background composition.

\subsection{Dalitz plot analysis}

\begin{figure}[tb]
\centering
\small
\includegraphics[width=0.48\textwidth]{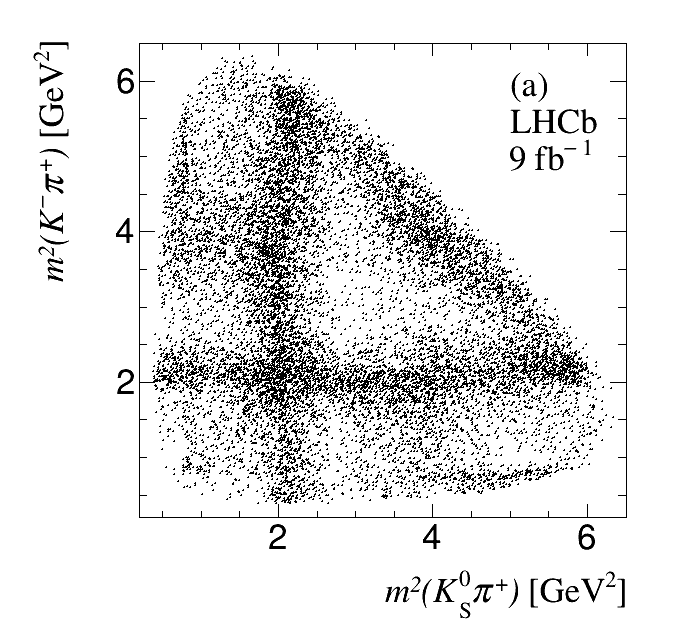}
\includegraphics[width=0.48\textwidth]{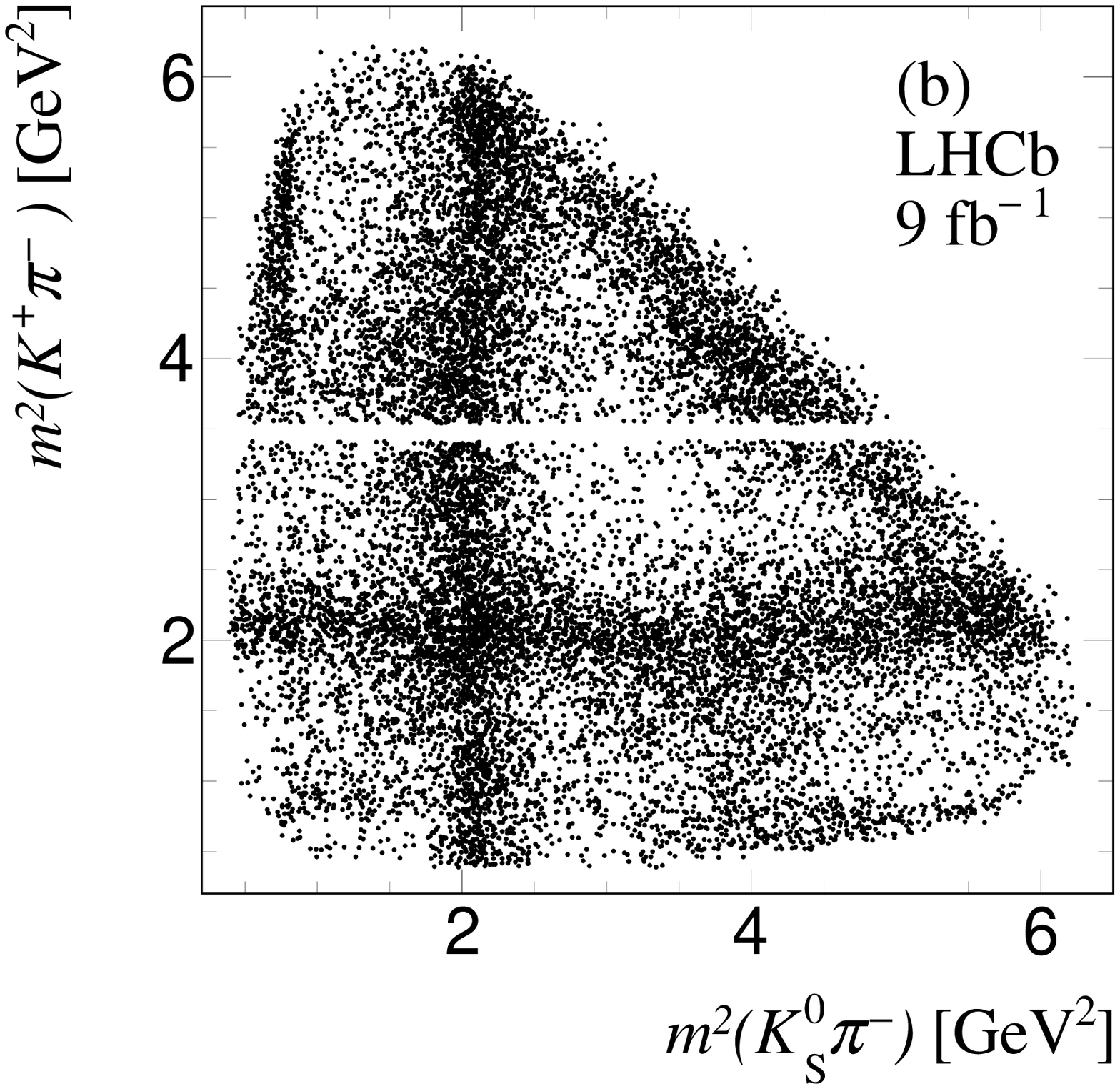}
\caption{\small\label{fig:fig13} Dalitz plot of $\etac \to \kskpi$ decays for (a) \bkskkpip and (b) \kskkpim. The empty horizontal band in (b) is due to the removal of the $\Dzb \to \Kp \pim$ channel.}
\end{figure}

The \etac Dalitz plot is shown in Fig.~\ref{fig:fig13}, for (a) \bkskkpip and (b) \mbox{\bkskkpim} data. It is dominated by horizontal and vertical bands due to the presence of the $K^*_0(1430)^{+,0}$ resonances. 

\subsubsection{Dalitz plot analysis method}

An amplitude analysis of the \kskpi final state in the \etac mass region is performed, using unbinned maximum-likelihood fits.
The likelihood function is written as
\begin{eqnarray}    
 \mathcal{L} = \prod_{n=1}^N\bigg[P\cdot\epsilon(x'_n,y'_n)\frac{\sum_{i,j} c_i c_j^* A_i(x_n,y_n) A_j^*(x_n,y_n)}{\sum_{i,j} c_i c_j^* I_{A_i A_j^*}} 
+(1-P)\frac{\sum_{i} k_iB_i(x_n,y_n)}{\sum_{i} k_iI_{B_i}}\bigg]
\end{eqnarray}
\noindent where
$N$ is the number of events in the signal region. For the $n$-th event, $x_n=m^2(K^0 \pip)$ and $y_n=m^2(\Km \pip)$. The signal purity ($P$) is obtained from the fit to the \kskpi invariant-mass spectrum (see Fig.~\ref{fig:fig4}) and listed in Table~\ref{tab:tab5}. The efficiency $\epsilon(x'_n,y'_n)$ is parameterized as a function of $x'_n=m(\KS K)$ and $y'_n=\cos \theta_{\pi}$ and described in Sec.~\ref{sec:effy}. The complex signal-amplitude contribution for the $i$-th signal component is indicated as $A_i(x_n,y_n)$. The corresponding complex parameter $c_i$ is allowed to vary in the fit, except for the largest amplitude, the ($K \pi$ {\it S}-wave)$K$ (labeled in the following as \Kpisk) in the quasi-model-independent (QMI) analysis (Sec.~\ref{sec:qmi}) or the $K^*_0(1430)K$ in the isobar model analysis (Sec.~\ref{sec:iso}), which is taken as the reference, setting the modulus $|c_1|=1$ and phase $\phi_1=0$.
The background probability density function for the $i$-th background component is indicated as $B_i(x_n,y_n)$, and it is assumed that interference between signal and background amplitudes can be ignored.  The magnitude of the $i$-th background component is indicated by $k_i$ and is obtained by fitting to the sideband regions and interpolating linearly in the signal region. The expressions $I_{A_i A_j^*}=\int A_i (x,y)A_j^*(x,y) \epsilon(x', y')\ {\rm d}x{\rm d}y$ and 
$I_{B_i}=\int B_i(x,y)\ {\rm d}x{\rm d}y$ represent normalization
 integrals for signal and background. Numerical integration is performed on phase-space generated events~\cite{James:1968gu} with \etac signal and background generated according to their experimental distributions. For resonances with free parameters, integrals are re-computed at each minimization step.
Integrals corresponding to the background components, and fits involving amplitudes with fixed resonance parameters, are computed only once.
 Both the \KSLL and \KSDD data are included in the likelihood function, each dataset with its own efficiency and purity. The $A_i$ amplitudes are taken from Ref.~\cite{CLEO:2000fvk,BaBar:2010wqe}.  The standard Blatt--Weisskopf~\cite{Blatt:1952ije} form factors are used at the resonance vertices with radius $r$ fixed to 1.5$\gev^{-1}$.
 The decay of the \etac to $K \Kb \pi$ is governed by strong interactions with two allowed processes: $\etac \to a_L \pi$ and $\etac \to K^*_L \Kbar$, where $a_L$ indicates an isospin-1 spin-$L$ resonance and $K^*_L$ indicates an isospin-1/2 spin-$L$ resonance. Conservation of $G$-parity for the $\etac \to a_L \pi$ case implies $L$ to be even. For the decay $\etac \to K^*_L \Kbar$, $G$-parity relates the final states $K^*_L \Kbar$ and $\Kbar^*_L K$ as $(K^*_L \Kbar + G\Kbar^*_L K)/\sqrt{2}$. Since $G$-parity is positive, it follows that in the
Dalitz plot, the $K^*_L \Kbar$ and $\Kbar^*_L K$ bands will interfere constructively and all values of $L$ are in principle allowed. In particular, the $a_0(980)$ resonance is described by a coupled-channel Breit--Wigner as measured in Ref.~\cite{Abele:1998qd}.
In the following, all resonances listed in Ref.~\cite{Workman:2022ynf} and satisfying the above criteria are included or tested in the Dalitz-plot analysis.
The two-body invariant-mass resolution varies with invariant-mass and is approximately $7\mev$ in the $K \pi$ mass region between 1 and 2 \gev, much smaller than the width of the resonances present in this analysis. Therefore, resolution effects are ignored.
The efficiency-corrected fractional contribution $f_i$, due to resonant or non-resonant contribution $i$, is defined as follows:
\begin{equation}
f_i = \frac {|c_i|^2 \int |A_i(x,y)|^2 {\rm d}x {\rm d}y}
{\int |\sum_j c_j A_j(x,y)|^2 {\rm d}x {\rm d}y},
\end{equation}
where $x$ and $y$ indicate the Dalitz-plot variables and 
the integrals in the numerator and denominator are computed on a large sample of simulated events generated uniformly in the decay phase space.
The $f_i$ do not necessarily sum to 100\% because of interference effects. The uncertainty for each $f_i$ is evaluated by propagating the full covariance matrix obtained from the fit.
Interference fractions between amplitudes $i$ and $j$ are evaluated as:
\begin{equation}
f_{ij} = \frac {\int c_ic^*_j A_i(x,y) A^*_j(x,y) {\rm d}x {\rm d}y}
{\int |\sum_j c_j A_j(x,y)|^2 {\rm d}x {\rm d}y}.
\end{equation}

After the fit, a large simulated sample is prepared, where events are generated uniformly in the decay phase space~\cite{James:1968gu}. These events are weighted by the fitted likelihood function, normalized to the data-event yields and compared to the data on several invariant-mass and angular projections.
To assess the fit quality, the Dalitz plot is divided into a grid of $n \times n$
cells ($n$ depends on the event yields and Dalitz-plot size) and only those cells containing at least two simulated events are considered. A $\chi^2$ estimator is used, defined as 
$\chi^2 = \sum_{i=1}^{N_{\rm cells}} (N^i_{\rm obs}-N^i_{\rm exp})^2/\varepsilon^2$, where $N^i_{\rm obs}$ and $N^i_{\rm exp}$ are event yields from data and simulation, respectively. Here $\varepsilon=\sqrt{N^i_{\rm exp}}$ for cells containing more than 9 entries, while it is approximated as the average of the lower and upper Poisson 68\% CL errors for lower statistics cells. The figure of merit is defined as \chisqndf where ndf is computed as $N_{\rm cells}-n_{\rm par}$ and $n_{\rm par}$ is the number of free parameters. 

Two amplitude fits are then performed. The first employs a QMI description of the $K \pi$ {\it S}-wave (Sec.~\ref{sec:qmi}) and the second uses an isobar model (Sec.~\ref{sec:iso}).

\subsubsection{Fits to the \boldmath{\etac} sidebands}
The background model is obtained by fitting the \etac sidebands.
An unbinned maximum-likelihood fit is performed by inserting, one-by-one, all possible $a_0$ and $a_2$ resonances decaying to a $\KS K$ state and all possible $K^*$ resonances decaying to a $K \pi$ state~\cite{Workman:2022ynf} which could contribute to the \kskpi final state. To describe the fraction of background candidates associated to the \Bp decay (see Table~\ref{tab:tab5}), resonances are described by relativistic Breit--Wigner functions with appropriate Blatt--Weisskopf~\cite{Blatt:1952ije} form factors using angular terms just as for the \etac decay but, for $K^*$ resonances,  there is no symmetrization requirement for charged and neutral contributions. Interference is also allowed between all the contributing amplitudes. The fraction of background not associated to the \Bp decay is assumed to be uniform, except for a small contribution from 
$K^*(892)$ resonances, described only by relativistic Breit--Wigner functions.
Small open-charm contributions are described by simple Gaussian functions with parameters fitted to the data. 
No efficiency is included in the description of the sidebands.
The amplitudes are numerically integrated using a phase-space simulation generated according to the \kskpi invariant-mass distributions in the sidebands
shown in Fig.~\ref{fig:fig12}. 
Contributions which are uniform in the phase space of the decay, having fixed fractions $f_B$ (given in Table~\ref{tab:tab5}), are included.
Due to the limited sample size in the sideband regions, the \KSLL and \KSDD data are combined. 

\subsection{Fit using a QMI description of the \boldmath{$K \pi \ S$}-wave}
\label{sec:qmi}

To measure the $I=1/2$  $K \pi$ {\it S}-wave, the QMI technique described in Refs.~\cite{E791:2002xlc,BaBar:2015kii} is used.
In this fit, the $K \pi$ {\it S}-wave, being the largest contribution, is taken as the reference amplitude.
The $K \pi$ invariant-mass spectrum is divided into 37 equally-spaced mass intervals each $50\mev$ wide, and two new fit parameters are added for each interval:
the amplitude and the phase of the $K \pi$ {\it S}-wave. 
The amplitude is fixed to 1.0 and its phase to $\pi/2$ at an arbitrary point in the mass spectrum, chosen to be interval 16, corresponding to a mass of $1.45\gev$. The number of free parameters associated to the description of the QMI $K\pi$ {\it S}-wave is therefore 72.
The $K \pi$ {\it S}-wave amplitude in bin $j$ is written as

\begin{equation}
  A_{S\myhyphen\rm{wave}}^j = \frac{1}{\sqrt{2}}(a_j^{K \pi}e^{i\phi_j^{K \pi}} + a_j^{K^0 \pi}e^{i\phi_j^{K^0\pi}}),
  \label{eq:amp}
\end{equation}

\noindent where the amplitude $a_j^{K \pi}(m)=a_j^{K^0 \pi}(m)$ and the phase $\phi_j^{K \pi}(m) = \phi_j^{K^0 \pi}(m)$ at the $K \pi$ mass $m$.
All the $K^*_L(1430) \Kbar$ contributions with $L=0,2$ are symmetrized in the same way as for the {\it S}-wave amplitude.

The fit is initiated by performing a randomized scan for the $K \pi$ {\it S}-wave solution with arbitrary start values of parameters. In addition, expected contributions from known resonances such as $a_0(980)$, $a_2(1320)$, $K^*_2(1430)$, $a_0(1450)$, $a_2(1750)$ and $K^*_2(1900)$ are included. The $a_0(1700)$ resonance, recently observed in the Dalitz-plot analysis of $\etac \to \eta \pip \pim$~\cite{BaBar:2021fkz} is also included and found to be significant.
The presence of an $a_0(1950)$ resonance, for which some evidence was found in the \etac Dalitz-plot analysis Ref.~\cite{BaBar:2015kii}, is tested, but its contribution is found to be consistent with zero.
In the search for the best solution, the fit is iterated from the first found solution and additional resonances are added one-by-one; the process is completed when additional contributions give fit fractions consistent with zero.
Spin-1 $K^*$ resonances ($K^*(892)$, $K^*(1410)$ and $K^*(1680)$) are also tested but their contributions are found to be consistent with zero.
The inclusion of additional spin-0 and spin-2 resonance components with unconstrained resonance parameters leads to no improvement of the fit quality.

\begin{figure}[tb]
\centering
\small
\includegraphics[width=0.47\textwidth]{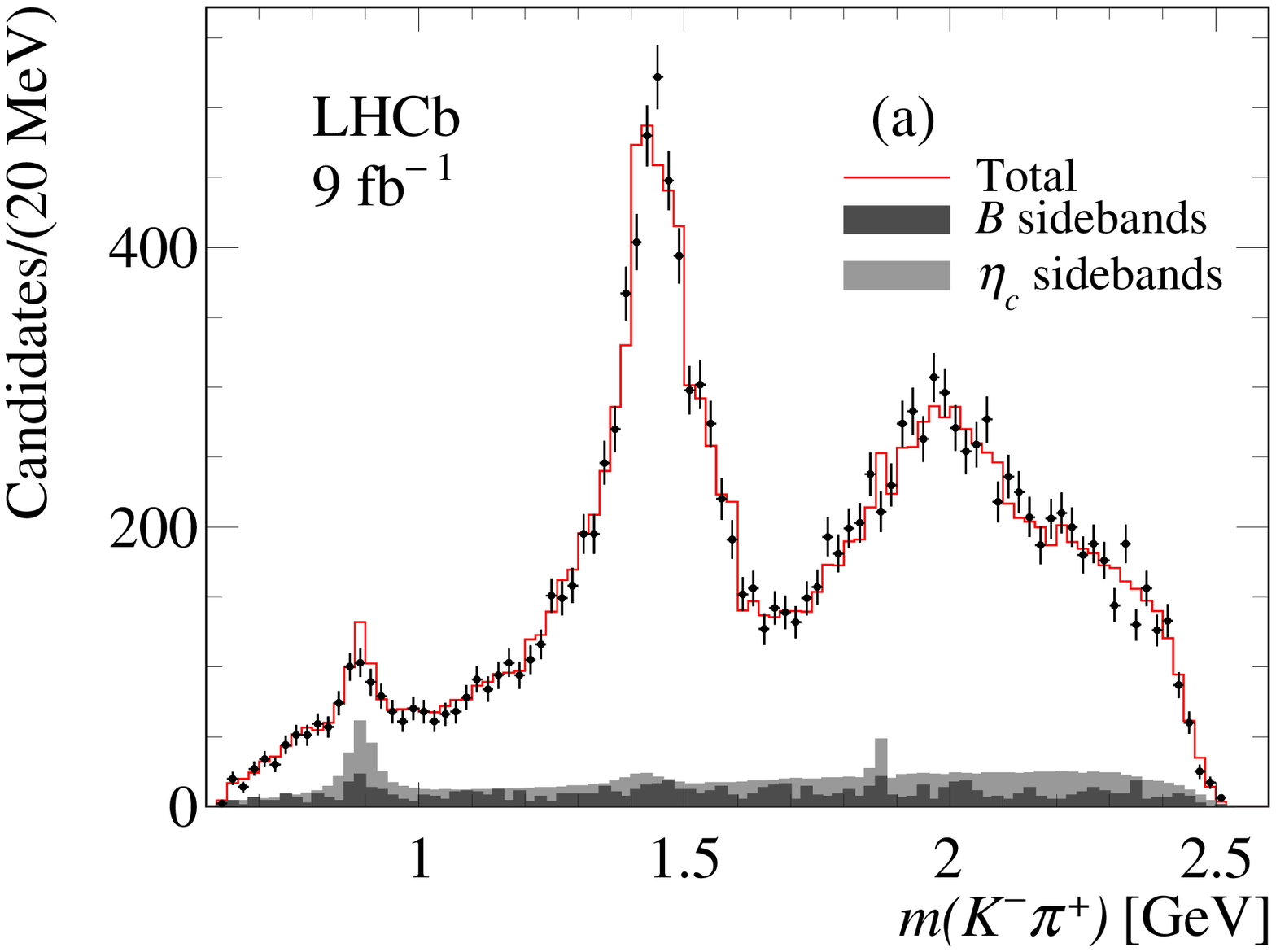}
\includegraphics[width=0.47\textwidth]{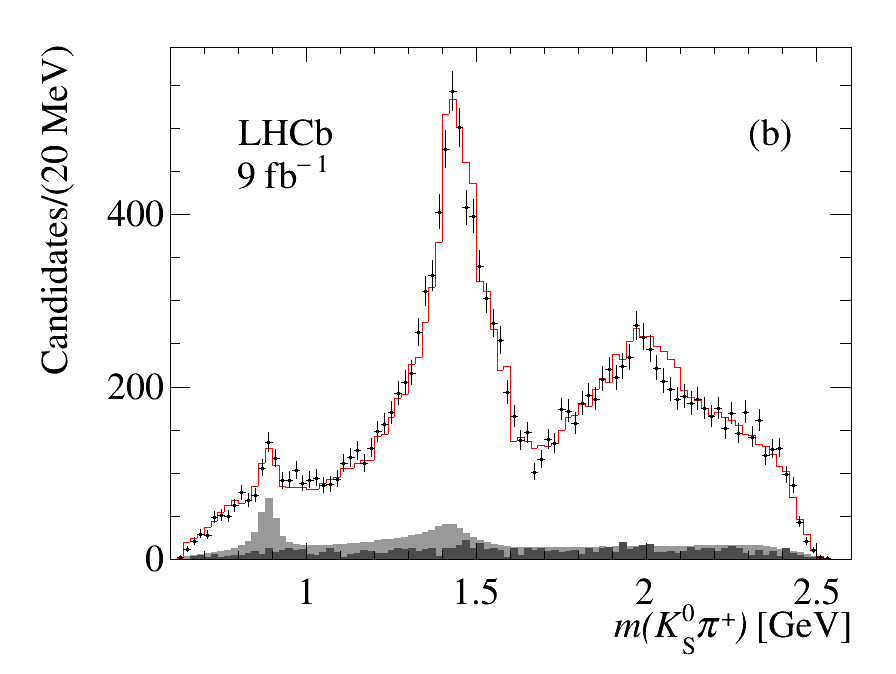}
\includegraphics[width=0.47\textwidth]{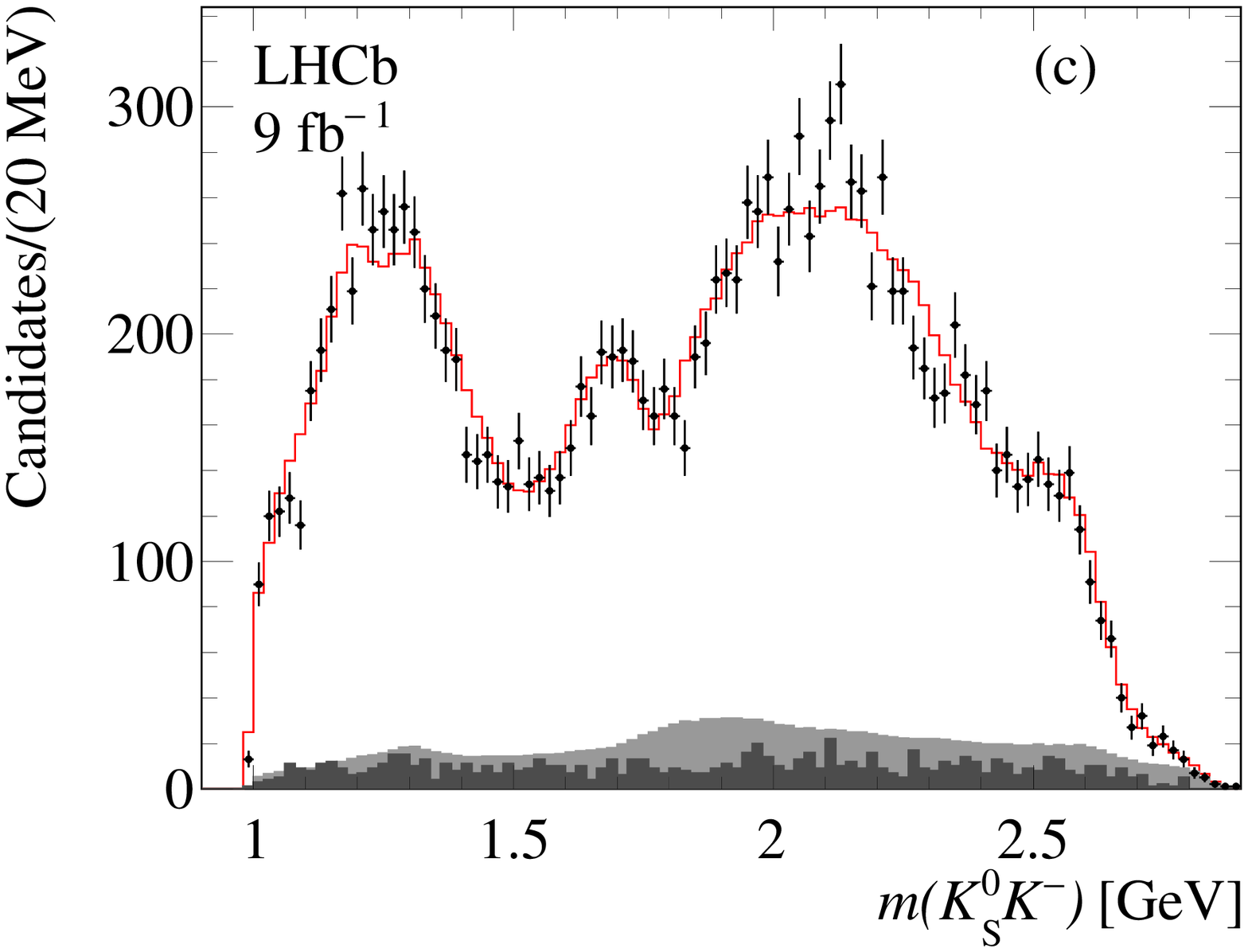}
\caption{\small\label{fig:fig14} Fit projections on the $\Km \pip$, $\KS \pip$ and $\KS \Km$ invariant-mass distributions from the Dalitz-plot analysis of the \etac decay using the QMI model for \bkskkpip data. }
\end{figure}

\begin{figure}[tb]
\centering
\small
\includegraphics[width=0.47\textwidth]{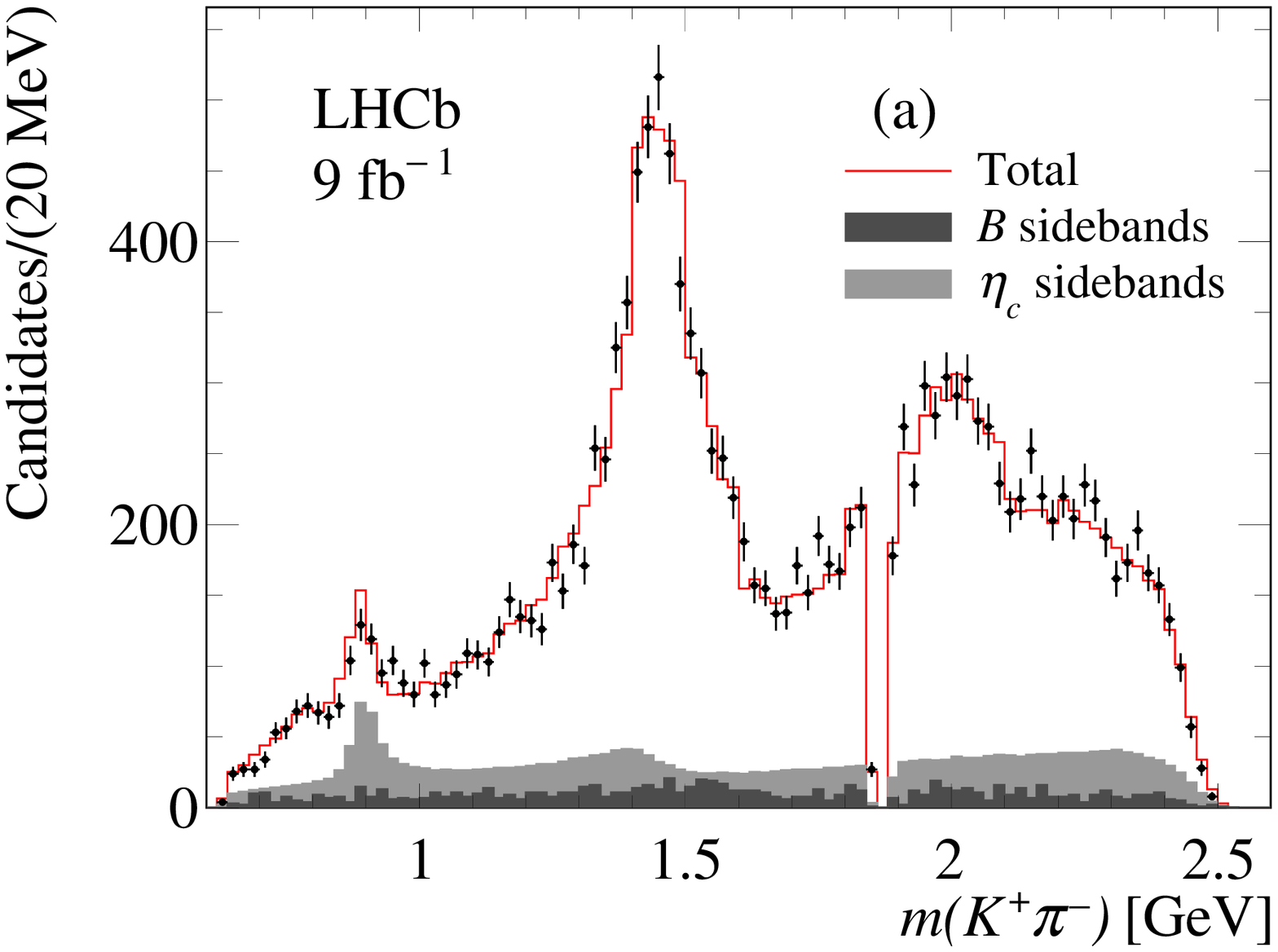}
\includegraphics[width=0.47\textwidth]{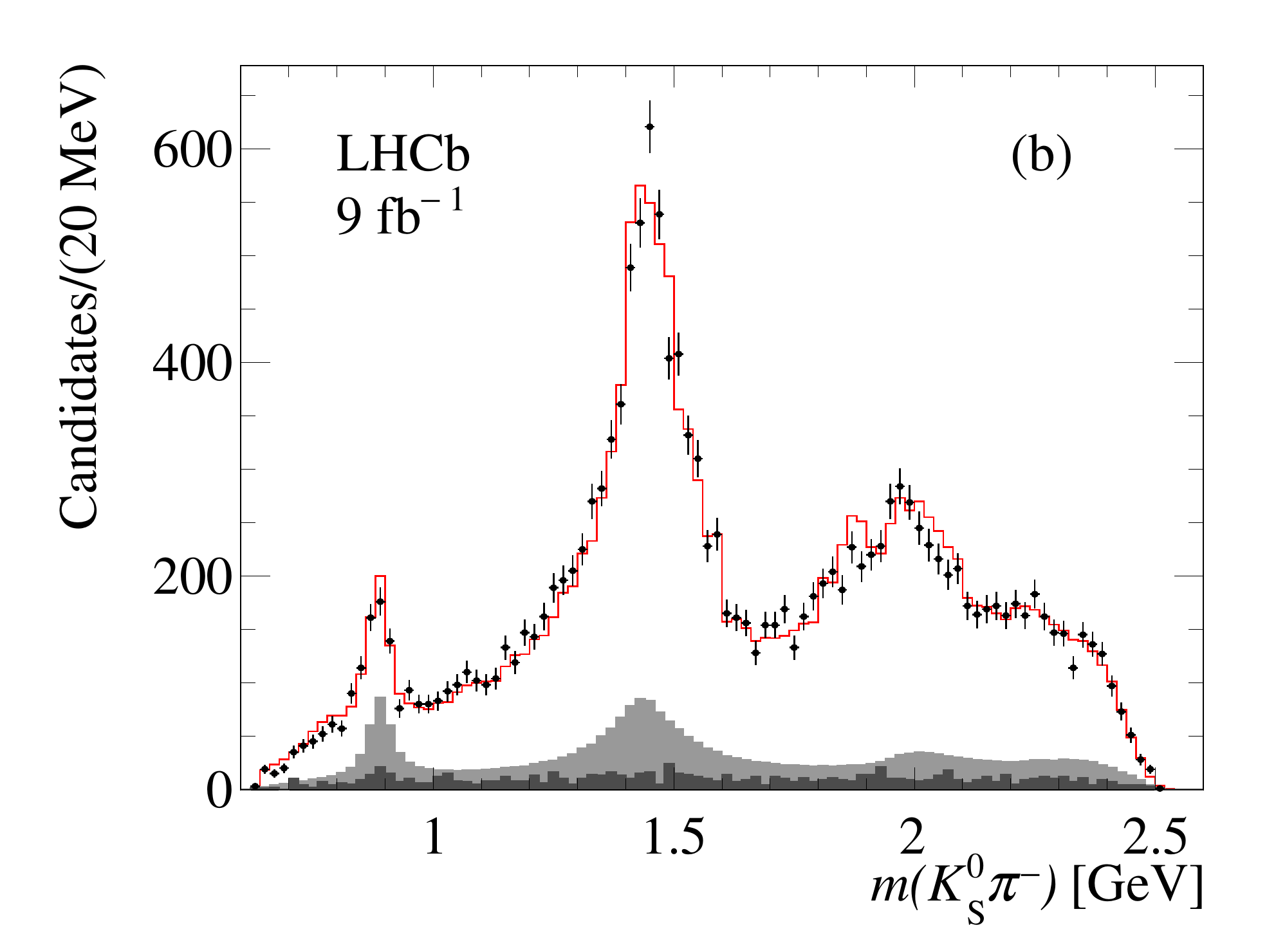}
\includegraphics[width=0.47\textwidth]{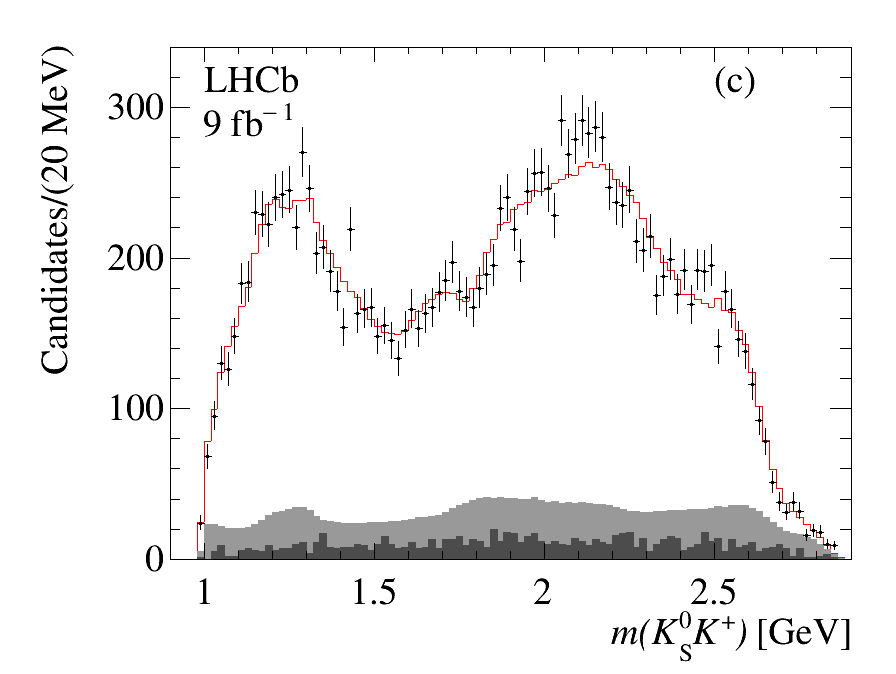}
\caption{\small\label{fig:fig15} Fit projections on the $\Kp \pim$, $\KS \pim$ and $\KS \Kp$ invariant-mass distributions from the Dalitz-plot analysis of \etac decay using the QMI model for the \bkskkpim data.}
\end{figure}

Figures~\ref{fig:fig14} and \ref{fig:fig15} show the Dalitz-plot projections with superimposed projections of the fit function for \bkskkpip and \bkskkpim data, respectively. The distributions of the fitted background, interpolated from the \etac and \Bp sidebands, are also included.
The fractional contributions and relative phases from the fit are given in Table~\ref{tab:tab6}.
There is a significant contribution from interference terms, evidenced by a sum of fractions much larger than 100\%. This is a common feature of Dalitz-plot
analyses dealing with several broad resonances having the same quantum numbers~\cite{BaBar:2006hyf}. 
Interferences between amplitudes are listed in Table~\ref{tab:tab24} for absolute values above 3\%.

\begin{table} [tb]
  \centering
  \caption{\small\label{tab:tab6} Results from the Dalitz-plot analysis of the \etac decay in (top) \bkskkpip, (center) \bkskkpim and (bottom) inverse-variance-weighted averages. The QMI model is used for the $K \pi$ {\it S}-wave.}
\begin{tabular}{lrr}
\hline\\ [-2.3ex]
Final state & Fraction [\%] &   Phase [rad] \alp \alp \all \cr
\hline\\ [-2.3ex]
& \bkskkpip & \cr
\hline\\ [-2.3ex]
\Kpisk& $120.6 \pm 2.4 \pm 5.4$ & 0. \alp \alp \alp \cr
$a_0(1450) \pi$ & $2.4 \pm 0.4 \pm 0.8$ & $2.48 \pm 0.07 \pm 0.09$ \cr
$K^*_2(1430) K$ & $16.6 \pm 0.8 \pm 0.9$ & $4.31 \pm 0.03 \pm 0.11$ \cr
$a_2(1320) \pi$ & $0.7 \pm 0.2 \pm 0.5$ & $4.18 \pm 0.10 \pm 0.27$ \cr
$a_0(980) \pi$ & $ 11.3 \pm 0.6 \pm 0.9$ & $-2.93 \pm 0.03 \pm 0.03$ \cr
$a_0(1700) \pi$ & $ 1.5 \pm 0.2 \pm 0.2$ & $2.00 \pm 0.08 \pm 0.14$ \cr
$K^*_2(1980) K$ & $2.8 \pm 0.3 \pm 1.0$ & $-0.08 \pm 0.07 \pm 0.14$ \cr
$a_2(1750) \pi$ & $ 0.2 \pm 0.1 \pm 0.1$ & $-3.56 \pm 0.20 \pm 0.24$ \cr
\hline\\ [-2.3ex]
Sum & $156.1 \pm 2.7 \pm 11.4$ & \cr
$\chisqndf=1706/(1597-17)=1.08$ & &\cr
\hline\\ [-2.3ex]
& \bkskkpim & \cr
\hline\\ [-2.3ex]
\Kpisk& $106.0 \pm 2.8 \pm 8.5$ & 0.  \alp \alp \alp \cr
$a_0(1450) \pi$ & $0.8 \pm 0.3 \pm 0.4$ & $1.64 \pm 0.14 \pm 0.47$ \cr
$K^*_2(1430) K$ & $17.8 \pm 0.9 \pm 1.0$ & $4.32 \pm 0.03 \pm 0.13$ \cr
$a_2(1320) \pi$ & $0.7 \pm 0.2 \pm 0.5$ & $4.22 \pm 0.11 \pm 0.93$ \cr
$a_0(980) \pi$ & $9.7 \pm 0.6 \pm 0.3$ & $-3.02 \pm 0.04 \pm 0.05$ \cr
$a_0(1700) \pi$ & $ 0.8 \pm 0.2 \pm 0.2$ & $2.10 \pm 0.11 \pm 0.24$ \cr
$K^*_2(1980) K$ & $6.3 \pm 0.6 \pm 1.9$ & $0.13 \pm 0.05 \pm 0.08$ \cr
$a_2(1750) \pi$ & $ 0.2 \pm 0.2 \pm 0.3$ & $-3.87 \pm 0.22 \pm 0.16$ \cr
\hline\\ [-2.3ex]
Sum & $143.7\pm 2.9 \pm 8.8$ & \cr
$\chisqndf=1686/(1589-17)=1.07$ &  &\cr
\hline\\ [-2.3ex]
& \bkskkpi & \cr
\hline\\ [-2.3ex]
\Kpisk & $114.4 \pm 1.8 \pm 4.6$& 0. \alp \alp \alp \cr
$a_0(1450) \pi$ & $1.4 \pm 0.2 \pm 0.4$ & $2.31 \pm 0.06 \pm 0.09$ \cr
$K^*_2(1430) K$ & $17.1 \pm 0.6 \pm 0.7$ & $4.32 \pm 0.02 \pm 0.08$ \cr
$a_2(1320) \pi$ & $0.7 \pm 0.1 \pm 0.4$ & $4.20 \pm 0.08 \pm 0.26$ \cr
$a_0(980) \pi$ & $ 10.5 \pm 0.4 \pm 0.4$ & $-2.97 \pm 0.02 \pm 0.03$\cr
$a_0(1700) \pi$ & $ 1.0 \pm 0.1 \pm 0.1$ & $2.04 \pm 0.06 \pm 0.12$\cr
$K^*_2(1980) K$ & $3.5 \pm 0.3 \pm 0.9$ & $0.06 \pm 0.04 \pm 0.07$\cr
$a_2(1750) \pi$ & $ 0.2 \pm 0.1 \pm 0.1$ & $-3.69 \pm 0.15 \pm 0.16$ \cr
\hline\\ [-2.3ex]
Sum & $148.8 \pm 2.0 \pm 4.8$ &  \cr
\hline
\end{tabular}
\end{table}

\begin{table} [tb]
  \centering
  \caption{\small\label{tab:tab24} Fractional interference contributions from the Dalitz-plot analysis of the \etac decay in \bkskkpip decays using the QMI model. Absolute values
  less than 3\% are not listed.}
  {\small
\begin{tabular}{llr}
\hline\\ [-2.3ex]
Amplitude 1 & Amplitude 2 & Fraction [\%] \cr
\hline\\ [-2.3ex]
\Kpisk & $a_0(1450) \pi$ & $-14.1 \pm 1.1$ \cr
\Kpisk & $K^*_2(1430) K$ & $-27.8 \pm 1.0$ \cr
\Kpisk & $a_0(980) \pi$ & $-9.2 \pm 1.3$ \cr
$a_0(980) \pi$ & $a_0(1450) \pi$ & $-3.7 \pm 0.3$ \cr
\hline
\end{tabular}
}
\end{table}


To test the fit quality, the efficiency-uncorrected Legendre-polynomial moments $P_L$ are computed in each $\Km\pip$, $\KS \pip$, and $\KS \Km$ mass interval by weighting each event by a $P_L(\cos \theta)$ function, where $\theta$ is the corresponding helicity angle~\cite{BaBar:2008myc}. These distributions are shown in Figs.~\ref{fig:fig26},~\ref{fig:fig27} and~\ref{fig:fig28} of Appendix~\ref{sec:app} as functions of the $\Km \pip$, $\KS \pip$, and $\KS \Km$ invariant masses, respectively. The extracted moment distributions are compared with the expected Legendre polynomial moment distributions obtained from simulated data weighted by the fit results.
Good agreement for all
the distributions is observed, which indicates that 
the fit is able to reproduce the local structures apparent
in the Dalitz plot.

Systematic uncertainties,  evaluated separately for the \bkskkpip and \mbox{\bkskkpim} final states, are summarized in Tables~\ref{tab:tab7}~and~\ref{tab:tab8}, respectively.  
The effect of the efficiency correction, labeled as ``Eff'' in Table~\ref{tab:tab7}, is evaluated by replacing the two-dimensional fitted efficiency $\epsilon(m_x(\KS K),\cos \theta_{\pi})) $ with the product of the fits to the efficiency projections $\epsilon=\epsilon_1(m_x(\KS K))\cdot\epsilon_2(\cos \theta_{\pi})$. The uncertainty related to trigger effects (Trig) is estimated by performing
separate Dalitz-plot analyses of the TOS and noTOS data samples using appropriate efficiency evaluations. The inverse-variance-weighted averages are compared with the values from the reference fit, and the absolute values of the deviations are taken as systematic uncertainties.
The uncertainty on the signal purity (Pur) is evaluated by performing Dalitz-plot analyses on datasets selected varying the BDT classifier requirement, which changes the purity by $\pm5\%$. Their inverse-variance-weighted averages are compared with the values from the reference fit, and the absolute deviations are taken as systematic uncertainties.
The radius $r$ of the Blatt--Weisskopf factor, by default fixed to $1.5\gev^{-1}$, is varied between 0.5 and $2.5\gev^{-1}$.
The uncertainty due to the background model (Back) is evaluated by performing 100 fits to the same dataset, where in each fit all the parameters describing the background model are randomly varied within their statistical uncertainties according to a Gaussian distribution. The absolute values of the averages of the differences with respect to the reference values are taken as systematic uncertainties on amplitudes and phases.
Possible fit biases are evaluated by generating, from the solution found by the fit, 100 pseudoexperiments having the same size as the data. The average of the differences between the reference solution and that from the pseudoexperiments gives an evaluation of the fit bias. It is found that this contribution is not significant and it is not included in the list of the systematic uncertainties.
The effect of varying the position of the sidebands, from which the background model is derived, is investigated and found to have negligible impact on the total systematic uncertainty, and is therefore ignored.
All the contributions are added in quadrature (Tot).

 \begin{table} [tb]
\centering
\caption{\small\label{tab:tab7} Systematic uncertainties  on (left) fractional contributions (\%) and (right) phases (rad) in the \etac Dalitz-plot analysis in the
  \bkskkpip decay using the QMI model for the $K \pi$ {\it S}-wave.}
 {\small
\begin{tabular}{l|ccccc|c|ccccc|c}
\hline\\ [-2.3ex]
Final state & Eff & Trig & Pur & $r$ & Back & Tot & Eff & Trig & Pur & $r$ & Back& Tot\cr
\hline\\ [-2.3ex]
\Kpisk& 3.27 & 0.89 & 1.03 & 4.01 & 0.48 & 5.37 & - & - & - & - & - & -\cr
$a_0(1450) \pi$ & 0.65 & 0.29 & 0.24 & 0.16 & 0.06 & 0.77 & 0.04 & 0.07 & 0.03 & 0.00 & 0.03 & 0.09\cr
$K^*_2(1430) K$ & 0.02 & 0.17 & 0.04 & 0.86 & 0.48& 1.00 & 0.08 & 0.06 & 0.03 & 0.05 & 0.01  & 0.11 \cr
$a_2(1320) \pi$ & 0.38 & 0.10 & 0.34 & 0.04 & 0.08 & 0.53 & 0.21 & 0.08 & 0.14 & 0.05 & 0.01 & 0.27 \cr
$a_0(980) \pi$ & 0.70 & 0.05 & 0.47 & 0.12 & 0.34 & 0.92 & 0.01 & 0.02 & 0.02 & 0.00 & 0.02 & 0.04 \cr
$a_0(1700) \pi$ & 0.12 & 0.05& 0.05 & 0.08 & 0.01 & 0.16 & 0.10 & 0.08 & 0.02 & 0.00 & 0.00 & 0.14 \cr
$K^*_2(1980) K$ & 0.04 & 0.08 & 0.19 & 0.96 & 0.16 & 1.00 & 0.10 & 0.06 & 0.04 & 0.08 & 0.01 & 0.14 \cr
$a_2(1750) \pi$ & 0.05 & 0.01 & 0.05 & 0.03 & 0.03 &0.09 & 0.04 & 0.14 & 0.05 & 0.11 & 0.06 & 0.20 \cr
\hline
\end{tabular}
}
 \end{table}

 \begin{table} [tb]
  \centering
  \caption{\small\label{tab:tab8} Systematic uncertainties on (left) fractional contributions (\%) and (right) phases (rad) in the \etac Dalitz-plot analysis in
  the \bkskkpim decay using the QMI model for the $K \pi$ {\it S}-wave.}    
  {\small
\begin{tabular}{l|ccccc|c|ccccc|c}
  \hline\\ [-2.3ex]
 Final state & Eff & Trig & Pur & $r$ & Back & Tot & Eff & Trig & Pur & $r$ & Back& Tot \cr
\hline\\ [-2.3ex]
\Kpisk& 6.06 & 2.94 & 2.80 & 4.39 & 0.20 & 8.52 & - & - & - & - & - & - \cr
$a_0(1450) \pi$ & 0.30 & 0.12 & 0.24 & 0.14 & 0.06 & 0.43 & 0.31 & 0.35 & 0.08 & 0.01 & 0.01 & 0.47 \cr 
$K^*_2(1430) K$ & 0.32 & 0.08 & 0.30 & 0.88 & 0.54 & 1.13 & 0.09 & 0.07 & 0.04 & 0.06 & 0.01 & 0.13 \cr
$a_2(1320) \pi$ & 0.32 & 0.09 & 0.30 & 0.08 & 0.11 & 0.47 & 0.27 & 0.13 & 0.19 & 0.05 & 0.86 & 0.93 \cr
$a_0(980) \pi$ & 0.01 & 0.17 & 0.02 & 0.22 & 0.26 & 0.46 & 0.02 & 0.02 & 0.03 & 0.02 & 0.02 & 0.06 \cr
$a_0(1700) \pi$ & 0.12 & 0.00 & 0.08 & 0.08 & 0.01 & 0.17 & 0.18 & 0.15 & 0.03 & 0.00 & 0.01 & 0.24 \cr
$K^*_2(1980) K$ & 0.46 & 0.11 & 0.43 & 1.77 & 0.22 & 1.89 & 0.04 & 0.01 & 0.06 & 0.05 & 0.01 & 0.08 \cr
$a_2(1750) \pi$ & 0.21 & 0.10 & 0.13 & 0.02 & 0.03 & 0.27 & 0.09 & 0.02 & 0.06 & 0.11 & 0.19 & 0.25 \cr 
\hline
\end{tabular}
}
 \end{table}

\begin{figure}[tb]
\centering
\small
\includegraphics[width=0.9\textwidth]{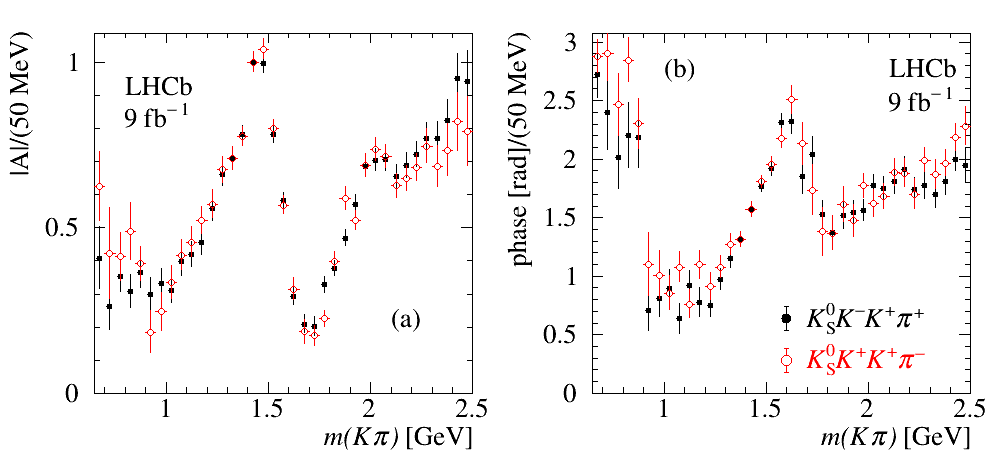}
\caption{\small\label{fig:fig16} Comparisons
between the fitted $K \pi$ {\it S}-wave (a) amplitude and (b) phase obtained from
the (black) \bkskkpip and (red) \bkskkpim data. Note that the point at 1.425~\gev is fixed at values 1.0 and 1.57 for amplitude and phase, respectively, and therefore there is no associated uncertainty. The plotted uncertainties are obtained averaging the uncertainties on the two adjacent mass measurements.}
 \end{figure}

 Figure~\ref{fig:fig16} shows the fitted $K \pi$ {\it S}-wave amplitude and phase obtained from the QMI analysis for both the \bkskkpip and \bkskkpim data, which agree well. The data show a dominance of the $K^*_0(1430)$ resonant contribution followed by a sharp drop and a subsequent
 signal of the $K^*_0(1950)$ resonance.
A contribution in the phase motion not continuous with the rest of the distribution is present in the threshold region of Fig.~\ref{fig:fig16}(b). This behavior may be attributed to the inability of the algorithm to obtain a correct phase motion
in this region of the phase space due to the absence, in this mass region, of additional significant resonant contributions in addition to that of the $K^*_0(1430)$ resonance.

 \begin{figure}[tb]
\centering
\small
\includegraphics[width=0.6\textwidth]{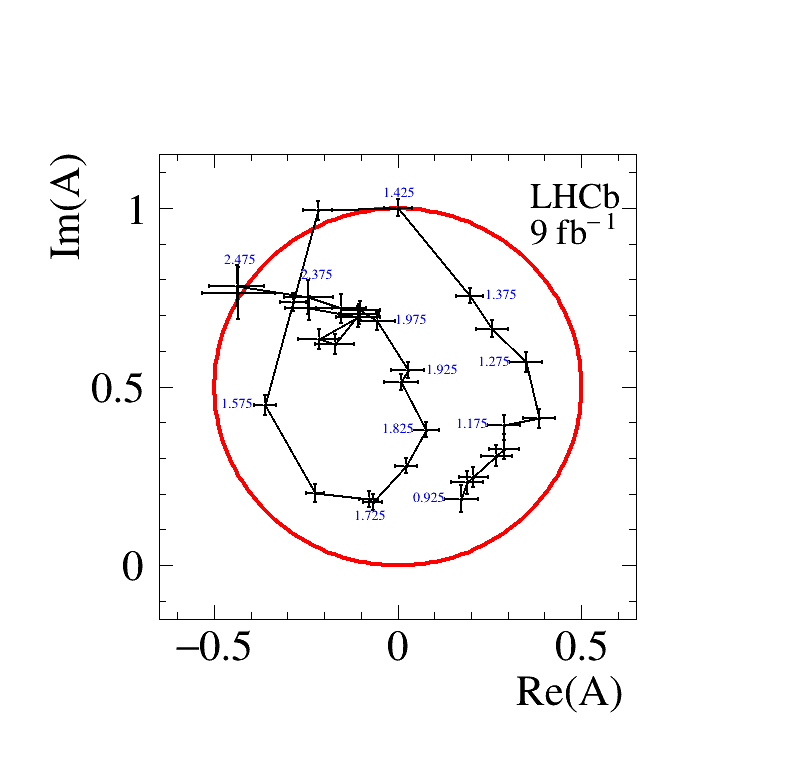}
\caption{\small\label{fig:fig17} Argand diagram for the $K \pi$ {\it S}-wave averaged over the QMI results from \mbox{\bkskkpip} and \mbox{\bkskkpim} data.}
\end{figure}
 
The inverse-variance-weighted average of the $K \pi$ {\it S}-wave from the \bkskkpip and \bkskkpim data is shown in terms of the real and imaginary parts of its amplitude in the Argand diagram of Fig.~\ref{fig:fig17}. The plot shows the expected anticlockwise phase motion due to the $K^*_0(1430)$ resonance and a second loop due to the presence of the $K^*_0(1950)$ resonance,
 observed here for the first time. Note that it was not possible to observe this behavior in Ref.~\cite{Aston:1987ir} due to the presence of two solutions above the mass of 1.6\gev.

Systematic uncertainties on the QMI $K \pi$ {\it S}-wave amplitudes are evaluated as follows.
The \bkskkpip and \bkskkpim datasets are each fitted independently using QMI for lower and higher purity selections, averaging the results; this is also done separately for TOS and noTOS selections.
The four solutions are compared with the reference solution and the absolute values of the deviations  are added in quadrature.
The effect on the QMI $K \pi$ {\it S}-wave due to the variation of the resonance parameters fixed to known values, in particular those for the $a_0(980)$ and $K^*_2(1430)$ resonances, is tested and found to be negligible.
The numerical values of the resulting QMI $K \pi$ {\it S}-wave amplitudes and phases are listed in Appendix~\ref{sec:app}, and this topic is further discussed in Sec.~\ref{sec:iso}.

\subsection{Isobar model \boldmath{\etac} Dalitz plot analysis}
\label{sec:iso}
The QMI method used in the previous section has allowed information to be obtained about the $K \pi$ {\it S}-wave which is produced by a combination of several scalar resonances. However, the procedure does not give insight into 
the parameters of the contributing resonances. An additional description of the data is therefore performed, known as the isobar model, which fits the data with a superposition of known resonances and additional (unknown or poorly known) states.
Both methods are expected to give similar descriptions of the data.

\begin{figure}[tb]
\centering
\small
\includegraphics[width=0.95\textwidth]{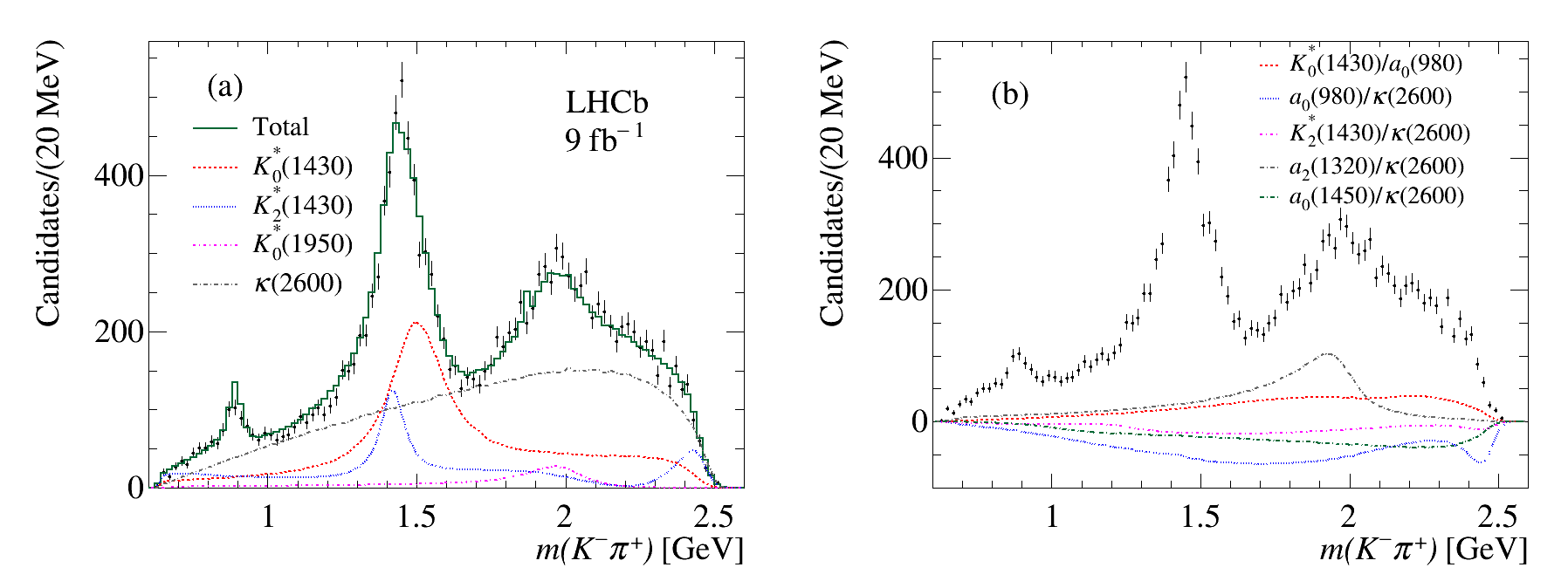}
\includegraphics[width=0.95\textwidth]{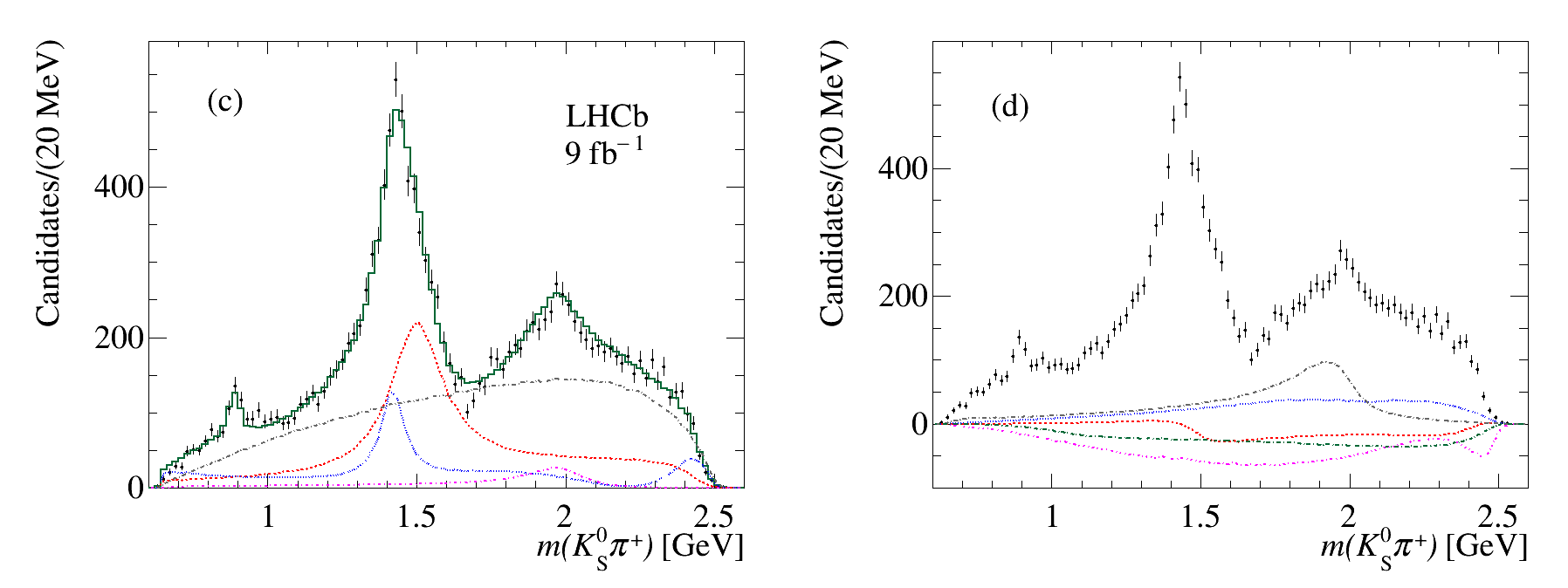}
\includegraphics[width=0.95\textwidth]{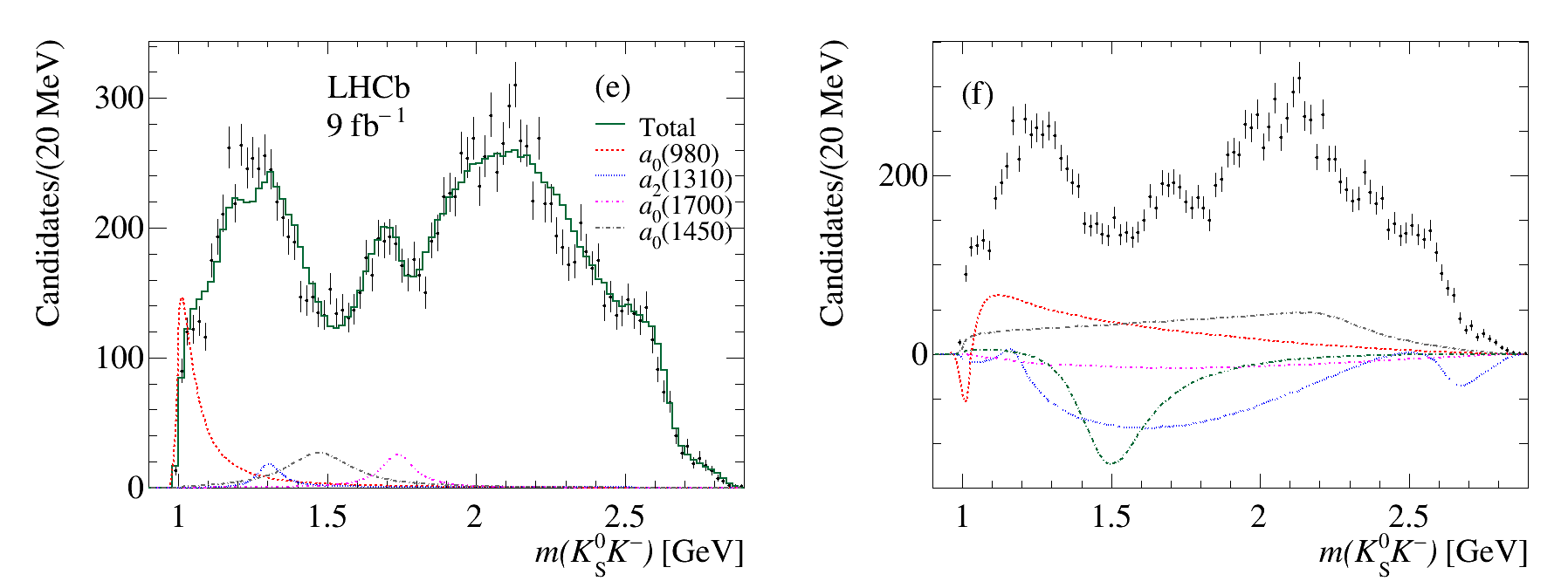}
\caption{\small\label{fig:fig18} Fit projections on the $\Km \pip$, $\KS \pip$ and $\KS \Km$ invariant-mass distributions from the Dalitz-plot analysis of the \etac decay using the isobar model for the \bkskkpip data. Panels (a, c, e) show the most important resonant contributions. 
To simplify the plots, only resonant contributions relative to that mass projection are shown. Panels (b, d, f) show the interference terms for contributions greater than 3\%.
The legend in (a) also applies to (c) and the legend in (b) also applies to (d) and (f).}
\end{figure}

The fitting method is the same as that described in Sec.~\ref{sec:qmi}, but in this case all the resonances are described by relativistic Breit--Wigner functions with parameters initially fixed to their default values~\cite{Workman:2022ynf}. The search for the best solution is performed by adding resonances one-by-one,
using the $K^*_0(1430)$ resonance as the reference amplitude, and considering as figures of merit, the likelihood and \chisqndf behavior, as discussed in Sec.~\ref{sec:qmi}.
It is found that, using only the list of the known resonances~\cite{Workman:2022ynf}, a poor description of the data is obtained.
A large improvement (increased likelihood $\Delta(2 \log \calL)=1338$ and decreased \chisqndf) is obtained by adding an additional scalar $K \pi$ resonance with free parameters, labeled in the following as $\kappa(2600)$ and
included in the fit using the simple non-relativistic BW function of Eq.~(\ref{eq:bw}).
Resonances having poorly known parameters, namely the $K^*_0(1430)$, $K^*_0(1950)$ and $a_0(1700)$ resonances, are allowed to have free parameters. 
The significance of the $K^*_2(1980)$ is found to be consistent with zero and its contribution is removed from the fit. 
The procedure is developed on the \bkskkpip data, because of the better quality of this final state due to reduced combinatorial background, and is then tested on the
\bkskkpim final state.
The fit projections are shown in Fig.~\ref{fig:fig18} for the \bkskkpip data; they are similar for the \bkskkpim data and are shown in Fig.~\ref{fig:fig29} of Appendix~\ref{sec:app}.

The resulting fitted parameters are listed in Table~\ref{tab:tab9} together with the estimated significances, evaluated using Wilks' theorem~\cite{Wilks:1938dza}.
A test is made for the presence of the low-mass $\kappa/K^*_0(700)$ resonance, with parameters fixed to average values~\cite{Workman:2022ynf}. The resulting likelihood variation is $\Delta(2\log\calL)=6.6$ for the difference
of two parameters which corresponds to a significance of 1.8$\sigma$. The same is true for the \bkskkpim decay where $\Delta(2\log\calL)=12.6$, corresponding to a significance of 2.9$\sigma$.
Replacing the broad $\kappa(2600)$ resonance with a uniform non-resonant contribution results in a decrease of the likelihood function by $\Delta(2\log\calL)=254$.

The fitted resonance contributions and relative phases are given in Table~\ref{tab:tab10}.
Interferences between amplitudes are listed in Table~\ref{tab:tab11} and shown in Fig.~\ref{fig:fig18} for absolute values above 3\%.
\begin{table}[tb]
  \centering
  \caption{\small\label{tab:tab9} Fitted parameters and significances of the resonances observed in the \etac Dalitz-plot analysis. Significances larger than 10$\sigma$ are evaluated as $\sqrt{\Delta(2 \log \calL)}$.}
\begin{tabular}{lccrc}
\hline\\ [-2.3ex]
Resonance & Mass [\mev]& $\Gamma$ [\mev] & $\Delta(2 \log \calL)$ & Significance \cr
\hline\\ [-2.3ex]
$K^*_0(1430)$ & $1493 \pm \al 4 \pm \al\al 7$ & $215 \pm \al 7 \pm \al 4$ & - & - \cr
$K^*_0(1950)$ & $1980 \pm 14 \pm \al 19$ & $229 \pm 26 \pm 16$ & 316 & 17.8$\sigma$ \cr
$a_0(1700)$ & $1736 \pm 10 \pm\al 12$ & $ 134 \pm 17 \pm 61$ & 161 & 12.7$\sigma$ \cr
$\kappa(2600)$ & $\allm 2662 \pm 59 \pm 201$ & $480 \pm 47 \pm 72$  & 1338 & 36.6$\sigma$ \cr
\hline
\end{tabular}
    \end{table}

\begin{table} [tb]
  \centering
  \caption{\small\label{tab:tab10} Results from the Dalitz-plot analysis of the \etac using the isobar model for (top) \bkskkpip decays, \bkskkpim decays (center) and their inverse-variance-weighted averages (bottom).}
  {\small
\begin{tabular}{lrr}
\hline
Final state & Fraction [\%] & Phase [rad] \alp \alp \all\cr
\hline\\ [-2.3ex]
 & \bkskkpip & \cr
 \hline\\ [-2.3ex]
$K^*_0(1430) K$ & $35.1 \pm 1.3 \pm 2.9$ & 0. \alp \alp \alp \cr
$a_0(980) \pi$ & $ 5.6 \pm 0.8 \pm 1.6$ & $-3.39 \pm 0.08 \pm 0.13$ \cr
$K^*_2(1430) K$ & $15.4 \pm 1.0 \pm 1.1$ & $3.53 \pm 0.03 \pm 0.10$ \cr
$a_2(1320) \pi$ & $1.1 \pm 0.2 \pm 0.3$ & $-2.90 \pm 0.11 \pm 0.26$ \cr
$K^*_0(1950) K$ & $3.9 \pm 0.4 \pm 0.3$ & $-0.46 \pm 0.06 \pm 0.63$ \cr
$a_0(1700) \pi$ & $ 1.7 \pm 0.3 \pm 0.4$ & $1.00 \pm 0.08 \pm 0.17$ \cr
$a_0(1450) \pi$ & $3.4 \pm 0.5 \pm 0.8$ & $-4.78 \pm 0.08 \pm 0.17$ \cr
$a_2(1750) \pi$ & $ 0.3 \pm 0.1 \pm 0.1$ & $2.43 \pm 0.17 \pm 0.17$ \cr
$\kappa(2600) K$ & $63.9 \pm 3.4 \pm 8.1$ & $-0.42 \pm 0.05 \pm 0.14$ \cr
\hline\\ [-2.3ex]
Sum & $130.5 \pm 4.0 \pm 8.9$ & \cr
$\chisqndf=1798/(1589-19)=1.15$ & &\cr
\hline\\ [-2.3ex]
 & \bkskkpim & \cr
 \hline\\ [-2.3ex]
$K^*_0(1430) K$ & $32.0 \pm 1.2 \pm 2.8$ & 0. \alp \alp \alp  \cr
$a_0(980) \pi$ & $ 4.9 \pm 0.6 \pm 1.0$ & $-3.37\pm 0.08 \pm 0.11$ \cr
$K^*_2(1430) K$ & $13.8 \pm 1.0 \pm 1.2$ & $3.56 \pm 0.03 \pm 0.11$ \cr
$a_2(1320) \pi$ & $1.2 \pm 0.2 \pm 0.3$ & $-2.82 \pm 0.11 \pm 0.24$ \cr
$K^*_0(1950) K$ & $3.4 \pm 0.4 \pm 0.3$ & $-0.42 \pm 0.06 \pm 0.64$ \cr
$a_0(1700) \pi$ & $ 0.7 \pm 0.2 \pm 0.2$ & $1.18 \pm 0.11 \pm 0.28$ \cr
$a_0(1450) \pi$ & $2.0 \pm 0.4 \pm 0.7$ & $-4.86 \pm 0.10 \pm 0.22$ \cr
$a_2(1750) \pi$ & $ 0.3 \pm 0.1 \pm 0.1$ & $2.24 \pm 0.18 \pm 0.17$ \cr
$\kappa(2600) K$ & $59.8 \pm 3.4 \pm 7.3$ & $-0.32 \pm 0.05 \pm 0.12$ \cr
\hline\\ [-2.3ex]
Sum & $118.1 \pm 2.7 \pm 8.0$ & \cr
$\chisqndf=1738/(1584-21)=1.11$ & &\cr
\hline\\ [-2.3ex]
 & $B \to \KS KK \pi$ & \cr
 \hline\\ [-2.3ex]
$K^*_0(1430) K$ & $33.4 \pm 0.9 \pm 2.0$ & 0. \alp \alp \alp\cr
$a_0(980) \pi$ & $ 5.1 \pm 0.5 \pm 0.8$ & $-3.38 \pm 0.06 \pm 0.08$ \cr
$K^*_2(1430) K$ & $14.6 \pm 0.7 \pm 0.8$ & $3.54 \pm 0.02 \pm 0.07$ \cr
$a_2(1320) \pi$ & $1.1 \pm 0.1 \pm 0.2$ & $-2.89 \pm 0.08 \pm 0.18$ \cr
$K^*_0(1950) K$ & $3.7 \pm 0.3 \pm 0.2$ & $-0.44 \pm 0.04 \pm 0.45$ \cr
$a_0(1700) \pi$ & $ 1.1 \pm 0.2\pm 0.2$ & $1.05 \pm 0.06 \pm 0.15$ \cr
$a_0(1450) \pi$ & $2.6 \pm 0.3 \pm 0.5$ & $-4.82 \pm 0.06 \pm 0.13$ \cr
$a_2(1750) \pi$ & $ 0.3 \pm 0.1 \pm 0.1$ & $2.33 \pm 0.12 \pm 0.11$ \cr
$\kappa(2600) K$ & $61.8 \pm 2.4 \pm 5.4$ & $-0.37 \pm 0.03 \pm 0.09$ \cr
\hline\\ [-2.3ex]
Sum & $123.7 \pm 2.7 \pm 4.7 $ & \cr
\hline
\end{tabular}
}
\end{table}
\begin{table} [tb]
  \centering
  \caption{\small\label{tab:tab11} Fractional interference contributions from the Dalitz-plot analysis of the \etac decay in \bkskkpip decays using the isobar model. Absolute values
  less than 3\% are not listed.}
  {\small
\begin{tabular}{llr}
\hline\\ [-2.3ex]
Amplitude 1 & Amplitude 2 & Fraction [\%] \cr
\hline\\ [-2.3ex]
$K^*_0(1430) K$ & $a_0(980) \pi$ & $-6.1 \pm 0.5$ \cr
$a_0(980) \pi$ & $a_0(1450) \pi$ & $-3.6 \pm 0.4$ \cr
$a_0(980) \pi$ & $\kappa(2600) K$ & $14.4 \pm  1.9$ \cr
$K^*_2(1430) K$ & $K^*_0(1950) K$ & $-3.7 \pm  0.3$ \cr
$K^*_2(1430) K$ & $\kappa(2600) K$ & $-26.2 \pm 1.1$ \cr
$a_2(1320) \pi$ & $\kappa(2600) K$ & $-6.5 \pm  0.7$ \cr
$K^*_0(1950) K$ & $\kappa(2600) K$ & $18.6 \pm  1.5$ \cr
$a_0(1700) \pi$ & $\kappa(2600) K$ & $-4.6 \pm  0.7$ \cr
$a_0(1450) \pi$ & $\kappa(2600) K$ & $-14.5 \pm 1.0$ \cr
\hline
\end{tabular}
}
\end{table}
Systematic uncertainties on fractions and relative phases are evaluated as in Sec.~\ref{sec:qmi} with the addition of an alternative model for the $K^*_0(1430)$ resonance:
a simplified coupled-channel Breit--Wigner function, which ignores the small $K \eta$ coupling. Its lineshape is parameterized as

\begin{equation}
        {\rm BW}(m) = \frac{1}{m_0^2 - m^2 - i(\rho_1(m)g^2_{K \pi} + \rho_2(m)g^2_{K \etapr})},
\end{equation}
\noindent
where $m_0$ is the resonance mass, $g_{K \pi}$ and $g_{K \etapr}$ are the couplings to the $K \pi$ and $K \etapr$ final states, and \mbox{$\rho_j(m)=2p/m$} are the respective Lorentz-invariant
phase-space factors, with $p$ the decay-particle momentum in the $K^*_0(1430)$ rest frame. 
The $\rho_2(m)$ function becomes imaginary below the $K \etapr$ threshold. The Dalitz-plot analysis is insensitive to the $g_{K \etapr}$
parameter, which has been fixed to $g^2_{K\etapr}=0.197\gevgevcccc$~\cite{BaBar:2021fkz}.
A similar method is used for the evaluation of the systematic uncertainties on the resonance parameters.

\begin{figure}[tb]
\centering
\small
\includegraphics[width=0.45\textwidth]{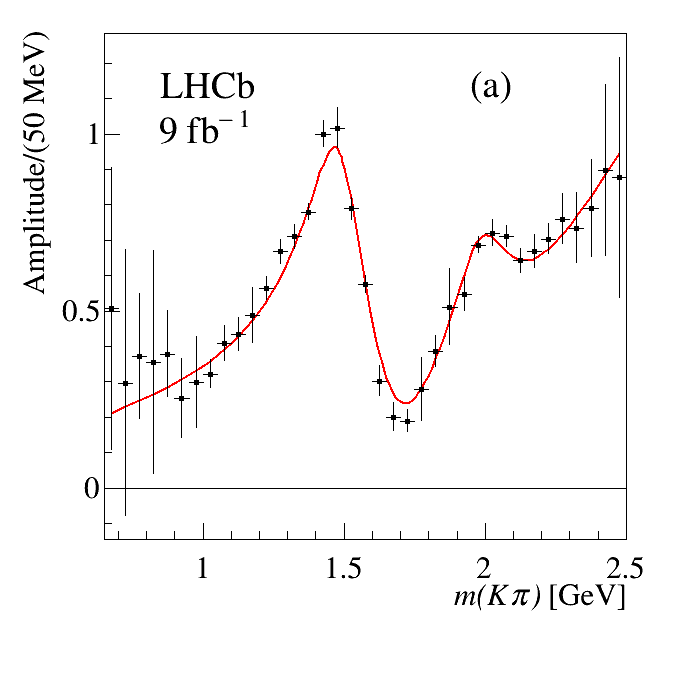}
\includegraphics[width=0.45\textwidth]{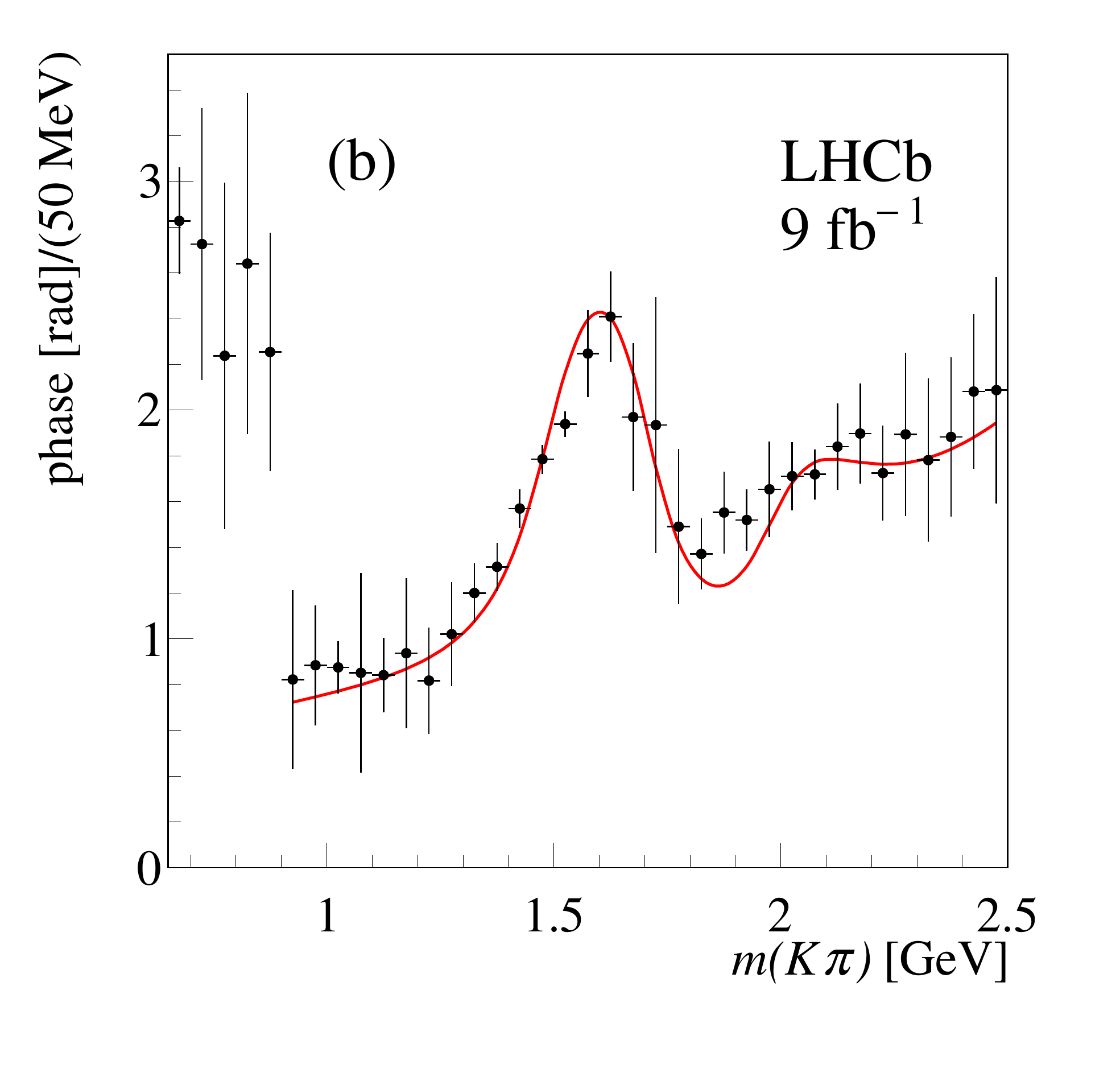}
\includegraphics[width=0.45\textwidth]{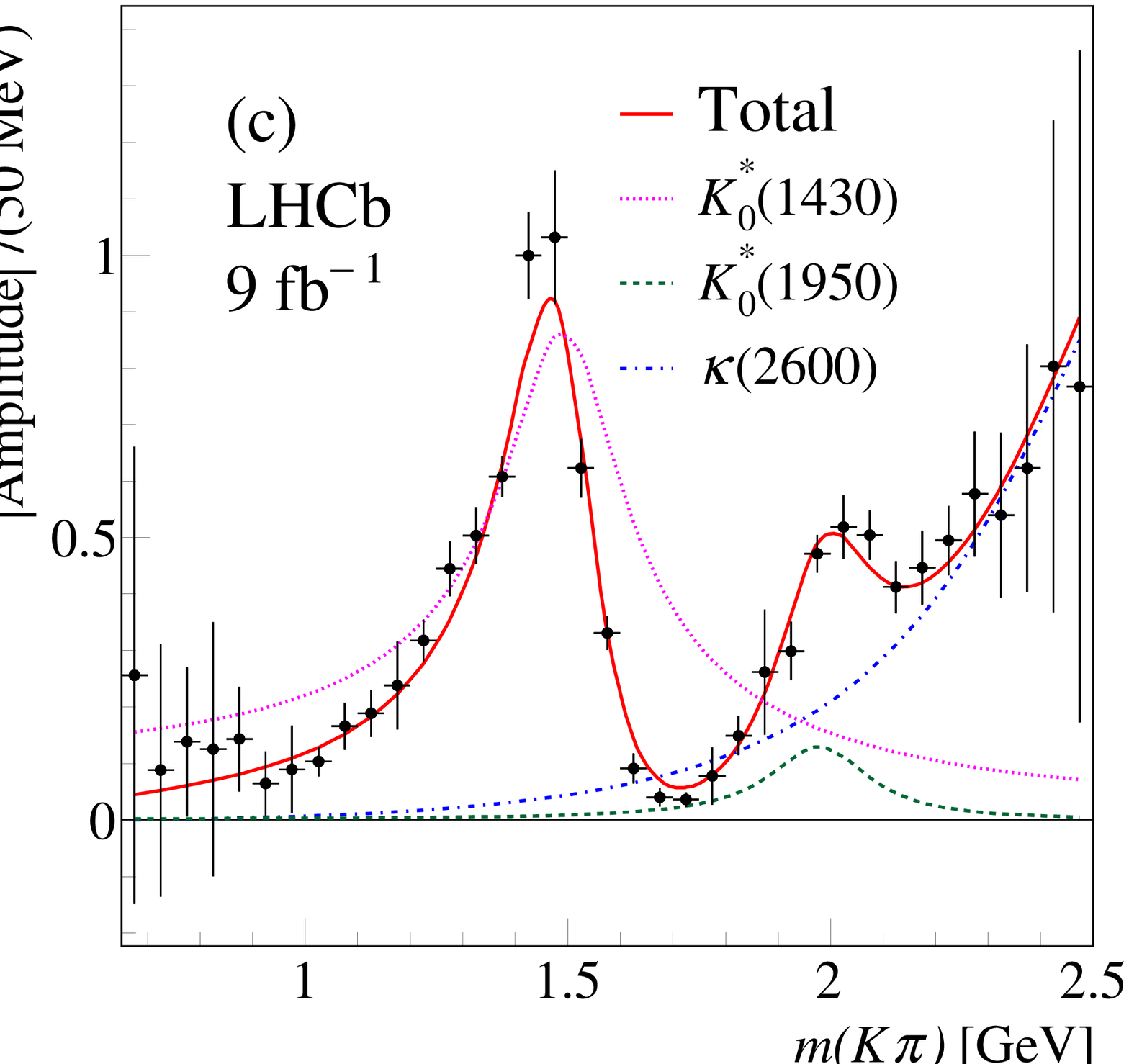}
\caption{\small\label{fig:fig19} Fit to the (a) amplitude, (b) phase and (c) squared amplitude from the \etac QMI analysis using the results from the isobar-model analysis. The black points are obtained using the inverse-variance-weighted averages of the \bkskkpip and \bkskkpim data. The error bars are computed as the quadratic sum of the statistical and systematic uncertainties. The phase measurements in the $K \pi$ mass region below 0.90\gev are not included in the fit.}
\end{figure}

The isobar model allows the separation of the resonance composition of the $K\pi$ {\it S}-wave obtained from the QMI method. A fit to the $K\pi$ {\it S}-wave amplitude and phase is performed using the sum of three interfering spin-0 relativistic Breit--Wigner amplitudes:
 \begin{equation}
  f(m) = c_1{\rm BW}_{K^*(1430)}(m)e^{i\phi_1} + c_2{\rm BW}_{K^*(1950)}(m)e^{i\phi_2} + c_3{\rm BW}_{\kappa(2600)}(m)e^{i\phi_3},
\end{equation}
where $m$ indicates the $K \pi$ mass. In this fit, the mass and width of the three resonances are fixed to the results obtained in the isobar model analysis while their coefficients and relative
phases are left as free parameters. The fitted $K\pi$ {\it S}-wave used in the fit is obtained from the inverse-variance-weighted averages from
the \bkskkpip and \bkskkpim data.
The QMI amplitude and phase, together with the squared amplitude, are shown in Fig.~\ref{fig:fig19},
where the statistical and systematic uncertainties are added in quadrature and the phase measurements in the $K \pi$ mass region below 0.90\gev are not included in the fit. The amplitude and phase of the data point at a mass of 1.425\gev are fixed in the QMI analysis and the reported uncertainties are obtained from the averages of the two adjacent mass values.
The fit has a p-value of 63\% and is shown in Fig.~\ref{fig:fig19}, together with the composition of the fitted resonances. The relative phases are
$\phi_1=0.66 \pm 0.03$ rad, $\phi_2=-0.08 \pm 0.09$ rad and $\phi_3=0.61\pm0.05$ rad. 
These values can be used to compare relative phases with respect to the $K^*_0(1430)$ resonance, yielding
$\phi_2 - \phi_1=-0.74 \pm 0.09$ rad and $\phi_3 - \phi_1=-0.05 \pm 0.06$ rad. These differences are  
within $0.65\sigma$ and $2.8\sigma$, respectively,  with respect to the values obtained from the isobar model listed in 
Table~\ref{tab:tab10}.
This test shows that the $K \pi$ {\it S}-wave obtained using the QMI method can be interpreted as due to the sum of the three interfering $K^*_0$ resonances observed with the isobar model. This result is similar to what was observed
in the $\pi \pi$ final state in Ref.~\cite{WA76:1991kef}, where the $f_0(980)$ resonance was found to interfere with the broad $\sigma/f_0(500)$ resonance.

\section{Dalitz-plot analysis of \boldmath{\etactwo} decay to \boldmath{$\KS K \pi$}}
\label{sec:etactwo}

In this section a Dalitz-plot analysis of the \etactwo decay to the final state
\begin{equation}
  \begin{split}
  \Bp & \to \etactwo \Kp,\\
  \etactwo & \to \KS \Km \pip,
  \end{split}
\end{equation}
is described. Figure~\ref{fig:fig5} shows the \kskpi mass spectra for the \bkskkpip and \bkskkpim data. The \etactwo signal is accompanied by a significant background contribution which is especially large in the \bkskkpim data by virtue of the two indistinguishable particles in the final state. Therefore, only the \bkskkpip data are used for the \etactwo Dalitz-plot analysis. In addition to the background from $\Dz \to \KS \Kp \Km$ (see Sec.~\ref{sec:evsel}), additional backgrounds involving the spectator \Kp are removed. A significant contribution from $\phi(1020) \to \Kp \Km$ decays is removed by selecting events outside the $1.01<m(\Kp \Km)<1.03$ \gev mass region, and a small contribution from $\Dsp \to \Kp \Km \pip$ decays is removed in the $\pm 2\sigma$ ($\sigma=7.8\mev$) mass region close to the known \Ds mass.
Finally, a significant contribution from $\Dsm \to \KS \Km$ decays is removed within $\pm 2.5 \sigma$ ($\sigma=7.8\mev$) from the known \Ds mass~\cite{Workman:2022ynf}.

\begin{figure}[tb]
\centering
\small
\includegraphics[width=0.70\textwidth]{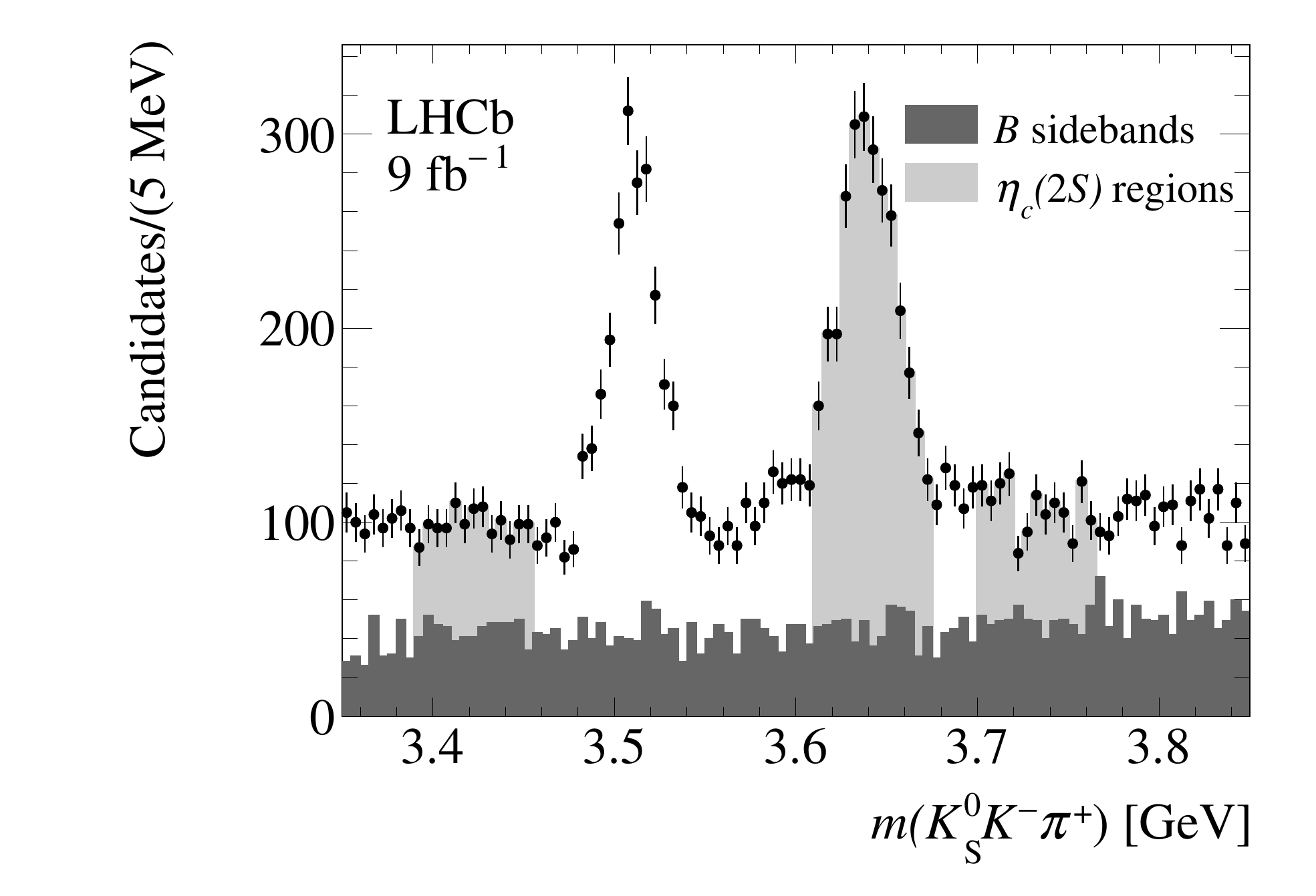}
\caption{\small\label{fig:fig20} Invariant \kskpi mass distribution in the region of the \etactwo resonance for \mbox{\bkskkpip}. 
The \etactwo signal and sideband regions are indicated in light gray, while the incoherent background estimated from the \Bp sidebands is shown in dark gray.}
\end{figure}

\begin{table} [tb]
  \centering
  \caption{\small\label{tab:tab12} Candidate events in the signal region, purities, and \Bp background contributions to the $\etactwo\to \kskpi$ Dalitz-plot analysis separated for \KS types. The incoherent background fractions $f_B$ in the low and high sidebands refer to the sum of the \KSLL and \KSDD data.}
  {\small
\begin{tabular}{lrccc}
\hline\\ [-2.3ex]
\KS type & Candidates & Purity [\%] & low $f_B$ [\%] & high $f_B$ [\%]\cr
\hline\\ [-2.3ex]
\KSLL &  781 & $63.0 \pm  1.7$ & & \cr
\KSDD &  2008 &  $55.2 \pm 1.1$ & &\cr
Combined \KS & & & $54.0 \pm 1.4$ & $55.1 \pm 1.4$ \cr
\hline
\end{tabular}
 }
\end{table}

\begin{figure}[tb]
\centering
\small
\includegraphics[width=0.60\textwidth]{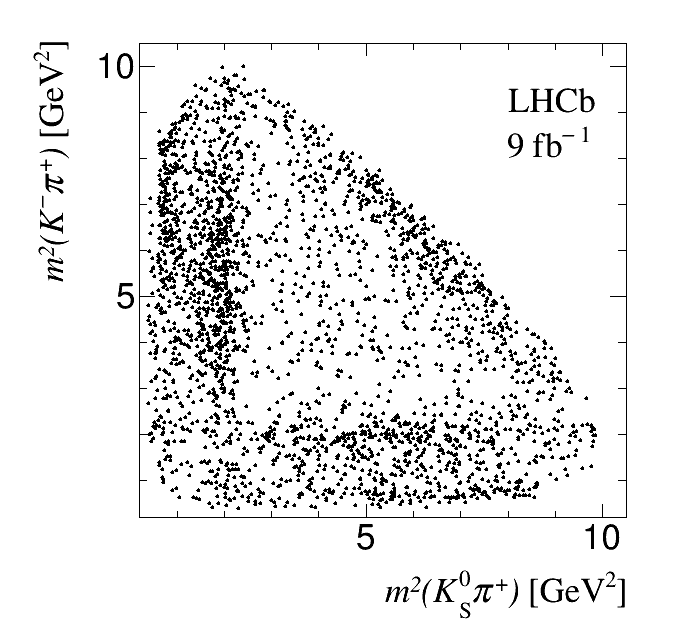}
\caption{\small\label{fig:fig21} Dalitz plot of $\etactwo \to \kskpi$ decays for the \bkskkpip sample.}
\end{figure}

Figure~\ref{fig:fig20} shows the resulting $\KS \Km \pip$  invariant-mass spectrum in the \etactwo region with the signal
[$3.61\text{--}3.67$]\gev and sideband [$3.42\text{--}3.45$]\gev, [$3.70\text{--}3.73$]\gev regions indicated.
Table~\ref{tab:tab12} shows the \etactwo event yields in the signal region, purities and background compositions.
Figure~\ref{fig:fig21} shows the \etactwo Dalitz plot. The distribution is dominated by the $K^*_0(1430)$ resonance with bands due to the
$K^*(892)$ resonance originating from background. 

\begin{figure}[tb]
\centering
\small
\includegraphics[width=0.48\textwidth]{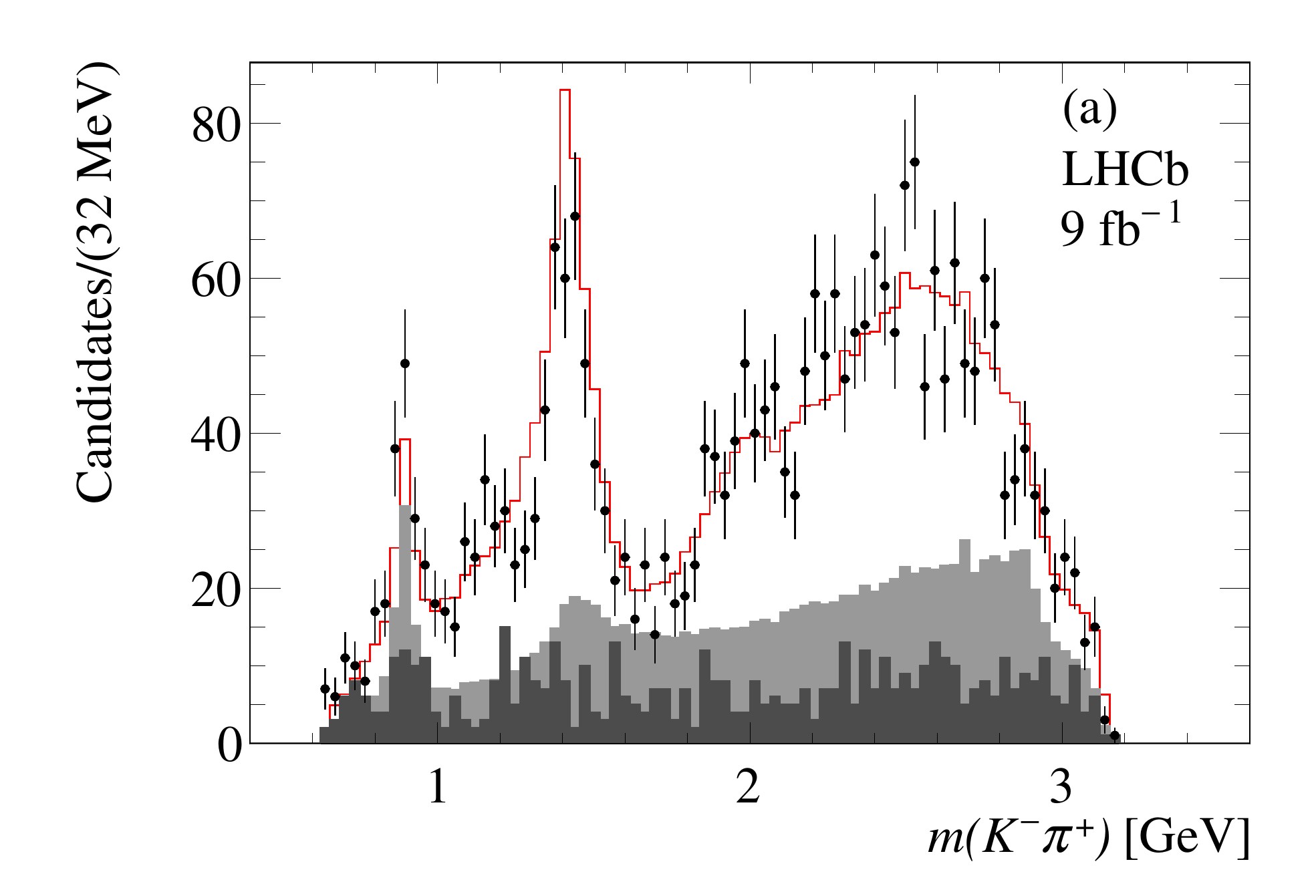}
\includegraphics[width=0.48\textwidth]{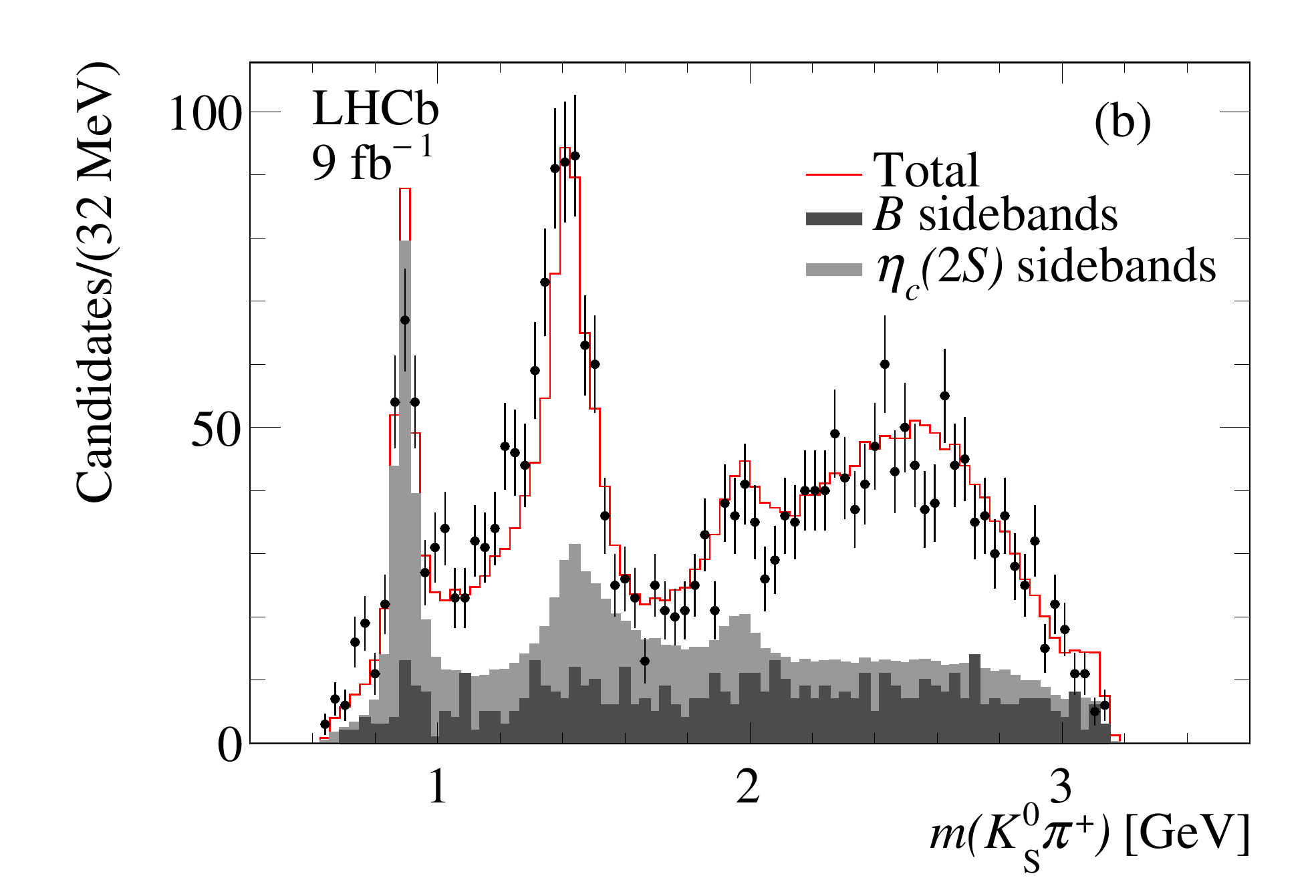}
\includegraphics[width=0.48\textwidth]{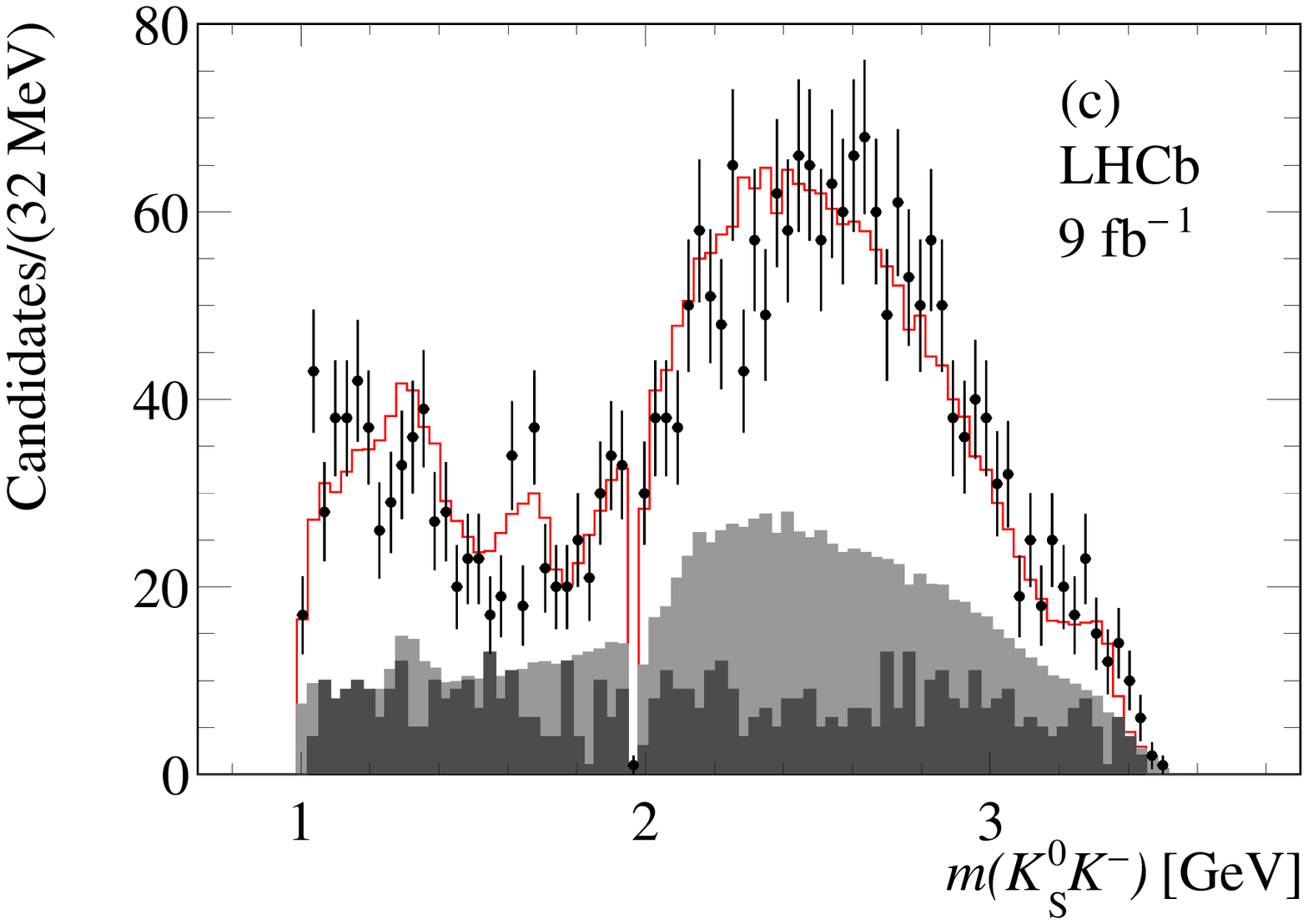}
\caption{\small\label{fig:fig22} Dalitz-plot projections of the $\etactwo \to \kskpi$ decays from the \bkskkpip sample. The solid red line shows the fit result, the light gray shaded area is the contribution from the fitted sidebands and the dark shaded gray area is the contribution from the \Bp sidebands. The empty region in the $\KS \Km$ invariant-mass projection is due to the removal of the contribution from $\Dsp \to \KS \Km$ decays.}
\end{figure}

The decay of the \etactwo to \kskpi is expected to be very similar to that of the \etac, except for the different size of the available phase space.
Therefore, the Dalitz-plot analysis follows closely the method used for the \etac analysis, except that, due to the limited statistics and the significant background, only the isobar model is used. In addition, all resonance parameters are fixed to those extracted from the \etac isobar model. The efficiency model used for the \etactwo Dalitz-plot analysis has been described in Sec.~\ref{sec:effloc}.
The resulting fitted amplitudes and phases are listed in Table~\ref{tab:tab13}.
The Dalitz-plot projections, together with the fitted background interpolated from the sidebands and the superimposed fit,  are shown in Fig.~\ref{fig:fig22}.
Large interference effects are observed, evidenced by the high value of the sum of the fractions.
Systematic uncertainties are evaluated in a similar way as for the \etac Dalitz-plot analysis.
In particular, the effect of the efficiency model is evaluated by replacing the fitted efficiencies by two-dimensional binned numerical maps. The effects of the uncertainties on the resonance parameters are evaluated by varying masses and widths by their statistical uncertainties according to a Gaussian distribution and averaging the absolute values of the deviations from the reference fit.
The effects of the different sources of systematic uncertainties on the fitted fractions and phases are listed in Table~\ref{tab:sys_par}. 
\begin{table} [ht]
  \centering
  \caption{\small\label{tab:sys_par} Systematic uncertainties on (left) fractional contributions (\%) and (right) phases in the \etactwo Dalitz-plot analysis in \bkskkpip decay using the isobar model.}
  {\small
\begin{tabular}{l|ccccc|c|ccccc|c}
\hline
Final state & Pur & Par & $r$ & Back & Eff & Tot & Pur & Par & $r$ & Back & Eff & Tot\cr
\hline\\ [-2.3ex]
$K^*_0(1430) K$ & 1.92 & 1.25 & 0.17 & 0.91 & 3.26 & 4.1 & -& - & - & - & - & -\cr
$K^*_2(1430) K$ & 2.47 &  0.43 & 1.24 & 0.86 & 3.30 & 4.4 & 0.06 & 0.02 & 0.04 & 0.02 & 0.01 & 0.08\cr
$K^*_0(1950) K$ &  0.60 & 0.22 & 0.37 & 0.18 & 0.74 & 1.1 &  0.03 & 0.09 & 0.08 & 0.21 & 0.01 & 0.24 \cr
$a_0(1700)^- \pip$ &  0.32 & 0.05 & 0.29 & 0.13 & 0.02 & 0.45 & 0.18 & 0.05 & 0.11 & 0.87 & 0.12 & 0.90\cr
$a_0(1450)^- \pip$ & 0.94 & 0.04 & 0.19 & 0.32 & 0.19 & 1.03 & 0.10 & 0.04 & 0.09 & 0.53 & 0.09 & 0.56\cr
$a_2(1750)^- \pip$ & 0.53 & 0.15 & 0.42 & 0.11 & 0.79 & 1.05& 0.03 & 0.04 & 0.08 & 0.38 & 0.03 & 0.39 \cr
$\kappa(2600) K$ & 4.33 & 3.28 & 4.97 & 0.01 & 2.40 & 7.74 & 0.04 & 0.05 & 0.04 & 0.02 & 0.01 & 0.08 \cr
\hline
\end{tabular}
}
\end{table}
Interferences between amplitudes are listed in Table~\ref{tab:tab23} for absolute values above 5\%.

\begin{table} [tb]
  \centering
  \caption{\small\label{tab:tab13} Results from the Dalitz-plot analysis of $\etactwo \to \kskpi$ in the \bkskkpip final state using the isobar model.}
  {\small
\begin{tabular}{lrr}
\hline\\ [-2.3ex]
Final state & Fraction [\%] & Phase [rad] \alp \alp \all\cr
\hline\\ [-2.3ex]
$K^*_0(1430) K$ & $25.5 \pm 3.3 \pm 4.1$ & 0. \alp \alp \alp \cr
$K^*_2(1430) K$ & $24.5 \pm 3.3 \pm 
4.4$ & $3.10 \pm 0.11 \pm 0.08$ \cr
$K^*_0(1950) K$ & $3.7 \pm 1.3 \pm 
1.1$ & $-0.82 \pm 0.17 \pm 0.24$ \cr
$a_0(1700)^- \pip$ & $ 1.7 \pm 1.1 \pm 0.5$ & $1.22 \pm 0.32 \pm 0.90$ \cr
$a_0(1450)^- \pip$ & $7.8 \pm 1.9 \pm 1.0$ & $1.86 \pm 0.14 \pm 0.56$ \cr
$a_2(1750)^- \pip$ & $ 4.9 \pm 1.4 \pm 1.1$ & $-1.75 \pm 0.15 \pm 0.39$ \cr
$\kappa(2600) K$ & $124.2 \pm 9.0 \pm 7.7$ & $-0.91 \pm 0.10 \pm 0.08$ \cr
\hline\\ [-2.3ex]
Sum & $192.3 \pm 10.9 \pm 10.0$ & \cr
\chisqndf=578/(591-13)=1.00 & & \cr
\hline
\end{tabular}
}
\end{table}

\begin{table} [tb]
  \centering
  \caption{\small\label{tab:tab23} Fractional interference contributions from the Dalitz plot analysis of the \etactwo decay in \bkskkpip decays using the isobar model. Absolute values
  less than 5\% are not listed.}
  {\small
\begin{tabular}{llc}
\hline\\ [-2.3ex]
Amplitude 1 & Amplitude 2 & Fraction[\%]\cr
\hline\\ [-2.3ex]
$K^*_0(1430) K$ & $\kappa(2600) K$ & $-23.1 \pm 4.7$ \cr
$K^*_2(1430) K$ & $\kappa(2600) K$ & $-40.3 \pm 5.0$ \cr
$K^*_0(1950) K$ & $\kappa(2600) K$ & \al \allm $16.0 \pm 4.9$ \cr
$a_0(1700) \pi$ & $\kappa(2600) K$ & \al$-6.1 \pm 1.8$ \cr
$a_0(1450) \pi$ & $\kappa(2600) K$ & $-21.1 \pm 3.0$ \cr
$a_2(1750) \pi$ & $\kappa(2600) K$ & $-15.8 \pm  3.8$ \cr
\hline
\end{tabular}
}
\end{table}

\section{Study of the \boldmath{\chicone} decay to \boldmath{\kskpi}}
\label{sec:chicone}

In this section a study is performed of the \chicone decay in the final states
\begin{equation}
  \begin{split}
  \Bp & \to \chicone \Kp,\\
  \chicone & \to \KS \Km \pip,
  \end{split}
\end{equation}
and
\begin{equation}
  \begin{split}
  \Bp & \to \chicone \Kp,\\
  \chicone & \to \KS \Kp \pim.
  \end{split}
\end{equation}

\begin{figure}[tb]
\centering
\small
\includegraphics[width=0.48\textwidth]{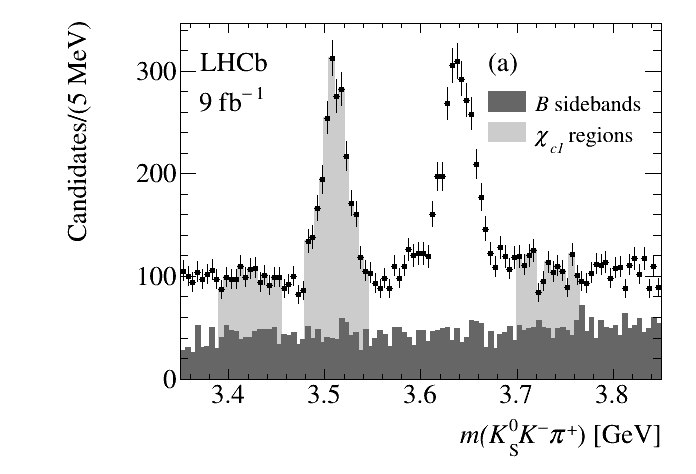}
\includegraphics[width=0.48\textwidth]{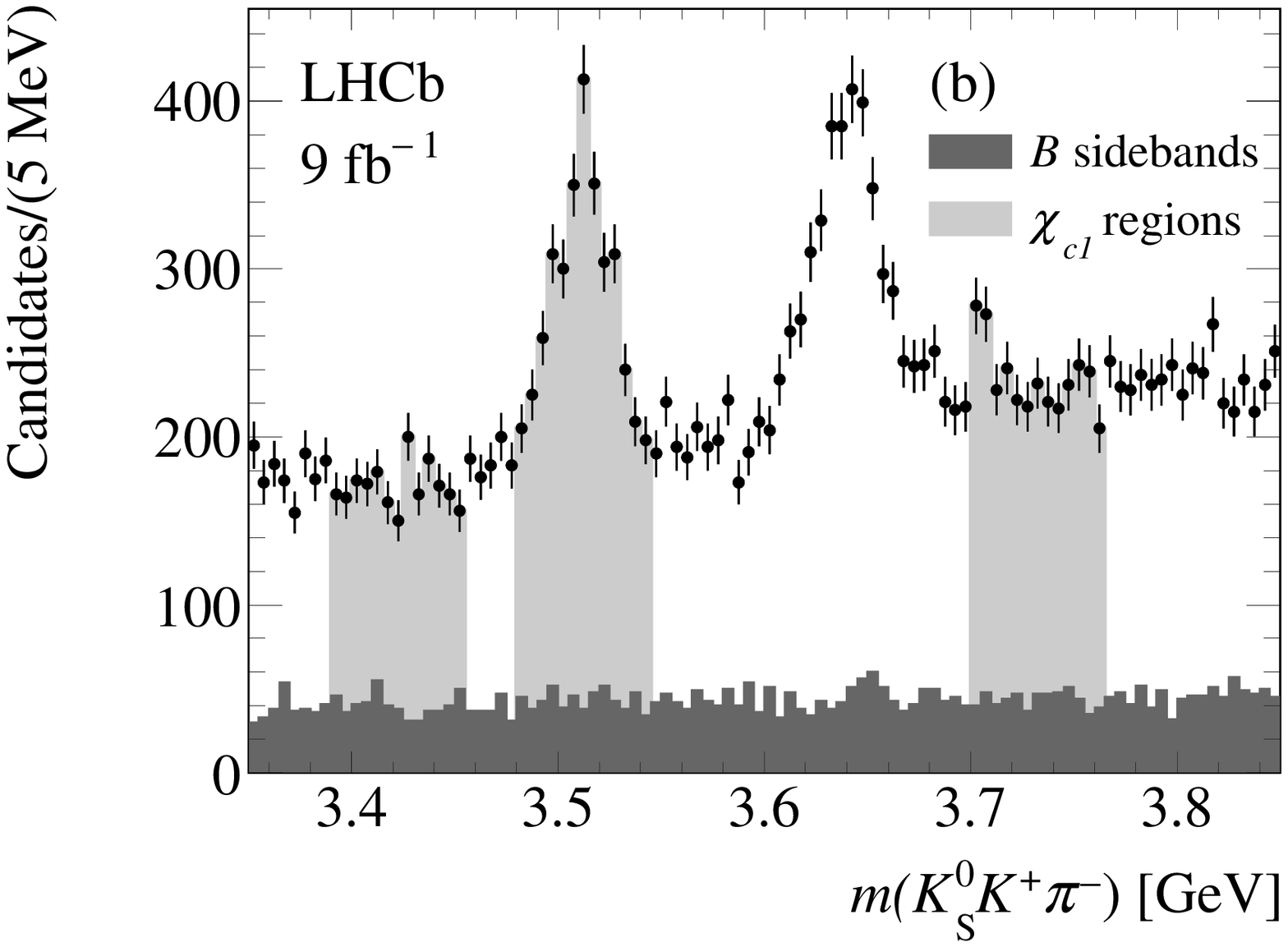}
\caption{\small\label{fig:fig23}  Invariant \kskpi mass distribution in the region of the \chicone/\etactwo resonances for (a) \bkskkpip and (b) \bkskkpim decays. In light gray are indicated the \chicone signal region and sidebands. In dark gray is indicated the incoherent background estimated from the \Bp sidebands.}
\end{figure}

The \kskpi invariant-mass spectrum in the $\chicone\text{--}\etactwo$ mass region, summed over the \KSLL and \KSDD data, is shown in Fig.~\ref{fig:fig23} for the \bkskkpip and \mbox{\bkskkpim} final states. The signal region is $[3.48\text{--}3.54]\gev$ and the lower and upper sidebands are $[3.39\text{--}3.45]\gev$ 
 and $[3.70\text{--}3.76]\gev$, respectively. Table~\ref{tab:tab14} reports the \chicone event yields, purities and background compositions, separated for the \KSLL and \KSDD data,
for the \bkskkpip and \bkskkpim decays.

\begin{table} [tb]
  \centering
  \caption{\small\label{tab:tab14} Candidate events and purities of \chicone for the \bkskkpip and \mbox{\bkskkpim} final states separated for the \KSLL and \KSDD data. The incoherent background fractions $f_B$ values for the lower and higher sideband regions are computed for the sum of the \KSLL and \KSDD data.}
  {\small
\begin{tabular}{llrccc}
\hline\\ [-2.3ex]
\Bp decay mode & \KS type & Candidates & Purity [\%]&  low $f_B$ [\%] & high $f_B$ [\%] \cr
\hline\\ [-2.3ex]
\kskkpip & \KSLL &  695 & $61.3 \pm  1.8$ & \cr
& \KSDD &  1726 &  $48.4 \pm 1.2$ &  & \cr
 & Combined \KS & & & $46.0 \pm 2.4$ & $44.9 \pm 2.2$ \cr
 \hline\\ [-2.3ex]
\kskkpim & \KSLL &  1031 & $34.8 \pm  1.5$ & &  \cr
& \KSDD &  2443 &  $36.2 \pm 1.0$ & &\cr
 & Combined \KS & & & $23.1 \pm 1.2$ & $18.6 \pm 0.9$ \cr
\hline
\end{tabular}
}
\end{table}

\begin{figure}[tb]
\centering
\small
\includegraphics[width=0.6\textwidth]{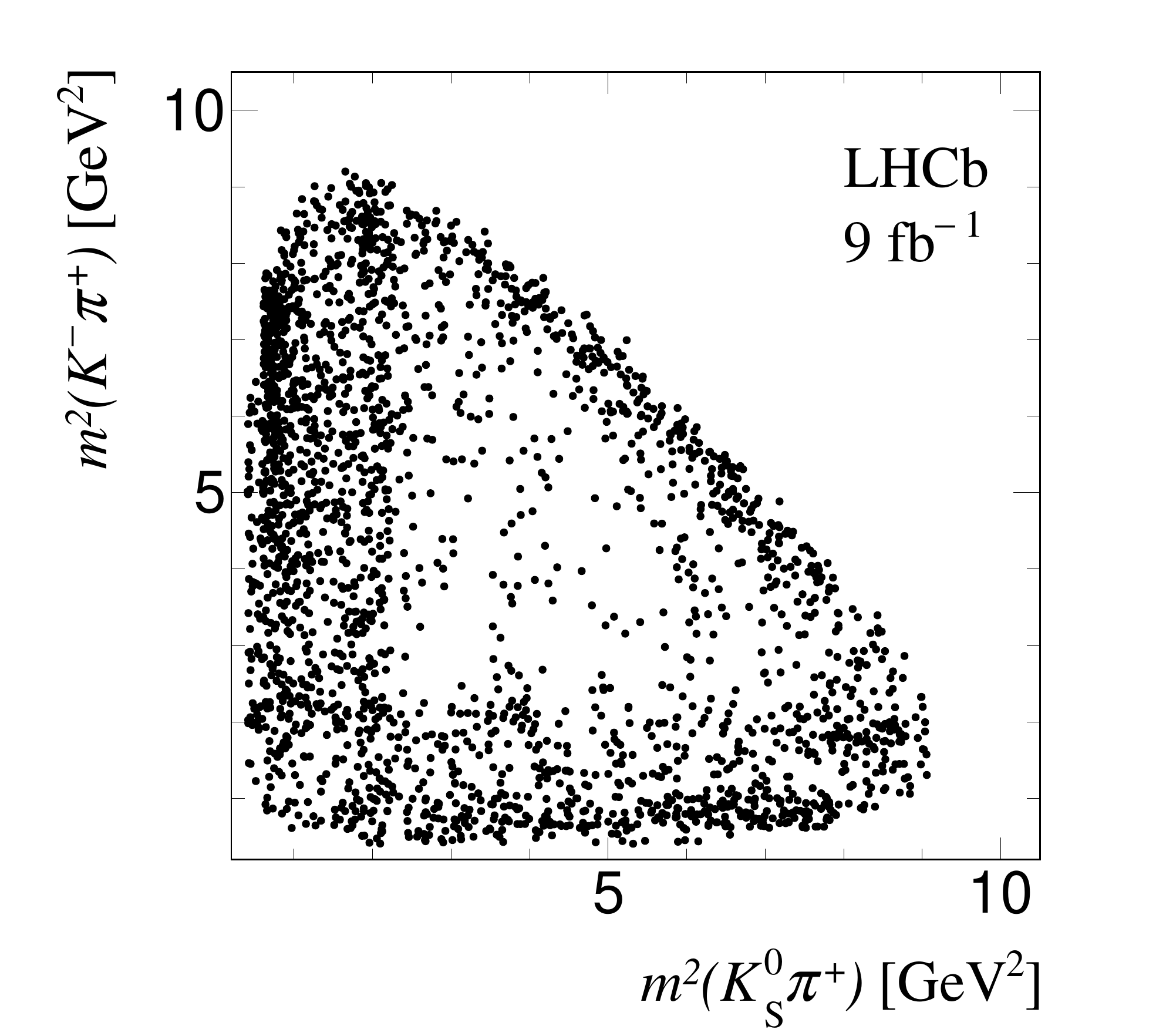}
\caption{\small\label{fig:fig24} Dalitz plot of $\chicone \to \KS \Km \pip$ decays for the \bkskkpip sample.}
\end{figure}

The \chicone Dalitz plot for \bkskkpip data, shown in Fig.~\ref{fig:fig24}, is dominated by horizontal and vertical bands due to the presence of $K^*(892)$ and $K^*_2(1430)$ resonances. 
A Dalitz-plot analysis of the \chicone decay to the \kskpi state is not feasible with the present dataset due to its limited sample size and the high level
of background; therefore a simplified approach, similar to that used in Ref~\cite{Ablikim:2006vm}, is taken where fits are applied to the $K \pi$ invariant-mass projections.
The efficiency distribution as a function of the $K \pi$ and $\KS \pi$  invariant masses is computed using the method described in Sec.~\ref{sec:effloc} and expressed in terms of $m_x(\KS K)$ and $\cos \theta_{\pi}$. 
Each event in the \chicone mass region is weighted by the inverse of the efficiency. The ratios of the unweighted over weighted $K \pi$ and $\KS \pi$ invariant-mass distributions are then computed, yielding the efficiency as functions of the $K \pi$ and $\KS \pi$ invariant masses.  These efficiencies are consistent with being uniform, so no efficiency correction is used in fitting to the $K \pi$ or $\KS \pi$ mass projections.

\begin{figure}[tb]
\centering
\small
\includegraphics[width=0.45\textwidth]{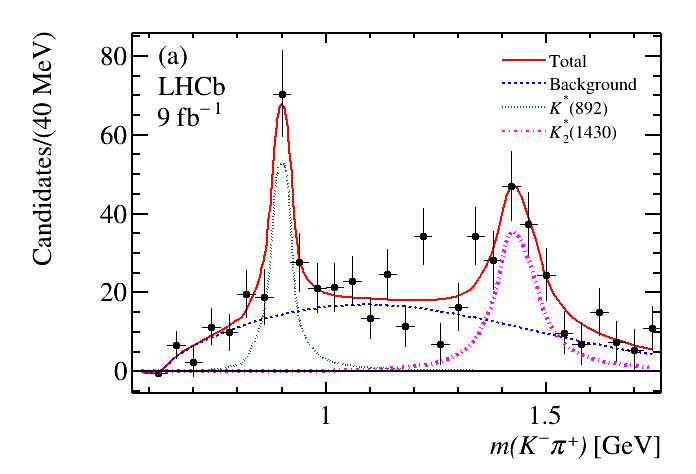}
\includegraphics[width=0.45\textwidth]{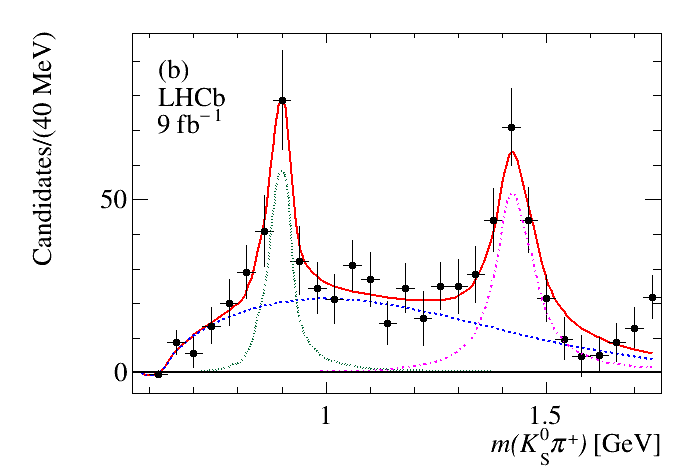}
\includegraphics[width=0.45\textwidth]{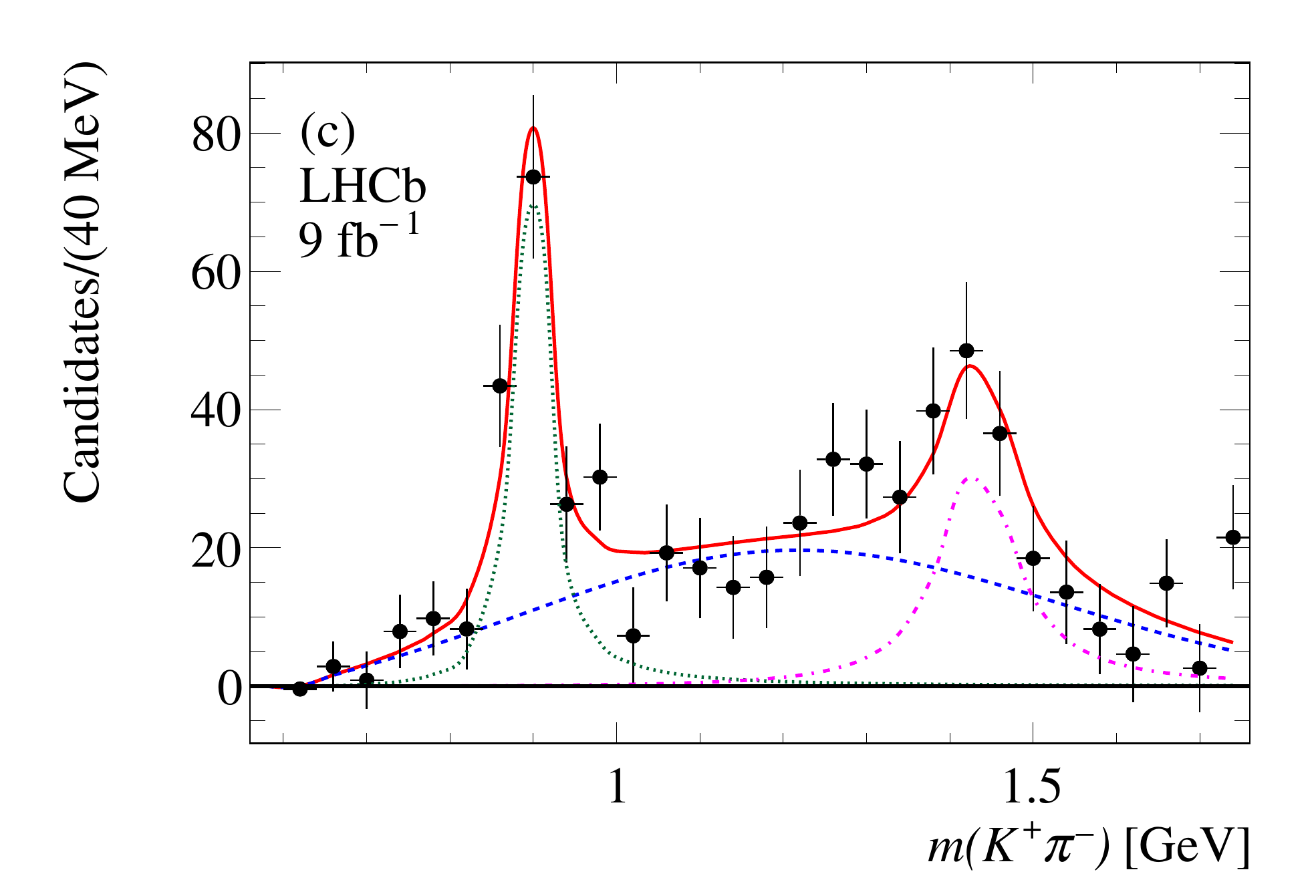}
\includegraphics[width=0.45\textwidth]{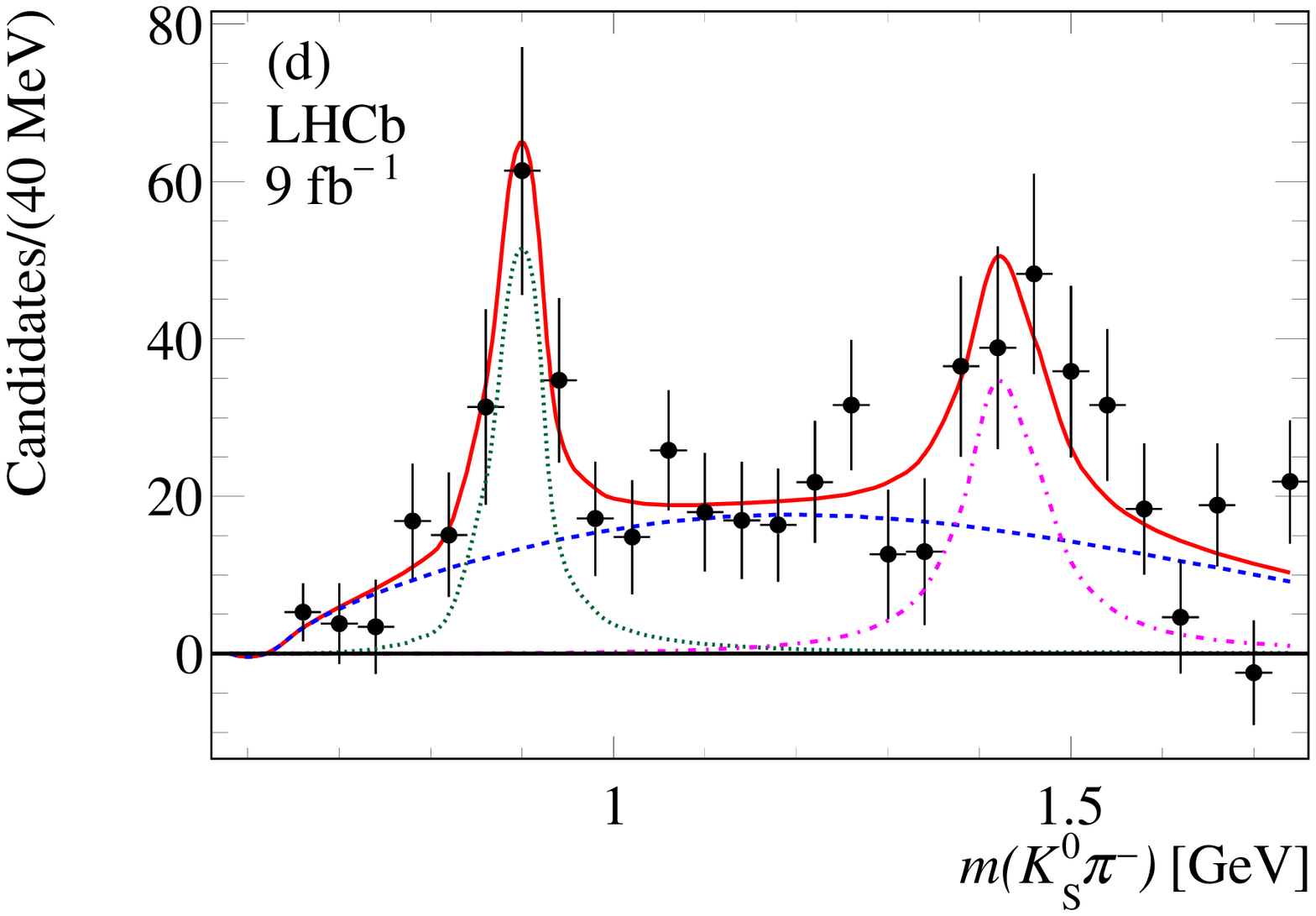}
\caption{\small\label{fig:fig25}Background-subtracted $K \pi$ and $\KS \pi$ invariant-mass spectra from $\chicone \to \kskpi$ decays for the (a)-(b) \bkskkpip and (c)-(d) \bkskkpim data. The results of the fits are overlaid.}
\end{figure}

The background contribution is evaluated using the sidebands shown in Fig.~\ref{fig:fig23}. For each \KS category,
the  $K \pi$ and $\KS \pi$ invariant-mass distributions from the \chicone  signal region and from the sidebands are obtained. The $K \pi$ and $\KS \pi$  invariant-mass distributions in the sidebands are normalized to the \chicone fitted purity,
and subtracted from the invariant-mass spectra in the \chicone signal region, obtaining the distributions shown in Fig.~\ref{fig:fig25}. Prominent $K^*(892)$ and $K^*_2(1430)$ resonances are observed.
A fit to the $K \pi$ and $\KS \pi$ invariant-mass distributions is performed using an empirical and flexible background shape in addition to two relativistic
Breit--Wigner functions describing the $K^*(892)$ and $K^*_2(1430)$ resonances with parameters fixed to their known values~\cite{Workman:2022ynf}.
The background is parameterized as
\begin{eqnarray}
B(m) = & P(m)e^{a_1m+a_2m^2} &{\rm for} \ m<m_0, \textrm{and}\nonumber\\
B(m) = & P(m)e^{b_0+b_1m+b_2m^2} &{\rm for} \ m>m_0,
\end{eqnarray}
where $P(m)$ is the two-body phase space
\begin{equation}
    P(m) = \frac{1}{2m}\sqrt{[m^2-(m_K - m_{\pi})^2][m^2-(m_K + m_{\pi})^2]}
\end{equation}
and $m_0, a_1, a_2, b_0, b_1,b_2$ are free parameters. The two functions and their first derivatives are required to be continuous at $m=m_0$, and this constraint reduces the number of freely varying parameters to four.
Figure~\ref{fig:fig25} shows the fits to the background-subtracted  $K \pi$ and $\KS \pi$ invariant-mass distributions in the \chicone signal region for both the \mbox{\bkskkpip} and \bkskkpim data. Table~\ref{tab:tab15} reports 
candidate yields and fractional contributions obtained from the fits. 

\begin{table} [tb]
  \centering
  \caption{\small\label{tab:tab15} Results from the fits to the $K \pi$ and $\KS \pi$ background-subtracted invariant-mass distributions in the \chicone decay for the (top) \bkskkpip and (bottom) \kskkpim data.}
  {\small
\begin{tabular}{lcccc}
\hline
Resonance &  p-value [\%] & \chicone Yield & Events & Fraction \cr
\hline
\bkskkpip & & & & \cr
\hline\\ [-2.3ex]
$\bar K^{*}(892)^0$ & 30 & $1262\pm54$ & $101 \pm 19$ & $0.080\pm 0.015$ \cr
$\bar K^{*}_2(1430)^0$      &   &   & $140 \pm 23$ & $0.111 \pm 0.019$ \cr
$K^{*}(892)^+$ & 56 &  & $125 \pm 26$ & $0.099 \pm 0.021$ \cr
$K^{*}_2(1430)^+$      &  &     & $188 \pm 26$ & $0.149\pm 0.022$ \cr
\hline\\ [-2.3ex]
\bkskkpim & & & & \cr
\hline\\ [-2.3ex]
$K^{*}(892)^0$ & 64 & $1244 \pm 72$ & $161 \pm 22$ & $0.129 \pm 0.019$ \cr
$K^{*}_2(1430)^0$&       &      & $138 \pm 31$ & $0.111 \pm 0.026$ \cr
$K^{*}(892)^-$ & 35 &       & $163 \pm 29$ & $0.131 \pm 0.025$ \cr
$K^{*}_2(1430)^-$&     &      & $163 \pm 36$ & $0.131 \pm 0.030$ \cr
\hline
\end{tabular}
 }
\end{table}
The  inverse-variance-weighted averages of the fractional $K^{*}(892)$ and $K^{*}_2(1430)$ contributions obtained from the fits to the \bkskkpip and \bkskkpim data are listed in Table~\ref{tab:tab16}.
The fractional $K^{*}(892)$ and $K^{*}_2(1430)$ contributions are converted into $\chicone \to K^* \bar K$ branching fractions (listed in Table~\ref{tab:tab16}) by multiplying the fractional contributions by the
known \chicone branching fraction~\cite{Workman:2022ynf},
\begin{equation}
  \calB(\chicone \to \bar{K^0} \Kp \pim) = (7.0 \pm 0.6) \times 10^{-3},
\end{equation}
and correcting for unseen $K^*$ decay modes. A comparison with Ref.~\cite{Ablikim:2006vm} shows good agreement, while improving
the statistical uncertainties and significances.

Systematic uncertainties are evaluated as follows. The deviations using different BDT classifier selections are evaluated by fitting the lower- and higher-purity datasets and averaging 
the absolute values of the deviations of the fitted fractions in comparing to the values obtained from the default fit.  
The differences resulting from the use of the \bkskkpip or \bkskkpim decay modes are also included as systematic uncertainties.
The Blatt--Weisskopf radius $r$ entering in the description of the $K^{*}(892)$ and the $K^{*}_2(1430)$ resonances, fixed at 1.5 $\gev^{-1}$, is varied between 0.5 and 2.5 $\gev^{-1}$.

\begin{table} [tb]
  \centering
  \caption{\small\label{tab:tab16} 
  Inverse-variance-weighted averages of the fractional contributions for \chicone decays to $K^*\Kbar$ resonances from the \bkskkpip and \bkskkpim final states.
  The branching fractions of the $\chicone \to \kskpi$ decays to $K^{*}(892)$ and the $K^{*}_2(1430)$ resonances are also included. The reported uncertainties are statistical, systematic and from the uncertainty on the \chicone branching fraction (see text)~\cite{Workman:2022ynf}.}
  {\small
\begin{tabular}{lcc}
\hline\\ [-2.3ex]
Decay mode & Fraction & Branching fraction ($\times 10^{-3}$) \cr
\hline\\ [-2.3ex]
$\calB(\chicone \to K^{*}(892)^0 \bar K^0)$ & $0.099\pm 0.012 \pm 0.004$ & $1.04 \pm 0.13 \pm 0.04 \pm 0.09$ \cr
$\calB(\chicone \to K^{*}_2(1430)^0 \bar K^0)$  & $0.111 \pm 0.015 \pm 0.005$ & $1.17 \pm 0.16 \pm 0.05 \pm 0.10$ \cr
$\calB(\chicone \to K^{*}(892)^+ \Km)$ & $0.112 \pm 0.016 \pm 0.013$ & $1.18 \pm 0.17 \pm 0.14 \pm 0.10$ \cr
$\calB(\chicone \to K^{*}_2(1430)^+ \Km)$ & $0.143\pm 0.018 \pm 0.006$  & $1.61 \pm 0.19 \pm 0.19 \pm 0.14$ \cr
\hline
\end{tabular}
 }
\end{table}

\section{Measurement of the branching fractions of \boldmath{\Bp} decays to charmonium resonances }
\label{sec:br}

This section is devoted to the extraction of the branching fractions
\begin{equation}
  \calB(\Bp \to (c \bar c) X^+),
  \label{eq:cc}
\end{equation}
where $c \bar c=\etac, \jpsi, \chiczero, \chicone, \chictwo, \etactwo$ and $X$ indicates a \Kp or a $\KS \pip$ system.
These measurements are performed with respect to the known \etac and \jpsi branching fractions~\cite{Workman:2022ynf}:
\begin{equation}
 \calB(\Bp \to \etac \Kp)\cdot\calB(\etac \to \kskpi)=(2.7 \pm 0.6) \times 10^{-5}, 
\end{equation}
and
\begin{equation}
\calB(\Bp \to \jpsi \Kp)\cdot\calB(\jpsi \to \kskpi)=(5.71 \pm 0.52) \times 10^{-6}.
\end{equation}
The entire dataset, selected as described in Sec.~\ref{sec:evsel} is used in this section.
Efficiency-corrected invariant-mass distributions are obtained by weighting each event by the inverse of the total efficiency described in Sec.~\ref{sec:effy}.
Fits are performed to the resulting efficiency-corrected invariant-mass spectra, and the extracted yields of \Bp decays and of various charmonium resonances are used to obtain efficiency-corrected ratios with respect to the \etac and \jpsi resonances. 
The procedure is performed separately for \KSLL and \KSDD data, which are subsequently combined by means of inverse-variance-weighted averages.

Prior to evaluating the \bkskkpip and \bkskkpim candidate yields, it is necessary to subtract 
all contributions from open-charm production. These are observed
in the two- or three-body invariant-mass spectra of the $B \to D_{(s)} X$ decays, where $X$ represents one or two additional particles. The charmed-meson vetoes from the previous sections are not applied. Since open-charm resonances may also be present in the \Bp-meson background, a
background subtraction is performed using the normalized \Bp-candidate invariant-mass sidebands. This is achieved by plotting the different two-body or three-body invariant-mass combinations in each of the two sideband regions, with both halves as wide as the \Bp signal region.
The resulting invariant-mass distributions are then subtracted from the corresponding ones in the signal region.
In performing fits to the two- or three-body mass spectra, the charmed mesons are described by two Gaussian functions,
sharing the same mean, with first- or second-order polynomials used to model the background.
Table~\ref{tab:tab17} summarizes the fit results for the \bkskkpip and \bkskkpim data, without efficiency corrections. 

\begin{table} [tb]
  \centering  \caption{\small\label{tab:tab17} Resulting yields and total charm fraction from the fits to the four-body, three-body and two-body invariant-mass spectra uncorrected for efficiency for the (top) \bkskkpip and (bottom) \bkskkpim data, separated by \KS category. The quoted uncertainties are statistical only.}
  {\small
\begin{tabular}{l|rr}
\hline
Resonance & LL \al\al\al& DD \al\al\al\cr
\hline\\ [-2.3ex]
\bkskkpip & $27446 \pm 228$ & $67515 \pm 426$ \cr
\hline\\ [-2.3ex]
$\Dsp \to \KS \Kp$ & $423 \pm \al 28$ & $994 \pm \al 49$ \cr
$\Dz \to \Kp \Km$ & $99 \pm \al 16$ & $185 \pm \al 26$  \cr
$\Dz \to \KS \Km \pip$ & $116 \pm \al 16$  & $319 \pm \al 28$  \cr
$\Dz \to \KS \Km \Kp$ & $4895 \pm \al 71$  & $11627 \pm 109$   \cr
\hline\\ [-2.3ex]
Sum of charm & $ 5533 \pm \al 80$ & $13125 \pm 125 $ \cr
Charm fraction (\%) & $20.2 \pm \all 0.3$ & $19.4 \pm \all 0.2$ \cr
\Bp no charm & $21913 \pm 242$  & $54390 \pm 444$  \cr
\hline\\ [-2.3ex]
\hline\\ [-2.3ex]
\bkskkpim & $20950 \pm 221$ & $47154 \pm 342$ \cr
\hline\\ [-2.3ex]
$\Dsp \to \KS \Kp$ & $321 \pm \al 17$ & $726 \pm \al 51$  \cr
$\Dz \to \Km \pip$ & $1553 \pm \al 49$ & $2798 \pm \al 68$ \cr
$\Dz \to \KS \Km \pip$ & $366 \pm \al 24$  & $390 \pm \al 30$  \cr
\hline\\ [-2.3ex]
Sum of charm & $ 2210 \pm \al 57$ & $3914 \pm \al 90 $  \cr
Charm fraction (\%) & $10.6 \pm \all 0.3$ & $8.3 \pm \all 0.2$ \cr
\Bp no charm & $18740 \pm 228$  & $43240 \pm 354$  \cr
\hline
\end{tabular}
}
\end{table}

The efficiency-corrected \kskpi and $\Kp \Km$ invariant mass spectra are fitted as described in Sec.~\ref{sec:charmrespar} to obtain efficiency-corrected yields for \etac, \jpsi (see Fig.~\ref{fig:fig4}), \chicone, \etactwo (see Fig.~\ref{fig:fig5}), \chiczero and \chictwo (see Fig.~\ref{fig:fig6}) resonances. 
The efficiency corrected \kskkpip and \kskkpim invariant-mass spectra are fitted as described in Sec.~\ref{sec:evsel} (see Fig.~\ref{fig:fig2}).
These yields are used to compute the ratios with respect to the \etac and \jpsi resonance yields,
labeled as 
\begin{equation}
\calR_1(X) = \frac{N(\Bp \to X \Kp)}{N(\Bp \to \etac \Kp)}
\end{equation}
and 
\begin{equation}
\calR_2(X)=\frac{N(\Bp \to X \Kp)}{N(\Bp \to \jpsi \Kp)},
\end{equation}
respectively, where $X$ labels the charmonium resonance.
Similar expressions are used to calculate the \Bp four-body decays ratios $\calR_1(\Bp)$ and $\calR_2(\Bp)$. 
Table~\ref{tab:tab18} lists the measurements of the efficiency-corrected relative ratios.

The following systematic uncertainties are considered.
In fitting for \Bp-candidate and charmonium-resonance 
yields, the fit model is varied to use Gaussian functions sharing or not sharing the same mean on a quadratic, linear or cubic polynomial background shape. The maximum deviation from the default fit is taken as the systematic uncertainty for each parameter.
The open-charm fraction is re-evaluated for efficiency-corrected invariant-mass distributions and the difference with respect to the default fit is considered as a systematic uncertainty.
In fitting the \kskpi invariant-mass spectra, the bin size is changed and the background shape is modified from a linear to quadratic function. 
To evaluate the size of the fit bias on the resonance yields, starting from the reference fits, 
400 randomized pseudoexperiments are generated according to Poissionian statistics, having the same sample size as the original ones, and then fitted. The resonance yields are compared with the reference fits. In all cases, the deviations from the reference fits are included as systematic uncertainties.
The entire procedure of fitting the invariant-mass spectra and evaluating the relative ratios is repeated dividing the data into the different trigger conditions, TOS and noTOS, whose results are then averaged and compared with the reference fit.
All the systematic uncertainties are added in quadrature to calculate the total systematic uncertainty. 

\begin{table} [tb]
  \centering
  \caption{\small\label{tab:tab18} Ratios of efficiency-corrected intermediate-resonance yields relative to the yield of the (top) \etac and (bottom) \jpsi resonances. The first uncertainty is statistical, the second systematic.}
\begin{tabular}{lrr}
\hline\\ [-2.3ex]
Resonance & \bkskkpip & \bkskkpim \cr
\hline\\ [-2.3ex]
$\calR_1(\Bp)$ & $4.14 \pm 0.04 \al \pm  0.25 \al$ & $3.40 \pm 0.04 \al \pm 0.09 \al $\cr
$\calR_1(\jpsi)$ &  $0.201 \pm 0.005 \pm 0.008$ & $0.200 \pm 0.005 \pm 0.013$ \cr
$\calR_1(\chicone)$ & $0.082 \pm 0.004 \pm 0.003$  & $0.073 \pm 0.004 \pm 0.007$\cr
$\calR_1(\etactwo)$ & $0.113 \pm 0.005 \pm 0.004$ & $0.109 \pm 0.007 \pm 0.016$\cr
$\calR_1(\chiczero)$ &  $0.107 \pm 0.006 \pm 0.009$ & \cr 
$\calR_1(\chictwo)$ & $0.011 \pm 0.003 \pm 0.001$ & \cr
\hline\\ [-2.3ex]
$\calR_2(\Bp)$ &   $20.6 \pm 0.5\al  \pm 0.6 \al$& $17.0 \pm 0.4 \al \pm 0.8 \al $ \cr
$\calR_2(\etac)$ & $4.95 \pm 0.12 \pm 0.23$& $5.26 \pm 0.24 \pm 0.29$\cr
$\calR_2(\chicone)$ &  $0.41 \pm 0.02 \pm 0.02$& $0.36 \pm 0.02 \pm 0.04$\cr
$\calR_2(\etactwo)$ &  $0.56 \pm 0.03 \pm 0.02$& $0.55 \pm 0.04 \pm 0.08$\cr
$\calR_2(\chiczero)$ &  $0.53 \pm 0.03 \pm 0.04$ & \cr 
$\calR_2(\chictwo)$ &  $0.06 \pm 0.01 \pm 0.01$ & \cr
\hline
\end{tabular}
\end{table}

Using the listed ratios it is possible to evaluate the branching fractions as
\begin{equation}
  \calB_1(X) = f_1 \cdot \calR_1(X)\cdot \calB(\Bp \to \etac \Kp)\cdot\calB(\etac \to \kskpi)
\end{equation}
for the \etac mode and
\begin{equation}
  \calB_2(X) = f_2 \cdot \calR_2(X)\cdot\calB(\Bp \to \jpsi \Kp)\cdot\calB(\jpsi \to \kskpi)
\end{equation}
for the \jpsi mode, where the $f_i$ are correction factors for unseen decay modes, including $K^0$ modes. 
The resulting measured branching fractions are listed in Table~\ref{tab:tab19} ($\calB_1$) and Table~\ref{tab:tab20} ($\calB_2$). 
For the modes involving \chiczero and \chictwo resonances, listed in Table~\ref{tab:tab21}, the $f_i$ parameters also include the corrections 
for their known branching fractions to the $\Kp \Km$ state, $(6.05 \pm 0.31)\times 10^{-3}$ and $(1.01 \pm 0.06) \times 10^{-3}$, respectively~\cite{Workman:2022ynf}.

\begin{table} [tb]
  \centering
\caption{\small\label{tab:tab19} Measured branching fractions with the \etac resonance as a reference using the (top) \bkzkkpip and (bottom) \bkzkkpim data. 
  The first uncertainty is statistical, the second systematic and the third due to the PDG uncertainty on the $\Bp \to \etac \Kp$ branching fraction.}
\begin{tabular}{lrc}
\hline\\ [-2.3ex]
Final state & $\calB_1$ $(\times 10^{-5})$\al \al \al \al \al \al  & PDG $(\times 10^{-5})$\cr
\hline\\ [-2.3ex]
\bkzkkpip & $32.28 \pm  0.33 \al \pm 1.97 \al \pm 7.17 \al $ \cr
\hline\\ [-2.3ex]
$\Bp \to \jpsi \Kp$ & $0.543 \pm 0.012 \pm 0.023 \pm 0.121$ & $0.571 \pm 0.052$ \cr
$\Bp \to \chicone \Kp$ & $0.222 \pm 0.010 \pm 0.009 \pm 0.049$ & $0.194 \pm 0.037$ \cr
$\Bp \to \etactwo \Kp$ & $0.306 \pm 0.014 \pm 0.010 \pm 0.068$& $0.34^{+0.23}_{-0.16}$ \cr
\hline\\ [-2.3ex]
\bkzkkpim & $26.56 \pm 0.31 \al \pm 0.68 \al \pm 5.90 \al $ \cr
\hline\\ [-2.3ex]
$\Bp \to \jpsi \Kp$ & $0.541\pm  0.014\pm 0.035\pm 0.120$ \cr
$\Bp \to \chicone \Kp$ & $0.196 \pm 0.011\pm 0.018\pm 0.044$ \cr
$\Bp \to \etactwo \Kp$ & $0.295\pm  0.019\pm 0.042\pm 0.066$ \cr
\hline
\end{tabular}
\end{table}

\begin{table} [tb]
\centering
\caption{\small\label{tab:tab20} Measured branching fractions with the \jpsi resonance as a reference using the (top) \bkzkkpip and (bottom) \bkzkkpim data. The first uncertainty is statistical, the second systematic, the third due to the PDG uncertainty on the $\Bp \to \jpsi \Kp$ branching fraction.}
\begin{tabular}{lrc}
\hline\\ [-2.3ex]
Final state & $\calB_2$ $(\times 10^{-5})$ \al \al \al \al \al \al  & PDG $(\times 10^{-5})$\cr
\hline\\ [-2.3ex]
\bkzkkpip & $34.01 \pm 0.74 \al \pm 0.91 \al \pm 3.10 \al $ & \cr
\hline
$\Bp \to \etac \Kp$ & $2.83 \pm 0.07\al  \pm 0.13 \al \pm 0.26 \al$ & $2.7 \pm 0.6$ \cr
$\Bp \to \chicone \Kp$ & $0.225 \pm 0.011 \pm 0.006\pm 0.021$ & $0.194 \pm 0.037$ \cr
$\Bp \to \etactwo \Kp$ & $0.327 \pm 0.017\pm0.015\pm0.030$ & $0.34^{+0.23}_{-0.16}$ \cr
\hline\\ [-2.3ex]
\bkzkkpim & $28.01 \pm 0.68 \al \pm 1.35 \al \pm 2.55 \al $ & \cr
\hline\\ [-2.3ex]
$\Bp \to \etac \Kp$ & $3.00\pm 0.14 \al \pm 0.16 \al \pm 0.27 \al$ & \cr
$\Bp \to \chicone \Kp$ & $0.206 \pm  0.012 \pm 0.023 \pm 0.019$ \cr
$\Bp \to \etactwo \Kp$ & $0.313\pm  0.021 \pm 0.044 \pm 0.028$ \cr
\hline
\end{tabular}
\end{table}

\begin{table} [tb]
  \centering  \caption{\small\label{tab:tab21} Measured branching fractions using (top) the \etac and (bottom) the \jpsi resonance as reference for \bkzkkpip data. 
    The first uncertainty is statistical, the second systematic, the third due to the PDG uncertainty on the
   $\Bp \to \etac \Kp$ or $\Bp \to \jpsi \Kp$ branching fraction. The PDG reports an upper limit $\Bp \to \chiczero K^{*+}<0.21 \times 10^{-3}$.}
\begin{tabular}{lrc}
\hline\\ [-2.3ex]
Final state & $\calB_1$ $(\times 10^{-3})$ \al \al \al \al \al \al  & PDG $(\times 10^{-3})$\cr
\hline\\ [-2.3ex]
$\Bp \to \chiczero K^0 \pip$ & $1.38 \pm 0.07 \pm 0.11 \pm 0.32$ &   \cr
$\Bp \to \chictwo K^0 \pip$ & $0.87 \pm 0.20 \pm 0.08 \pm 0.20$& $0.116\pm0.025$\cr
\hline\\ [-2.3ex]
Final state & $\calB_2$ $(\times 10^{-3})$ \al \al \al \al \al \al  & \cr
\hline\\ [-2.3ex]
$\Bp \to \chiczero K^0 \pip$ & $1.45 \pm 0.08 \pm 0.11 \pm 0.16$ & \cr
$\Bp \to \chictwo K^0 \pip$ & $ 0.92 \pm 0.21 \pm 0.08 \pm 0.10$  &  \cr
\hline
\end{tabular}
\end{table}

The \bkzkkpip and \bkzkkpim branching fractions are measured here for the first time.
Tension is found for the $\Bp \to \chictwo K^0 \pip$ branching fraction with respect to the PDG value~\cite{Workman:2022ynf}, measured only in Ref.~\cite{Belle:2015opn} to be $(0.116\pm  0.022 \pm 0.012)\times 10^{-3}$ using the $\chictwo \to \jpsi \gamma$ decay mode as a reference with $76.4 \pm 14.7$ events and a significance of $4.6\sigma$.
This measurement differs from that reported here by $2.6\sigma$ ($3.2\sigma$), using the \etac (\jpsi) meson as a reference
and adding the statistical and systematic uncertainties in quadrature.

\section{Summary}
\label{sec:summary}

A study is presented of \bkskkpip and \bkskkpim decays at proton-proton collision energies of 7, 8 and 13 \tev using the \lhcb detector. The \kskpi invariant-mass spectra from both \Bp decay modes show a rich spectrum of charmonium resonances. New  measurements of the \etac and \etactwo masses and widths are performed resulting in the best determinations from a single measurement. 
Branching fractions are measured  for \Bp decays to \etac, \jpsi, \etactwo and \chicone resonances. 
Evidence is also found for the $\Bp \to \chictwo K^0 \pip$ decay.
Furthermore, the first observation and branching fraction measurement of $\Bp \to \chiczero K^0 \pip$ decays is reported. The first measurements of \bkzkkpip and \bkzkkpim  branching fractions are also given. A Dalitz-plot analysis of \decay{\etac}{\kskpi} decay is undertaken using both a quasi-model-independent approach for the $K \pi$ {\it S}-wave and an isobar-model approach. The $K^*_0(1950) \to K \pi$ resonance is established and its parameters are measured. A new parametrization of the 
$K \pi$ {\it S}-wave is obtained, which includes $K^*_0(1430)$, $K^*_0(1950)$ and broad $\kappa(2600)$ resonances.
No evidence for the $\kappa(700)$ resonance is found in \etac decays. The $a_0(1700)$ resonance as well as its decay to $K^0 \bar K$ are confirmed, and its parameters are measured. The first Dalitz-plot analysis of the $\etactwo \to \kskpi$ decay is performed.
Finally, branching fractions of \decay{\chicone}{K^*(892)^0 \bar K^0}, \decay{\chicone}{K^{*}_2(1430)^0 \bar K^0}, \decay{\chicone}{K^{*}(892)^+ \Km} and \decay{\chicone}{K^{*}_2(1430)^+ \Km} decays are measured; these measurements are more precise than previous evaluations and correspond to the first observations of the two $\chicone \to K^*_2(1430)\Kbar$ decay modes.

\clearpage

\appendix
\section{Appendix}
\label{sec:app}

The inverse-variance-weighted averages of the numerical values of the $K \pi$ {\it S}-wave as a function of the $K \pi$ mass from the Dalitz plot analysis of \etac mesons in \bkskkpip and \bkskkpim decays are listed in Table~\ref{tab:tab22}.

\begin{table} [b]
  \centering
  \caption{\small\label{tab:tab22} Inverse-variance averaged numerical values of the measured $K \pi$ {\it S}-wave as a function of the $K \pi$ mass.
    The first uncertainty is statistical, the second systematic. The values at $K \pi$ mass of 1.425 \gev (in italics) are fixed.}
  {\small
\begin{tabular}{ccc}
\hline\\ [-2.3ex]
$K \pi$ mass [$\gev$] & Amplitude & Phase [rad] \cr
\hline\\ [-2.3ex]
0.675 & $0.506 \pm 0.071 \pm 0.393$  &  $2.827 \pm 0.117 \pm 0.202$ \cr
0.725 & $0.297 \pm 0.065 \pm 0.370$  &  $2.726 \pm 0.195 \pm 0.562$ \cr
0.775 & $0.372 \pm 0.041 \pm 0.173$  &  $2.238 \pm 0.189 \pm 0.734$ \cr
0.825 & $0.354 \pm 0.045 \pm 0.314$  &  $2.641 \pm 0.158 \pm 0.729$ \cr
0.875 & $0.378 \pm 0.036 \pm 0.117$  &  $2.254 \pm 0.171 \pm 0.492$ \cr
0.925 & $0.254 \pm 0.041 \pm 0.105$  &  $0.823 \pm 0.148 \pm 0.362$ \cr
0.975 & $0.299 \pm 0.036 \pm 0.125$  &  $0.885 \pm 0.127 \pm 0.230$ \cr
1.025 & $0.322 \pm 0.032 \pm 0.024$  &  $0.875 \pm 0.110 \pm 0.031$ \cr
1.075 & $0.408 \pm 0.032 \pm 0.039$  &  $0.852 \pm 0.094 \pm 0.427$ \cr
1.125 & $0.434 \pm 0.030 \pm 0.037$  &  $0.842 \pm 0.087 \pm 0.138$ \cr
1.175 & $0.488 \pm 0.029 \pm 0.074$  &  $0.937 \pm 0.088 \pm 0.317$ \cr
1.225 & $0.564 \pm 0.029 \pm 0.017$  &  $0.818 \pm 0.077 \pm 0.220$ \cr
1.275 & $0.667 \pm 0.027 \pm 0.025$  &  $1.021 \pm 0.067 \pm 0.217$ \cr
1.325 & $0.710 \pm 0.024 \pm 0.025$  &  $1.201 \pm 0.063 \pm 0.113$ \cr
1.375 & $0.780 \pm 0.021 \pm 0.010$  &  $1.315 \pm 0.047 \pm 0.095$ \cr
{\it 1.425} & $1.000$ & $1.570$\cr
1.475 & $1.016 \pm 0.022 \pm 0.054$  &  $1.786 \pm 0.040 \pm 0.049$ \cr
1.525 & $0.789 \pm 0.020 \pm 0.026$  &  $1.939 \pm 0.049 \pm 0.028$ \cr
1.575 & $0.575 \pm 0.019 \pm 0.018$  &  $2.248 \pm 0.060 \pm 0.181$ \cr
1.625 & $0.302 \pm 0.022 \pm 0.038$  &  $2.409 \pm 0.083 \pm 0.179$ \cr
1.675 & $0.200 \pm 0.023 \pm 0.033$  &  $1.970 \pm 0.117 \pm 0.301$ \cr
1.725 & $0.189 \pm 0.022 \pm 0.025$  &  $1.935 \pm 0.127 \pm 0.544$ \cr
1.775 & $0.279 \pm 0.019 \pm 0.089$  &  $1.491 \pm 0.103 \pm 0.324$ \cr
1.825 & $0.387 \pm 0.019 \pm 0.040$  &  $1.371 \pm 0.097 \pm 0.123$ \cr
1.875 & $0.512 \pm 0.022 \pm 0.106$  &  $1.553 \pm 0.092 \pm 0.154$ \cr
1.925 & $0.547 \pm 0.021 \pm 0.042$  &  $1.520 \pm 0.083 \pm 0.105$ \cr
1.975 & $0.686 \pm 0.024 \pm 0.005$  &  $1.654 \pm 0.070 \pm 0.197$ \cr
2.025 & $0.720 \pm 0.025 \pm 0.030$  &  $1.712 \pm 0.076 \pm 0.129$ \cr
2.075 & $0.710 \pm 0.026 \pm 0.016$  &  $1.719 \pm 0.073 \pm 0.081$ \cr
2.125 & $0.642 \pm 0.026 \pm 0.025$  &  $1.841 \pm 0.080 \pm 0.172$ \cr
2.175 & $0.668 \pm 0.028 \pm 0.040$  &  $1.898 \pm 0.081 \pm 0.203$ \cr
2.225 & $0.704 \pm 0.029 \pm 0.033$  &  $1.725 \pm 0.083 \pm 0.190$ \cr
2.275 & $0.760 \pm 0.035 \pm 0.063$  &  $1.894 \pm 0.081 \pm 0.348$ \cr
2.325 & $0.735 \pm 0.040 \pm 0.091$  &  $1.782 \pm 0.090 \pm 0.346$ \cr
2.375 & $0.789 \pm 0.050 \pm 0.130$  &  $1.883 \pm 0.081 \pm 0.339$ \cr
2.425 & $0.896 \pm 0.058 \pm 0.236$  &  $2.081 \pm 0.077 \pm 0.329$ \cr
2.475 & $0.876 \pm 0.070 \pm 0.332$  &  $2.088 \pm 0.108 \pm 0.483$ \cr
\hline
\end{tabular}
}
\end{table}

\begin{figure}[tb]
\centering
\small
\includegraphics[width=0.95\textwidth]{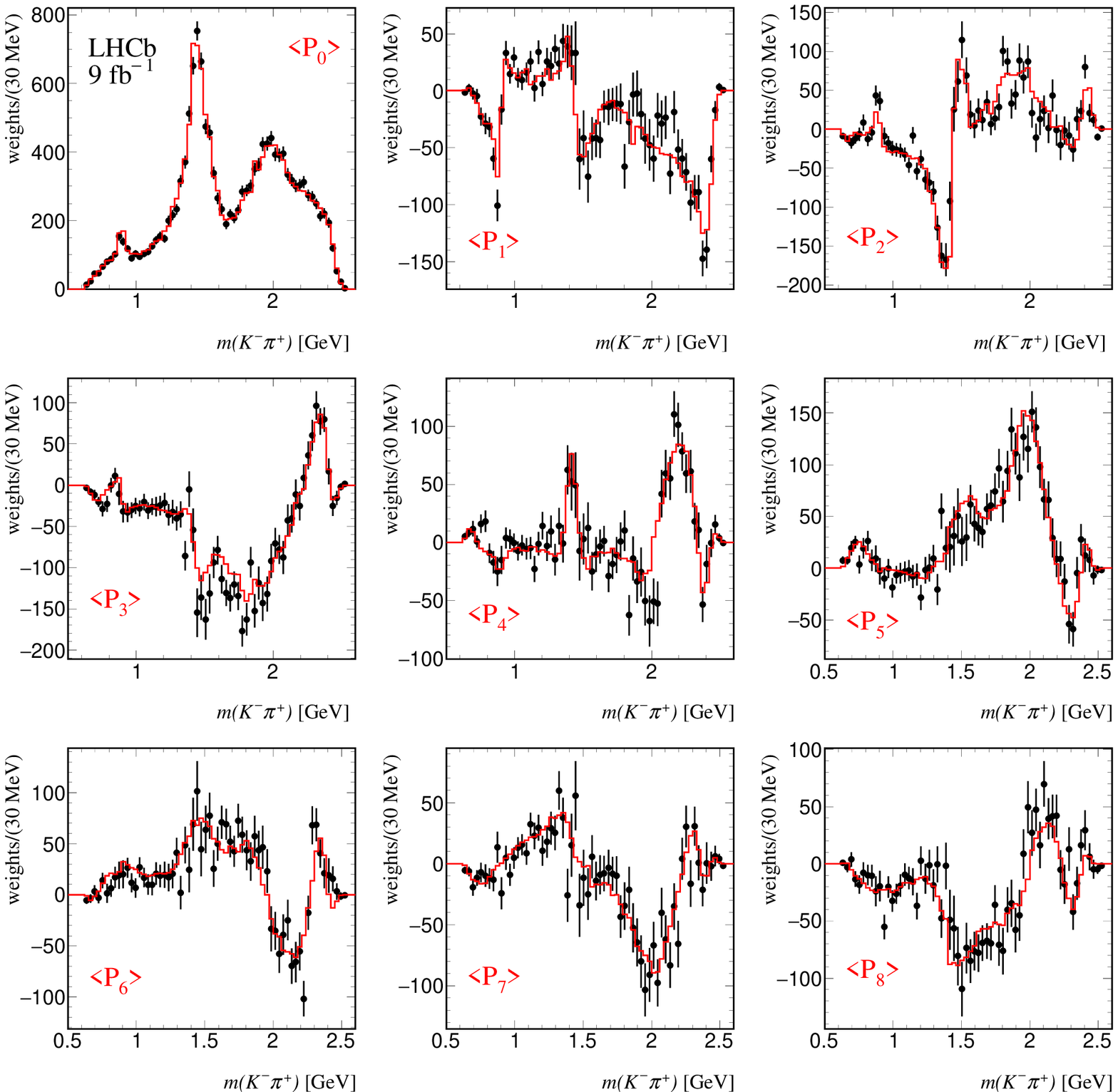}
\caption{\small\label{fig:fig26} 
Projected $\Km \pip$ invariant mass spectra weighted by Legendre polynomial moments $P_L(\cos\theta_{\KS})$ for $\etac \to \KS \Km \pip$ from \bkskkpip data. The
drawn curves result from the Dalitz plot fit using the QMI method described in the text.}
\end{figure}

\begin{figure}[tb]
\centering
\small
\includegraphics[width=0.95\textwidth]{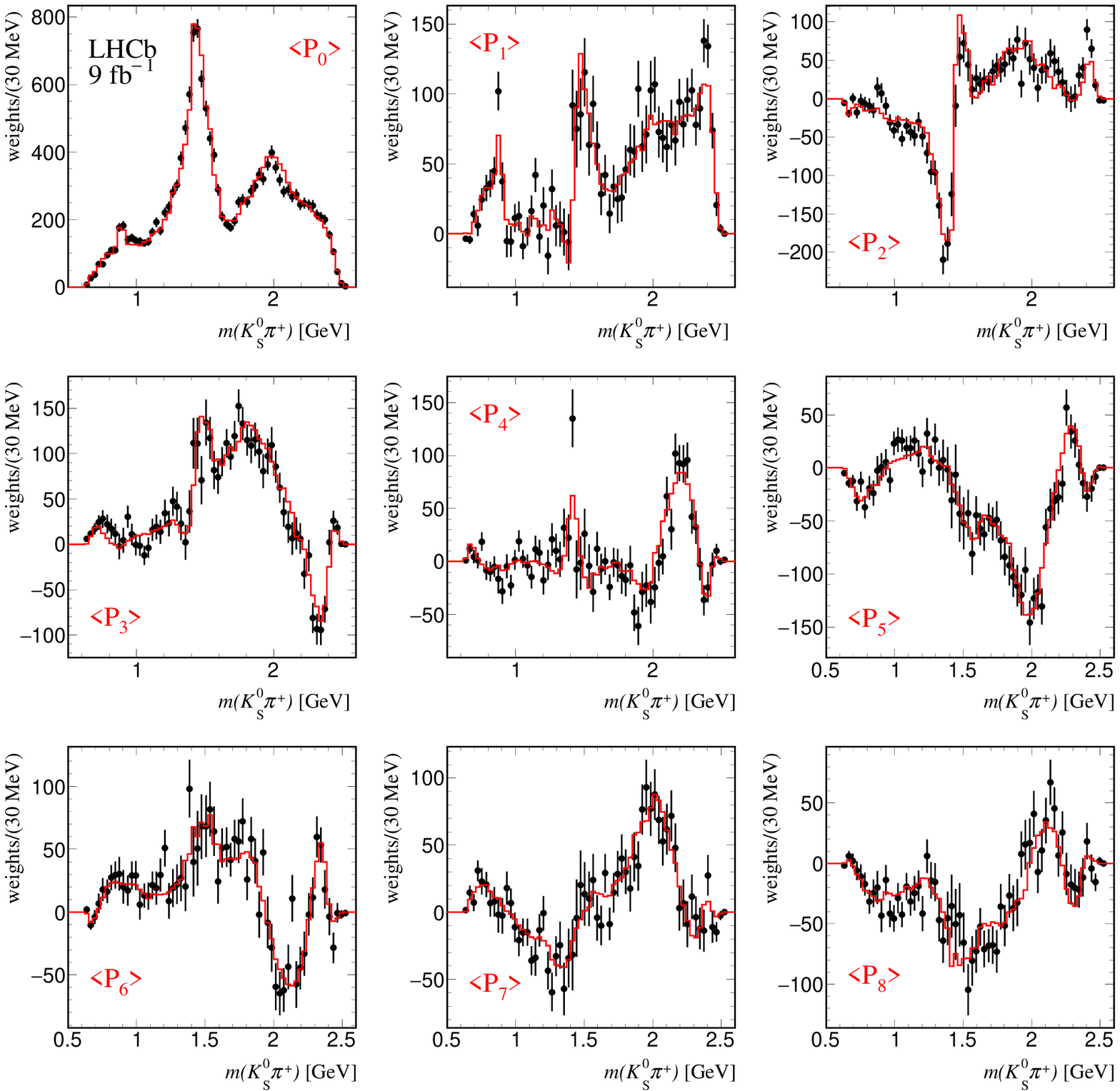}
\caption{\small\label{fig:fig27} 
Projected $\KS \pip$ invariant mass spectra weighted by Legendre polynomial moments $P_L(\cos\theta_{\Km})$ for $\etac \to \KS \Km \pip$ from \bkskkpip data. The
drawn curves result from the Dalitz plot fit using the QMI method described in the text.}
\end{figure}

\begin{figure}[tb]
\centering
\small
\includegraphics[width=0.95\textwidth]{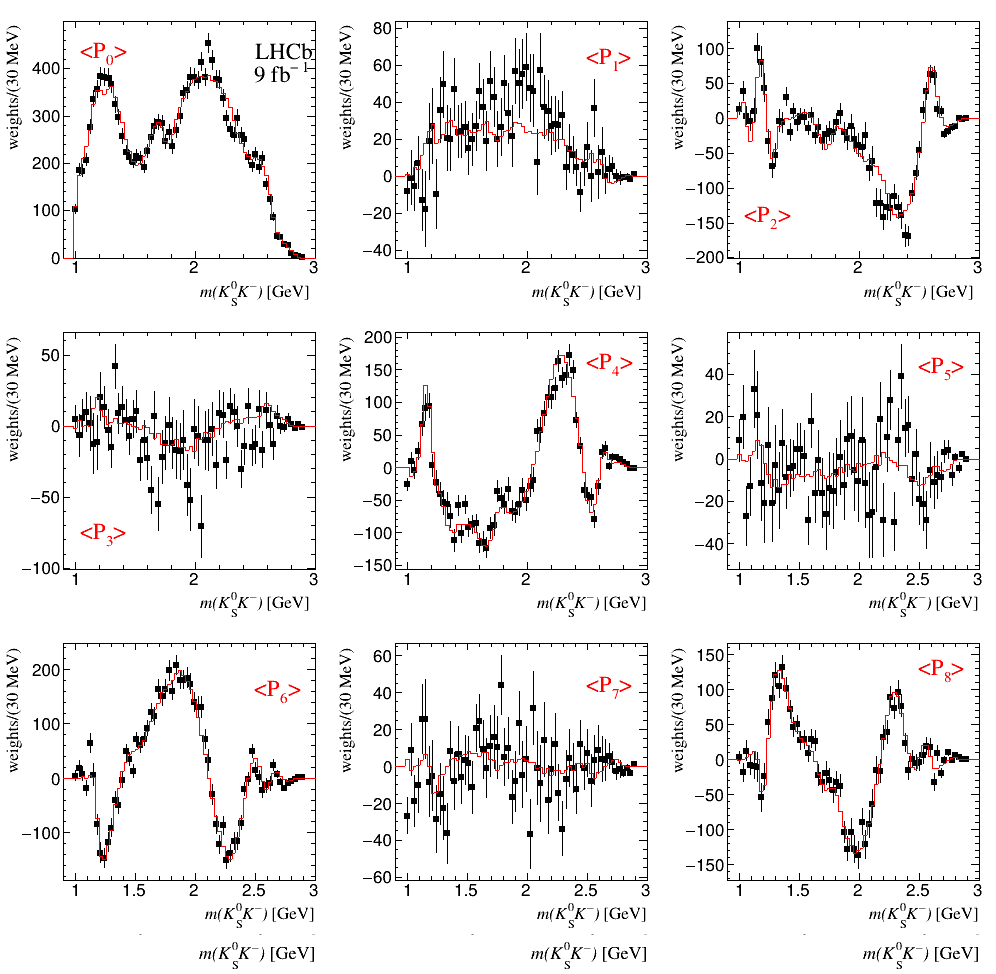}
\caption{\small\label{fig:fig28} 
Projected $\KS \Km$ invariant mass spectra weighted by Legendre polynomial moments $P_L(\cos\theta_{\pip})$ for $\etac \to \KS \Km \pip$ from \bkskkpip data. The
drawn curves result from the Dalitz plot fit using the QMI method described in the text.}
\end{figure}

Figures~\ref{fig:fig26}, ~\ref{fig:fig27}, and ~\ref{fig:fig28} show the $\Km \pip$, $\KS \pip$ and $\KS \Km$ invariant mass distributions weighted by Legendre polynomial moments computed as functions of $\cos \theta_{\KS}$, $\cos\theta_{\Km}$ and $\cos\theta_{\pip}$, respectively, from the Dalitz plot analysis of the \etac decay to \kskpi using the QMI approach to the fit to \bkskkpip data.

\begin{figure}[tb]
\centering
\small
\includegraphics[width=0.6\textwidth]{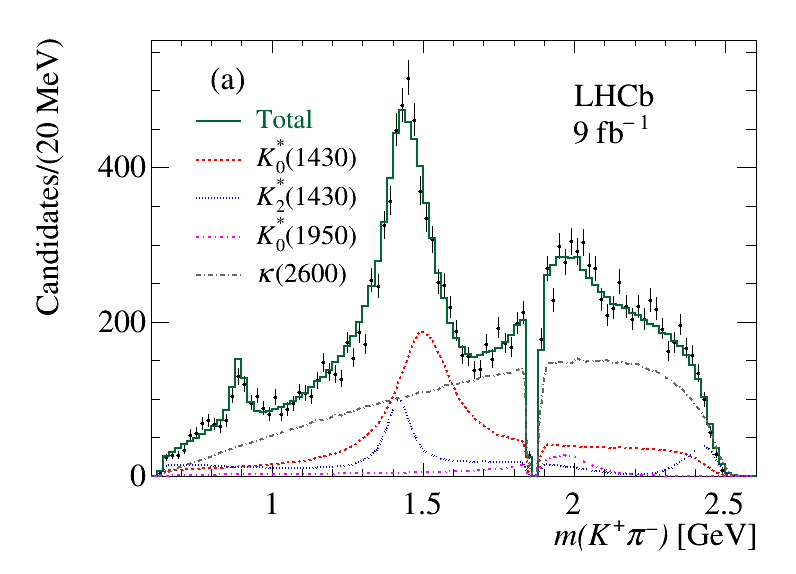}
\includegraphics[width=0.6\textwidth]{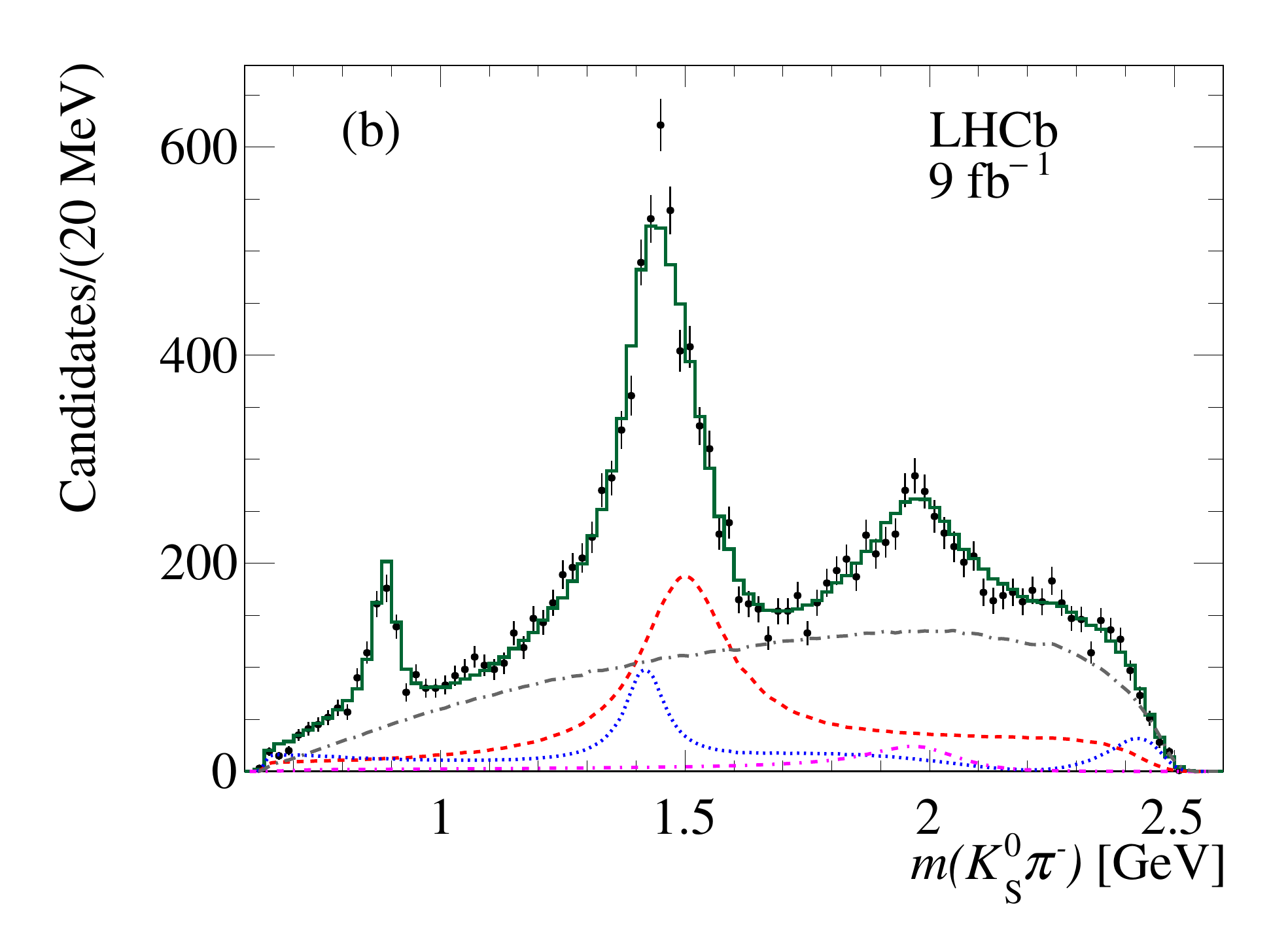}
\includegraphics[width=0.6\textwidth]{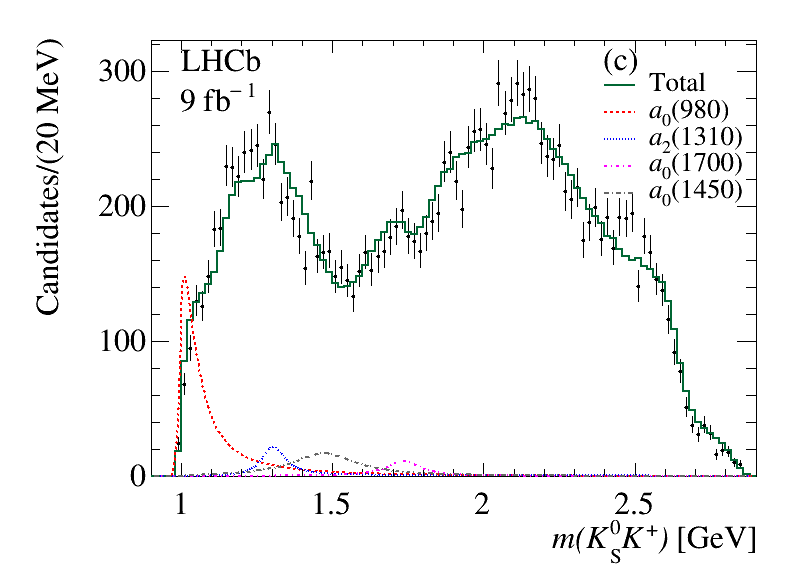}
\caption{\small\label{fig:fig29} Fit projections on the (a) $\Kp \pim$, (b) $\KS \pim$ and (c) $\KS \Kp$ invariant mass distributions from the Dalitz-plot analysis of the \etac decay using the isobar model for \bkskkpim data. The curves show the most important resonant contributions. 
To simplify the plots only resonant contributions relative to that mass projection are shown. The legend in (a) applies also to (b).}
\end{figure}

Figure~\ref{fig:fig29} shows the mass projections from the \etac Dalitz plot analysis using the QMI model from \bkskkpim 
data.

\clearpage

% Do not include this in any draft (just for information in the template)
%\input{acknowledgements_intro}
% Comment this in for paper drafts; do not include this in analysis note, conference and figure reports
\section*{Acknowledgements}
%
% These Acknowledgements valid from 3-May-2019
%
\noindent We express our gratitude to our colleagues in the CERN
accelerator departments for the excellent performance of the LHC. We
thank the technical and administrative staff at the LHCb
institutes.
We acknowledge support from CERN and from the national agencies:
CAPES, CNPq, FAPERJ and FINEP (Brazil); 
MOST and NSFC (China); 
CNRS/IN2P3 (France); 
BMBF, DFG and MPG (Germany); 
INFN (Italy); 
NWO (Netherlands); 
MNiSW and NCN (Poland); 
MEN/IFA (Romania); 
%MSHE (Russia); 
MICINN (Spain); 
SNSF and SER (Switzerland); 
NASU (Ukraine); 
STFC (United Kingdom); 
DOE NP and NSF (USA).
We acknowledge the computing resources that are provided by CERN, IN2P3
(France), KIT and DESY (Germany), INFN (Italy), SURF (Netherlands),
PIC (Spain), GridPP (United Kingdom), 
%RRCKI and Yandex LLC (Russia), 
CSCS (Switzerland), IFIN-HH (Romania), CBPF (Brazil),
Polish WLCG  (Poland) and NERSC (USA).
We are indebted to the communities behind the multiple open-source
software packages on which we depend.
Individual groups or members have received support from
ARC and ARDC (Australia);
Minciencias (Colombia);
AvH Foundation (Germany);
EPLANET, Marie Sk\l{}odowska-Curie Actions and ERC (European Union);
A*MIDEX, ANR, IPhU and Labex P2IO, and R\'{e}gion Auvergne-Rh\^{o}ne-Alpes (France);
Key Research Program of Frontier Sciences of CAS, CAS PIFI, CAS CCEPP, 
Fundamental Research Funds for the Central Universities, 
and Sci. \& Tech. Program of Guangzhou (China);
%Key Research Program of Frontier Sciences of CAS, CAS PIFI,
%Thousand Talents Program, and Sci. \& Tech. Program of Guangzhou (China);
%RFBR, RSF and Yandex LLC (Russia);
GVA, XuntaGal, GENCAT and Prog.~Atracci\'on Talento, CM (Spain);
SRC (Sweden);
the Leverhulme Trust, the Royal Society
 and UKRI (United Kingdom).

%\input{supplementary}

%\input{appendix}

% This should be taken out in the final paper
%\input{supplementary-app}

\addcontentsline{toc}{section}{References}
%\setboolean{inbibliography}{true}
\bibliographystyle{LHCb}
\bibliography{main,standard,LHCb-PAPER,LHCb-CONF,LHCb-DP,LHCb-TDR}

\newpage
% LHCb collaboration author list
% Data extracted on April 28th, 2023 at 10:08am for paper reference LHCb-PAPER-2022-051
\centerline
{\large\bf LHCb collaboration}
\begin
{flushleft}
\small
R.~Aaij$^{32}$\lhcborcid{0000-0003-0533-1952},
A.S.W.~Abdelmotteleb$^{50}$\lhcborcid{0000-0001-7905-0542},
C.~Abellan~Beteta$^{44}$,
F.~Abudin{\'e}n$^{50}$\lhcborcid{0000-0002-6737-3528},
T.~Ackernley$^{54}$\lhcborcid{0000-0002-5951-3498},
B.~Adeva$^{40}$\lhcborcid{0000-0001-9756-3712},
M.~Adinolfi$^{48}$\lhcborcid{0000-0002-1326-1264},
P.~Adlarson$^{77}$\lhcborcid{0000-0001-6280-3851},
H.~Afsharnia$^{9}$,
C.~Agapopoulou$^{13}$\lhcborcid{0000-0002-2368-0147},
C.A.~Aidala$^{78}$\lhcborcid{0000-0001-9540-4988}
% ...

{\footnotesize \it

% ...
$^{9}$Universit{\'e} Clermont Auvergne, CNRS/IN2P3, LPC, Clermont-Ferrand, France\\
% ...
$^{13}$LPNHE, Sorbonne Universit{\'e}, Paris Diderot Sorbonne Paris Cit{\'e}, CNRS/IN2P3, Paris, France\\
% ...
$^{32}$Nikhef National Institute for Subatomic Physics, Amsterdam, Netherlands\\
% ...
$^{40}$Instituto Galego de F{\'\i}sica de Altas Enerx{\'\i}as (IGFAE), Universidade de Santiago de Compostela, Santiago de Compostela, Spain\\
% ...
$^{44}$Physik-Institut, Universit{\"a}t Z{\"u}rich, Z{\"u}rich, Switzerland\\
% ...
$^{48}$H.H. Wills Physics Laboratory, University of Bristol, Bristol, United Kingdom\\
% ...
$^{50}$Department of Physics, University of Warwick, Coventry, United Kingdom\\
% ...
$^{53}$School of Physics and Astronomy, University of Glasgow, Glasgow, United Kingdom\\
$^{54}$Oliver Lodge Laboratory, University of Liverpool, Liverpool, United Kingdom\\
% ...
$^{62}$Syracuse University, Syracuse, NY, United States\\
% ...
$^{77}$Department of Physics and Astronomy, Uppsala University, Uppsala, Sweden, associated to $^{53}$\\
$^{78}$University of Michigan, Ann Arbor, MI, United States, associated to $^{62}$\\
% ...
}
\end{flushleft}

\end{document}